\documentclass[twocolumn,  prx]{revtex4} 

\usepackage{amsmath}    
\usepackage{graphicx}   
\usepackage{verbatim}   
\usepackage{color}      
\usepackage{subfigure}  
\usepackage{hyperref}   
\usepackage{makecell}   
\usepackage{textgreek}  
\usepackage{rotating}

\graphicspath{{afig/}{fig/}{tex_fig/}}

\newcommand{\db}{\mathbf{m}} 
\newcommand{\Mn}{\text{Mn}}  
\newcommand{\Pe}{\text{Pe}}  
\newcommand{\LR}{\text{LR}}  

\begin{document}

\title{On the Apparent Yield Stress in Non-Brownian Magnetorheological Fluids}
\author{Daniel V{\aa}gberg}
\affiliation{Delft University of Technology, Process \& Energy Laboratory, Leeghwaterstraat 44, 2628 CA Delft, The Netherlands}
\author{Brian P. Tighe}
\affiliation{Delft University of Technology, Process \& Energy Laboratory, Leeghwaterstraat 44, 2628 CA Delft, The Netherlands}

\date{\today}

\begin{abstract}
We use simulations to probe the flow properties of dense two-dimensional magnetorheological fluids. Prior results from both experiments and simulations report that the shear stress $\sigma$ scales with strain rate $\dot \gamma$ as $\sigma \sim \dot \gamma^{1-\Delta}$, with values of the exponent ranging between $2/3 <\Delta \le 1$. However it remains unclear what properties of the system select the value of $\Delta$, and in particular under what conditions the system displays a yield stress ($\Delta = 1$). To address these questions, we perform simulations of a minimalistic model system in which particles interact via long ranged magnetic dipole forces, finite ranged elastic repulsion, and viscous damping. We find a surprising dependence of the apparent exponent $\Delta$ on the form of the viscous force law. For experimentally relevant values of the volume fraction $\phi$ and the dimensionless Mason number $\Mn$ (which quantifies the competition between viscous and magnetic stresses),  models using a Stokes-like drag force show $\Delta \approx 0.75$ and no apparent yield stress. When dissipation occurs at the contact, however, a clear yield stress plateau is evident in the steady state flow curves. In either case, increasing $\phi$ towards the jamming transition suffices to induce a yield stress. We relate these qualitatively distinct flow curves  to clustering mechanisms at the particle scale. For Stokes-like drag, the system builds up anisotropic, chain-like clusters as $\Mn$ tends to zero (vanishing strain rate and/or high field strength). For contact damping, by contrast, there is a second clustering mechanism due to inelastic collisions. 
\end{abstract}

\maketitle

\begin{figure}[b!]
 \begin{center}
  \includegraphics[scale=0.45] {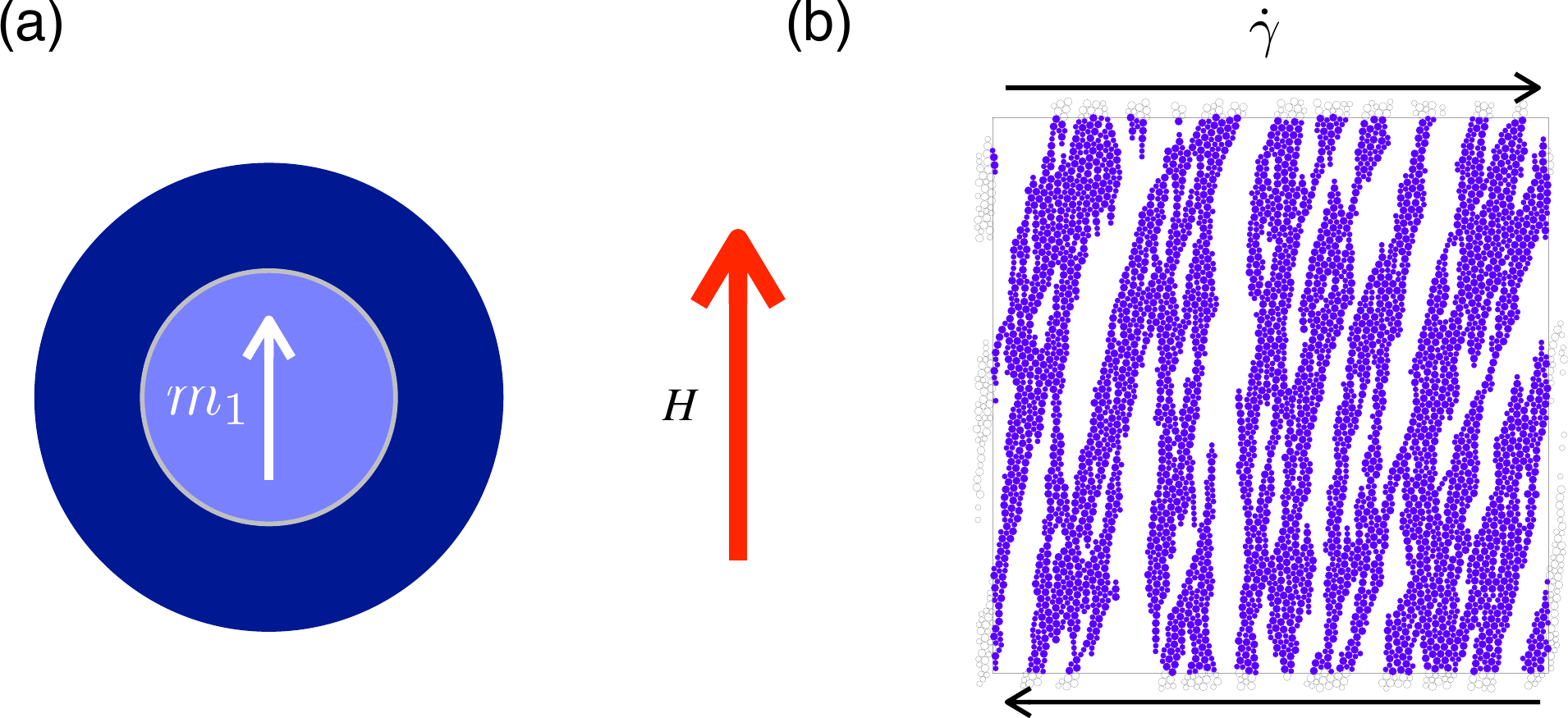}
 \end{center}
  \caption{(a) Particles have a magnetic core in side a non-magnetic shell. The core develops a dipole moment $m_1$ (white arrow) in the presence of a magnetic field $H$. The shell resists deformation elastically. (b) Snapshot of shear flow under Lees-Edwards periodic boundary conditions. Shear is applied transverse to the applied field.  }
  \label{fig:particle}
\end{figure}

Magnetorheological (MR) fluids consist of magnetizable particles suspended in a viscous carrier fluid. An external magnetic field $\mathbf{H}$
induces magnetic moments in the particles, which then rearrange to form chain-like structures, as illustrated in Fig.~\ref{fig:particle}. Chain formation dramatically enhances the stress $\sigma$ needed to maintain a strain rate $\dot \gamma$, and by varying $\mathbf{H}$ it is possible to tune the viscosity of the suspension, with applications to damping and switching. 
An excellent introduction to the fundamental physics and engineering applications of MR fluids can be found in recent review articles by Vicente et al. \cite{vicente2011review} and Ghaffari et al. \cite{ghaffari2015review} and references therein. Here we numerically study non-Brownian MR fluids in steady shear flow. 

Steady state rheology is commonly characterized in terms of the enhancement of the shear viscosity $\eta={\sigma}/{\dot\gamma}$ over its value $\eta_0$ at zero field. The ratio $\eta/\eta_0$ is governed by the dimensionless Mason number $\Mn \propto \dot \gamma/H^2$ (discussed in detail below), which quantifies the relative strengths of viscous and magnetic stresses in the system; magnetic interactions dominate when $\Mn$ tends to zero. In practice, the empirical fitting function
\begin{equation}
  \frac{\eta}{\eta_0}=1+\left( \frac{\Mn^\ast}{\Mn}\right)^{\Delta} 
\label{eqn:enhance}
\end{equation}
is often found to give a good description of the viscosity enhancement in MR fluids. Here $\Mn^\ast$ is a function of the volume fraction $\phi$; it vanishes as $\phi \rightarrow 0$ and determines the crossover between the Newtonian flow regime $\eta/\eta_0 \sim 1$ at high $\Mn$ and the magnetically dominated regime $\eta/\eta_0 \sim \Mn^{-\Delta}$ at low $\Mn$. The exponent $\Delta$ controls the rate at which viscosity diverges as the Mason number decreases. 

The value $\Delta = 1$ is an important reference case, as Eq.~(\ref{eqn:enhance}) is then equivalent to the flow curve of a Bingham plastic
\begin{equation}
\sigma(\dot \gamma) = \sigma_y + A \dot \gamma \,.
\label{eqn:bingham}
\end{equation}
The Bingham plastic has a nonzero dynamic yield stress $\sigma_y$, defined here as the  asymptote of the steady state flow curve $\sigma(\dot \gamma)$ in the limit $\dot \gamma \rightarrow 0$ (henceforth ``the yield stress''). 
Experiments are of course performed at small but finite strain rates, hence in practice the yield stress is also identified with an apparent plateau in the flow curve at the lowest accessible rates; i.e.~one assumes the plateau persists to asymptotically low strain rates.
If instead $\Delta < 1$, then the system has no yield stress and the stress vanishes slowly with the strain rate, $\sigma \sim \dot \gamma^{1-\Delta}$ as $\dot \gamma \rightarrow 0$.
Competing theoretical descriptions predict exponents $\Delta = 1$ \cite{marshall1989,klingenberg1990model,martin1996,vicente2004,gans1999nonlinear} and $\Delta = 2/3$ \cite{halsey1992}; for a discussion of the different models see \cite{bossis2002,ruiz2016,ghaffari2015review}. Numerous experimental and numerical studies have measured $\Delta$ values throughout this range in a number of magnetorheological (and electrorheological) systems under varying conditions; a summary of their results is given in table \ref{tab:previous_work}. 
Thermal motion has been shown to give $\Delta<1$ both in experiment \cite{gans1999nonlinear} and simulation \cite{baxter1996brownian}. However there is no effective way to predict {\em a priori} whether a given non-Brownian MR fluid will display a yield stress.
In the present work we address the presence or absence of a yield stress in a two-dimensional numerical model of athermal MR fluids in which energy is dissipated via viscous forces, in the absence of Coulomb friction. 

We model composite particles with a core shell structure as seen in Fig \ref{fig:particle}a.   
Similar core-shell structures have been used in experiment to change the surface properties of particles and to lower the effective density of the particles in order to avoid sedimentation \cite{you2007,choi2007,ko2009,fang2010,fang2009}. The core shell structure is also suitable to model the recent experiments of  \cite{cox2016}, which bridge the gap between conventional magnetic suspensions and amorphous magnetic solids. 

We consider two types of damping. The first, which we denote reservoir damping (RD) in accord with the terminology of Ref.~\cite{vagberg2014dissipation}, is a Stokesian drag with respect to the carrier fluid. The second, contact damping (CD), is applied to the relative velocity of particles in contact. Removing friction gives us a cleaner system that helps us to understand the underlying physics. We will ultimately argue that contact damping provides insight into the case of frictional contacts. The use of RD and CD models also allows us to make contact with the extensive literature \cite{durian1995,liu1998,ohern2003,olsson2007,heussinger2009,hatano09,tighe2010,tighe2011,ikeda2012,boschan2016,baumgarten2017} on the rheology of yield stress fluids close to the jamming transition.

Our main finding is that the form of the viscous force law has a dramatic influence on the viscosity enhancement in the magnetically dominated regime, as characterized by the exponent $\Delta$. For reservoir damping we find no evidence of a yield stress over a wide range in $\phi$ and for Mason numbers as low as $10^{-6}$; instead the exponent $\Delta \approx 0.75$ gives an excellent description of the rheology. In sharp contrast, for contact damping there is a clear nonzero yield stress in the same range of Mason numbers, and so $\Delta = 1$. We relate this difference to clusters that form in the CD model at intermediate $\Mn$ due to inelastic collisions between particles, an effect that is absent in the RD model.

We further investigate the role of finite size effects and volume fraction, both of which we find to promote the emergence of a yield stress.

\begin{table}[tb]
  \caption{Previous work determining the exponent $\Delta$. These works have been performed under a wide variety of conditions, including variations in the type of particles, carrier fluid, and system geometry. For details about the specific parameters used in each experiment/simulation, readers are referred to the original articles.           }
  \centering
  \begin{tabular}{|l|c|c|}
    \hline
     Authors & Type & $\Delta$ \\ 
    \hline
       Marshall et al. \cite{marshall1989} & \makecell{experiment\\ER-fluid} & 1 \\ 
    \hline
      Halsey et al. \cite{halsey1992} & \makecell{experiment \\ER-fluid} &  $0.68$ -- $0.93$  \\
    \hline
       Felt et al. \cite{felt1996} & \makecell{experiment\\MR-fluid} & $0.74$ -- $0.83$ \\
    \hline
       Martin et al. \cite{martin1994} & \makecell{experiment\\ER-fluid} & $0.67$ \\ 
    \hline
      de Gans et al. \cite{gans1999nonlinear} & \makecell{experiment\\ inverse MR-fluid } & $0.8$ -- $0.9$ \\
    \hline
      de Gans et al. \cite{gans2000size} & \makecell{experiment\\ inverse MR-fluid } & $0.94\pm0.02$ \\
    \hline
      Volkova et al. \cite{volkova2000} & \makecell{experiment \\ a) MR-fluid \\b) inverse MR-fluid} &  \makecell{a) $0.86$ -- $0.97$ \\b) $0.74$ -- $0.87$} \\ 
    \hline
      Sherman et al. \cite{sherman2015} & \makecell{experiment \\MR-fluid} & 1 \\
    \hline 
      Bonnecaze and Brady \cite{bonnecaze1992dynamic} &  \makecell{simulation 2D} & 1 \\ 
    \hline
      Melrose \cite{melrose1992brownian} &
      simulation 3D &
      $0.8\pm0.05$  \\  
    \hline      
  \end{tabular}

  \label{tab:previous_work}
\end{table}

\begin{figure*}[tbp]
\begin{center}
\includegraphics[scale=0.5,angle=90,origin=bl,clip=true] {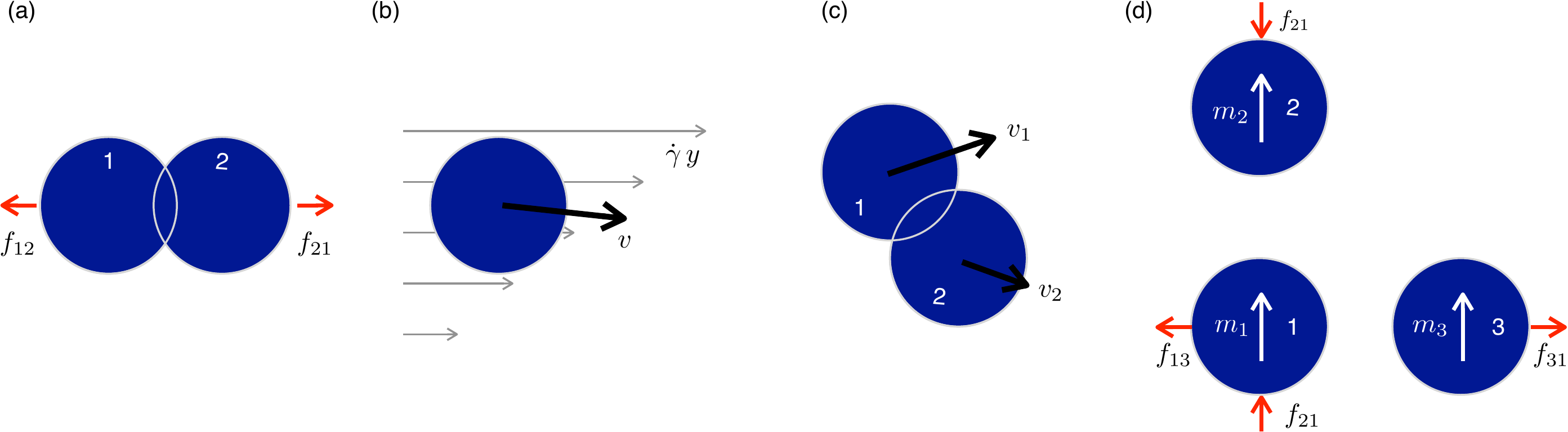}
\end{center}
  \caption{Elastic, viscous, and magnetic forces. (a) Particles experience repulsive elastic forces $\vec f_{ij} = -\vec f_{ji}$ proportional to their overlap.  (b) In the reservoir damping (RD) model, particles experience a Stokes drag force $\vec f \propto (\vec v - \dot \gamma y)$ with respect to a solvent that is assumed to have an affine velocity profile (gray arrows).  (c) In the contact damping (CD) model, particles in contact experience a viscous force $\vec f_{ij} \propto (\vec v_j - \vec v_i)$ opposed to their relative velocity. (d) Particles experience long range magnetic forces (red arrows) that are attractive when induced dipoles align end-to-end (particles 1 and 2), and repulsive when they are adjacent (1 and 3).}
  \label{fig:forces}
\end{figure*}

\section{Model}
The system comprises $N$ spherical particles confined to a plane. The particle distribution is a 50-50 bidisperse mixture with size ratio $1.4:1$. Each particle consists of an elastic non-magnetic outer layer and a hard inner core of a magnetically soft permeable material (Fig.~\ref{fig:particle}a). The diameter of the core is $d_i/2$, where $d_i$ is the diameter of particle $i$. The mass of each particle $m_i$ is directly proportional to its volume $V_i$ such that $m_i=V_i\rho$, where $\rho=1$ is the density of the material. For simplicity we assume the density is constant throughout the particle.   
We assume the particles are large enough so that thermal motion can be neglected and that there is no static friction.

The motion of each particle is governed by 
\begin{equation} \label{eq:eom}
  \ddot{\mathbf{r}}_i=\frac{1}{m_i} \left(\mathbf{f}^\text{\,e}_i+\mathbf{f}^\text{\,d}_i+\mathbf{f}^\text{\,m}_i\right) \,,
\end{equation}
where $\mathbf{r}_i$ is the position of particle $i$, $\mathbf{f}^\text{\,e}_i$ is the repulsive contact force, $\mathbf{f}^\text{\,d}_i$ is a dissipative force caused by the interaction between the particles and the surrounding liquid, and $\mathbf{f}^\text{\,m}_i$ is the magnetic dipole force. Since the particles are frictionless and do not have permanent dipole moments, there are no torques acting on the particles.

We drive the system by applying a uniform shear strain rate $\dot\gamma$ in the $\hat{\mathbf{x}}$ direction using Lees-Edwards boundary conditions\cite{evans1990}. The equations of motions were integrated using a Velocity-Verlet scheme modified to better handle dissipative forces \cite{vaagberg2017shear}.

The simulation is controlled by varying three parameters: the shear rate $\dot\gamma$, the external magnetic field $\mathbf{H}  = H \hat{\mathbf{y}}$ transverse to the flow direction, and the packing fraction 
\begin{equation}
\phi=\frac{1}{L^2}\sum_{i=1}^N\frac{\pi d_i^2}{4} \,,
\end{equation}
where $L$ is the side length of the quadratic simulation box.

\subsection{Interaction Forces}
Overlapping particles repel elastically (Fig.~\ref{fig:forces}a).
The elastic contact forces are given by the potential
\begin{equation}
 U_e(r_{ij})=\left\{ 
 \begin{array}{cl}
   \frac{k_e}{\alpha}\left(1-r_{ij}/d_{ij}\right)^\alpha &\mbox{ for $r_{ij}<d_{ij}$} \\
   0 &\mbox{ for $r_{ij}\ge d_{ij}$}
 \end{array} \right.
\label{eqn:elastic}
\end{equation}
where $r_{ij}=\left|\mathbf{r}_{ij}\right|=\left|\mathbf{r}_{i}-\mathbf{r}_{j}\right|$ is the distance between particle $i$ and $j$ and $d_{ij}$ is the sum of their radii. The constant $k_e=1$ sets the energy scale of the elastic interaction. For the parameter ranges studied here the particle overlaps are small, $d_{ij}-r_{ij}\ll d_{ij},$ so the contact interaction is limited to the outer shell;  it is therefore safe to neglect the particle core in the contact potential.

The potential (\ref{eqn:elastic}) produces an elastic force 
\begin{equation}
\mathbf{f}^{\text{e}}_i=-\sum_{j(i)} \frac{dU_e(r_{ij})}{dr_{ij}}\hat{\mathbf{r}}_{ij}=
-k_e\sum_{j(i)}\left(\frac{d_{ij}-r_{ij}}{d_{ij}^2}\right)^{\alpha-1}\hat{\mathbf{r}}_{ij} \,,
\end{equation}
where the sums run over the set of particles $j$ in contact with particle $i$. 
Using $\alpha=2$ gives the standard harmonic potential with corresponding force
\begin{equation}
\mathbf{f}^{\text{e}}_i=
-k_e\sum_{j(i)} \frac{d_{ij}-r_{ij}}{d_{ij}^2}\hat{\mathbf{r}}_{ij} \,.
\end{equation}

For the dissipative force $\mathbf{f}^{\text{d}}_i$ we use a viscous force proportional to the velocity difference between the particle velocity $\mathbf{v}_i$ and a reference velocity. We compare two different viscous dissipations (Fig.~\ref{fig:forces}b and c), by changing the definition of the reference velocity.

With the first viscous force law, which we denote reservoir dissipation (RD), the particle loses energy when moving relative to the carrier fluid.  We select the reference as $\mathbf{v}_\text{RD}=\dot\gamma y_i\hat{\mathbf{x}}$, the affine shear velocity. This gives
\begin{equation} \label{fRD}
\mathbf{f}^{\text{RD}}_i=-k_d\left(\mathbf{v}_i-\mathbf{v}_\text{RD}\right) \,,
\end{equation}
where the constant $k_d$ allows us to tune the strength of the dissipation. 

The second force law is a contact dissipation (CD), wherein the dissipation is proportional to the velocity difference of contacting particles
\begin{equation} \label{fCD}
  \mathbf{f}^{\text{\text{CD}}}_{ij}=-k_d\left(\mathbf{v}_i-\mathbf{v}_j\right) \,.
\end{equation}
To obtain the full dissipative force on particle $i$ one must sum over all particles $j$ in contact with $i$. 

We use $k_d=1$ for both the RD and CD dissipation. For RD this ensures the dynamics is overdamped for the studied parameter ranges. While the CD-dissipation is overdamped for contacting particles, it is highly sensitive to the average contact number and free particles do not dissipate energy. This mainly affects the behavior of dilute systems at high shear rates, which is not the limit we focus on here. 

The RD and CD force laws can be seen as two limiting cases: RD only considers the particle-carrier fluid interaction, while CD only considers the particle-particle interaction. The two force laws have been studied in detail for dense suspensions in the absence of dipole-interactions \cite{vagberg2014dissipation,vagberg2014universality,vaagberg2017shear}. In experimental systems both solvent and particle interactions affect the dissipation and a combination of CD- and RD-dissipation are usually needed to describe the behavior. Simulations are advantageous, in that they allow us to study these effects separately.

The magnetic interaction is modeled using point dipoles positioned at the center of each particle -- see Fig.~\ref{fig:forces}d. The dipole moments are induced in the particle core by the external field $\mathbf{H}$. 
The magnetic flux density $\mathbf{B}$ at a distance $r$ from a dipole $\mathbf{m}$ is given by
\begin{equation} \label{b_def}
  \mathbf{B}(\mathbf{r})=
  \frac{\mu_f}{4\pi}\left(\frac{3\mathbf{r}(\db\cdot\mathbf{r})}{r^5}
  -\frac{\mathbf{m}}{r^3} \right) \,,
\end{equation}
where $\mu_f$ is the permeability of the carrier fluid.
The potential energy between two dipoles $i$ and $j$ is given by 
\begin{equation} \label{b_pot_def}
  U_m(\mathbf{r}_{ij})=-\left(\mathbf{m}_j\cdot\mathbf{B}_i \right) \,,
\end{equation}
which gives the force 
\begin{equation} \label{fm_def}
  \mathbf{f}_{ij}^m=-\nabla U_m(r_{ij}) \,.
\end{equation}
Inserting \eqref{b_def} and \eqref{b_pot_def} into \eqref{fm_def} and evaluating gives 
the force from dipole $i$ acting on dipole $j$, 
\begin{align}
   \mathbf{f}_{ij}^m &= 
   \frac{3\mu_f}{4\pi r_{ij}^5}\Big[
   \db_i(\db_j\cdot\mathbf{r}_{ij})+
   \db_j(\db_i\cdot\mathbf{r}_{ij})  \nonumber \\
   & \,\,\,\,\,\,\, +
   \mathbf{r}_{ij}(\db_i\cdot\db_j)- 
   \frac{5}{r_{ij}^2}\mathbf{r}_{ij}
   (\db_i\cdot\mathbf{r}_{ij})
   (\db_j\cdot\mathbf{r}_{ij})
   \Big] \,.
\end{align}

The magnitude and direction of the induced dipole-moments are given by 
\begin{equation} 
  \db_i=V_{\rm ci}\mathbf{M}=V_{\rm ci}( 3\beta \mathbf{H} ) \,,
\end{equation}
where $V_{\rm ci}$ is the core-volume of particle $i$, and 
\begin{equation}
  \beta = \frac{\mu-1}{\mu+2} \,.
\end{equation}
Here $\mu=\mu_i/\mu_f =1000$ is the relative permeability and $\mu_i$ is the permeability of the core of particle $i$. The outer shell is assumed to have the same permeability as the carrier fluid. We  consider only direct induction from the external field, ignoring contributions form neighboring dipoles. This is justified by the core-shell structure of the particles, which keeps the magnetic cores separated. We refer to the appendix for a more detailed discussion.

\subsection{Stresses}

The shear stress $\sigma$ is a sum of four contributing terms
\begin{equation}
  \sigma=\sigma_e+\sigma_m+\sigma_d+\sigma_k \,.
\end{equation}
Each of the first three correspond to one of the forces in \eqref{eq:eom}. The additional term $\sigma_k$ is the kinetic stress.
In practice only $\sigma_e$ and $\sigma_m$ are important for the  rheology in the magnetically-dominated regime, as $\sigma_d$ and $\sigma_k$ are orders of magnitude lower and both go to zero in the quasistatic limit $\dot\gamma\rightarrow0$.

The first three stress terms are calculated by substituting $f_\circ$ with the corresponding force from equation \eqref{eq:eom} in the expression 
\begin{equation}
  \sigma_\circ = -\frac{1}{L^2}\sum_{i<j} r_{ijx} {f_\circ}_{ijy}
\end{equation}
Here the $x$ and $y$ subscripts indicate the $x$- and $y$-components of respective vector and $L$ is the length of the simulation box.
The kinetic stress $\sigma_k$ is calculated as 
\begin{equation}
  \sigma_k=-\frac{1}{L^2} \sum_i  v_{ix} v_{iy} m_i
\end{equation}
where $v_{ix}$ and $v_{iy}$ is the x- and y-component of the non-affine velocity of particle $i$.

\subsection{Dimensionless numbers}

Much of the observed rheology of MR-fluids can be described using four dimensionless numbers: the Mason number ($\Mn$), the Peclet number ($\Pe$),  Lambda ($\lambda$), and the volume fraction $\phi$. The first three characterize the relative strengths of magnetic, viscous, and thermal forces. As we consider non-Brownian particles, the Peclet number (viscous versus thermal forces) and Lambda (magnetic versus thermal forces) play no role in the present results. We are left with the  volume fraction and the
Mason number,  which vanishes when magnetic interactions dominate.

There is some flexibility when selecting the reference forces used to define the Mason number, which has led to competing conventions in the literature \cite{klingenberg2007}. 
We use microscopic properties to define $\Mn$. Assume there are two particles of the smaller species with diameter $d$ (core diameter $d/2$) placed at a distance $d$ such that their surfaces just touch. The dipole force  between these two particles when their dipole moments are aligned is $F_m=\frac{3\pi}{8}d^2\mu_f\beta^2H^2$.  
For reservoir damping the typical viscous force is $F_d=dk_d \dot\gamma$, while  for contact damping there is an additional dependence on the mean number of contacts per particle, $Z$, such that $F_d=Zdk_d\dot\gamma$. The Mason number  $\Mn \equiv F_d/F_m$ is therefore
\begin{equation}
  \Mn_{\text{RD}}=\frac{3\pi k_d \dot\gamma}{8d\mu_f\beta^2H^2}
\end{equation}
for the RD model and  
\begin{equation}
  \Mn_{\text{CD}}=Z\Mn_{\text{RD}}
\end{equation}
for the CD model. 

We report shear stresses in the dimensionless form
\begin{equation}
 \tilde\sigma=\frac{\sigma d^{D-3}}{\mu_f\beta H^2} \,.
\end{equation}
where $D$ is the dimensionality of the system. Because the presence or absence of a yield stress is a major focus of the present work, we present most rheological results in the form of a dimensionless flow curve, $\tilde{\sigma}(\Mn; \phi)$, as opposed to plotting the viscosity enhancement $\eta/\eta_0$. A yield stress is then clearly signaled by a plateau in $\tilde{\sigma}$ at low $\Mn$. When there is no yield stress, the stress vanishes as $\tilde \sigma \sim \Mn^{1-\Delta}$.

\subsection{Simulation}

The length of each simulation in total strain $\gamma$ varies from $\gamma=50$ at $\dot \gamma=0.05$ down to $\gamma=4$ at $\dot\gamma=10^{-8}$. Simulations are started  at high shear rates, and lower shear rate simulations are initialized using starting configurations obtained from the previous higher shear rate. In order to avoid transient effects the first 20\% of each run is discarded before calculating time-averaged quantities. For $N\ge1024$ we perform one simulation for each parameter value, while for $N=256$ and $N=64$ two, respectively five, independent runs are performed to improve statistics.  We study the parameter range $0.1\le \phi \le 0.86$, $10^{-8}\le\dot\gamma \le 10^{-1}$, $10^{-4}\le H \le 10^{-1}$ and $64 \le N\le 16384$, which allows us to probe Mason numbers in a window spanning 12 orders of magnitude for $N=256$ and 10 orders of magnitude for $N=4096$. Consequently, we cover a larger parameter space than any of the works referenced in table \ref{tab:previous_work}. 

For this work we are especially interested in the behavior at low Mason numbers. At $N=4096$ our lowest Mason number is $\Mn=5\times10^{-7}$, which is significantly lower than the the lowest values accessed by any of the simulations in table \ref{tab:previous_work} and comparable to or slightly lower than the lowest values accessed in experiment \cite{marshall1989,felt1996,martin1994,gans1999nonlinear}.

\begin{figure}[tb]
 \begin{center}
  \subfigure{\hspace{-0.2cm}\includegraphics[height=3.7cm] {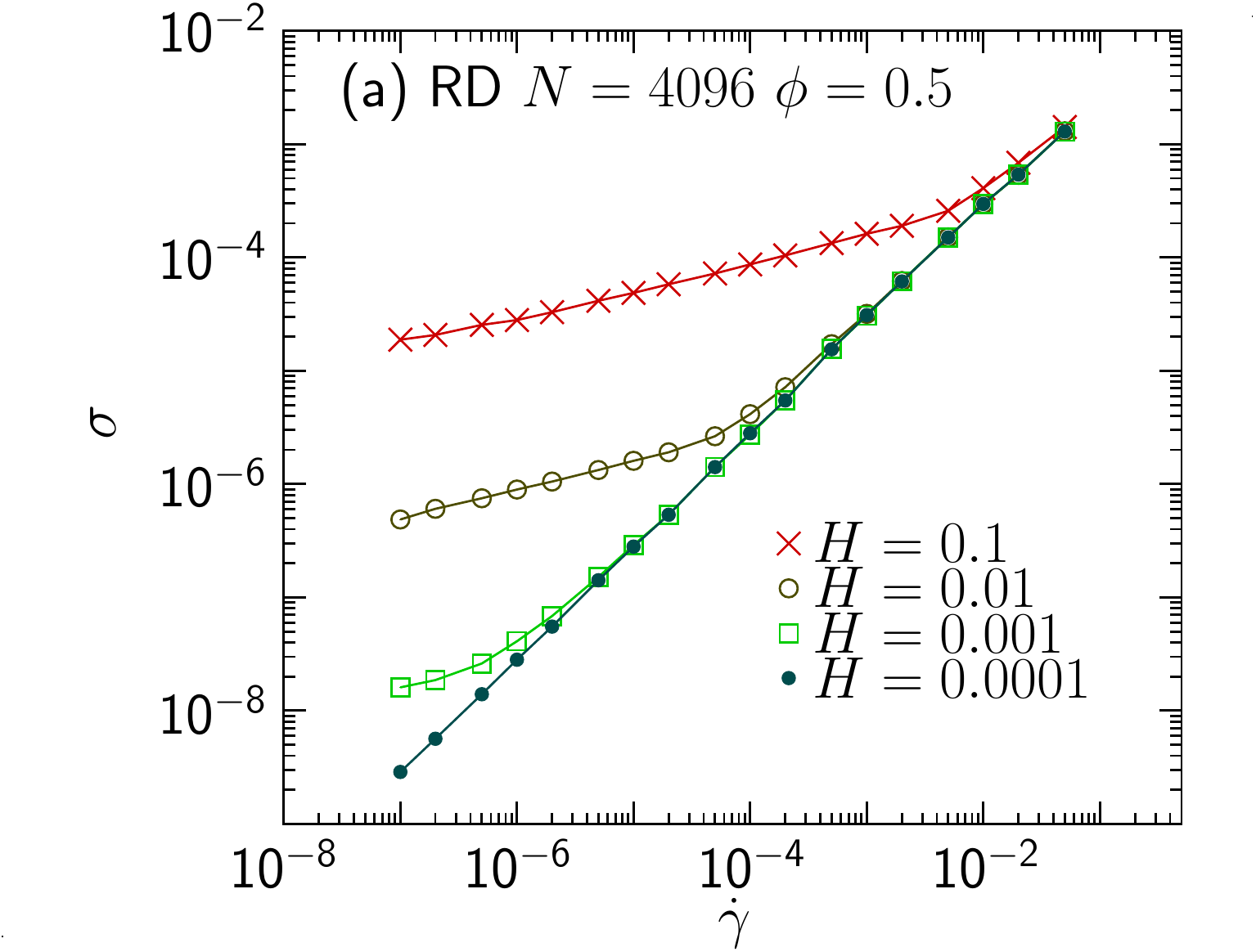}
             \hspace{-1.48cm}\includegraphics[height=3.7cm] {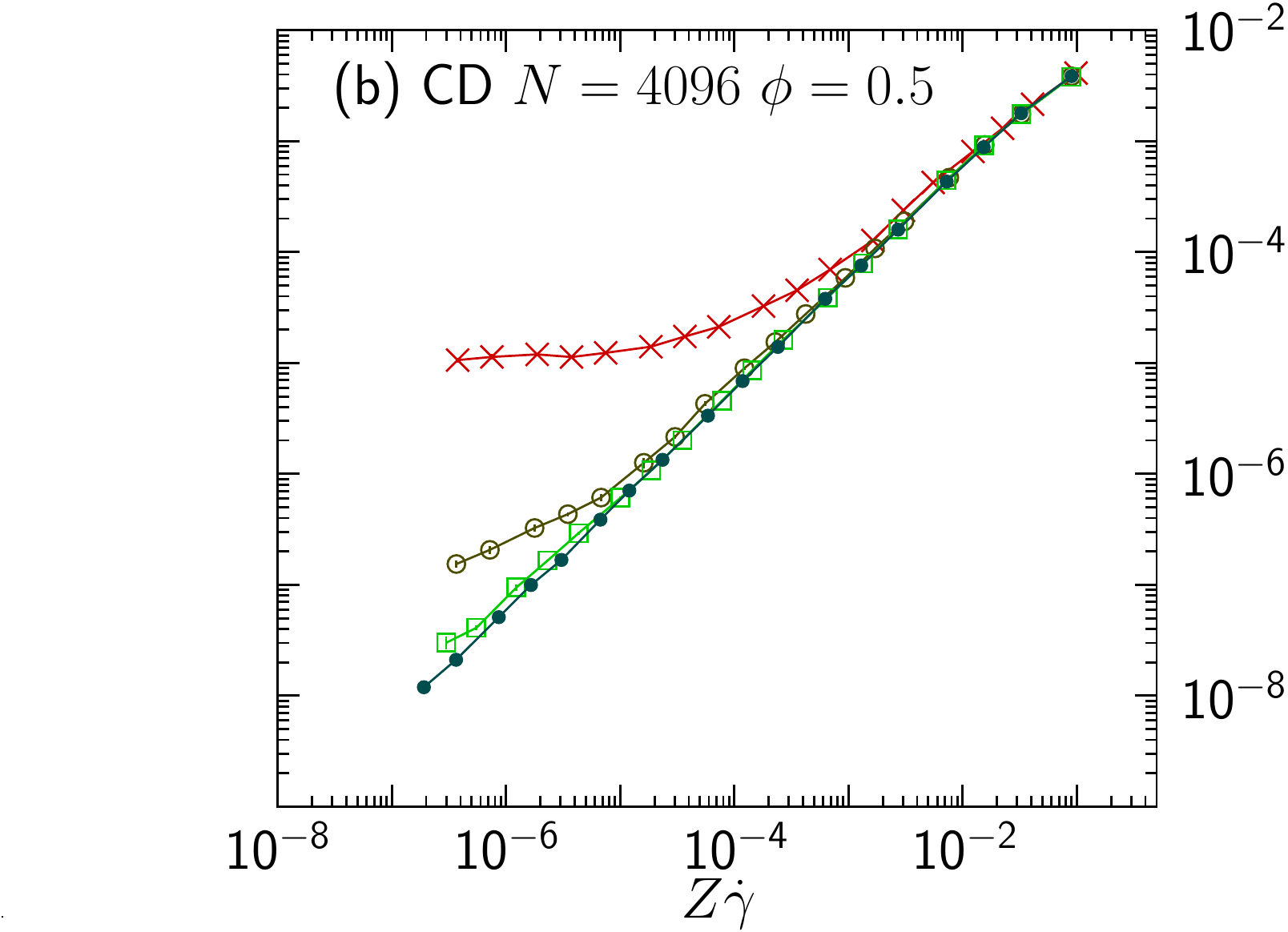}}

  \subfigure{\hspace{-0.2cm}\includegraphics[height=3.7cm] {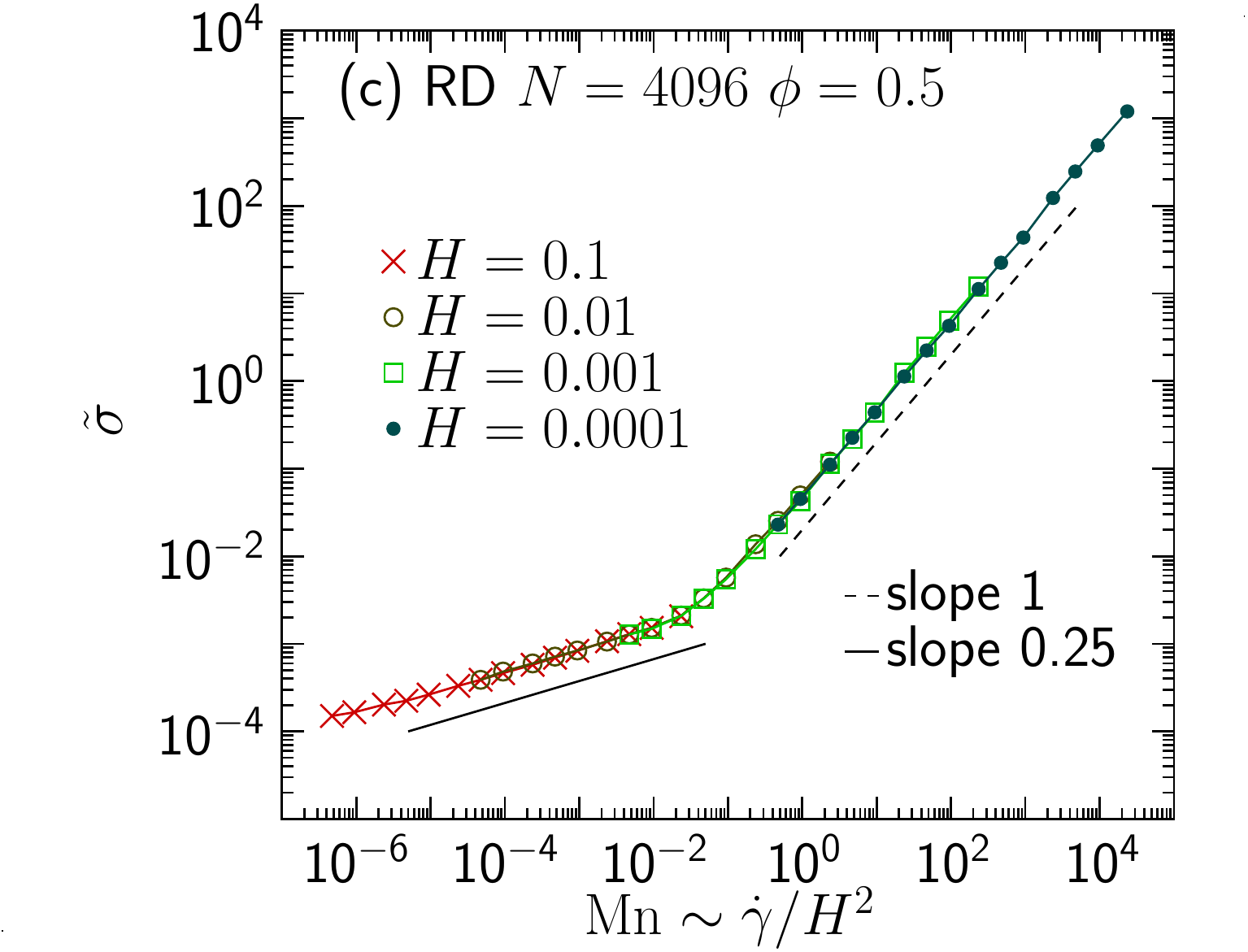}
             \hspace{-1.48cm}\includegraphics[height=3.7cm] {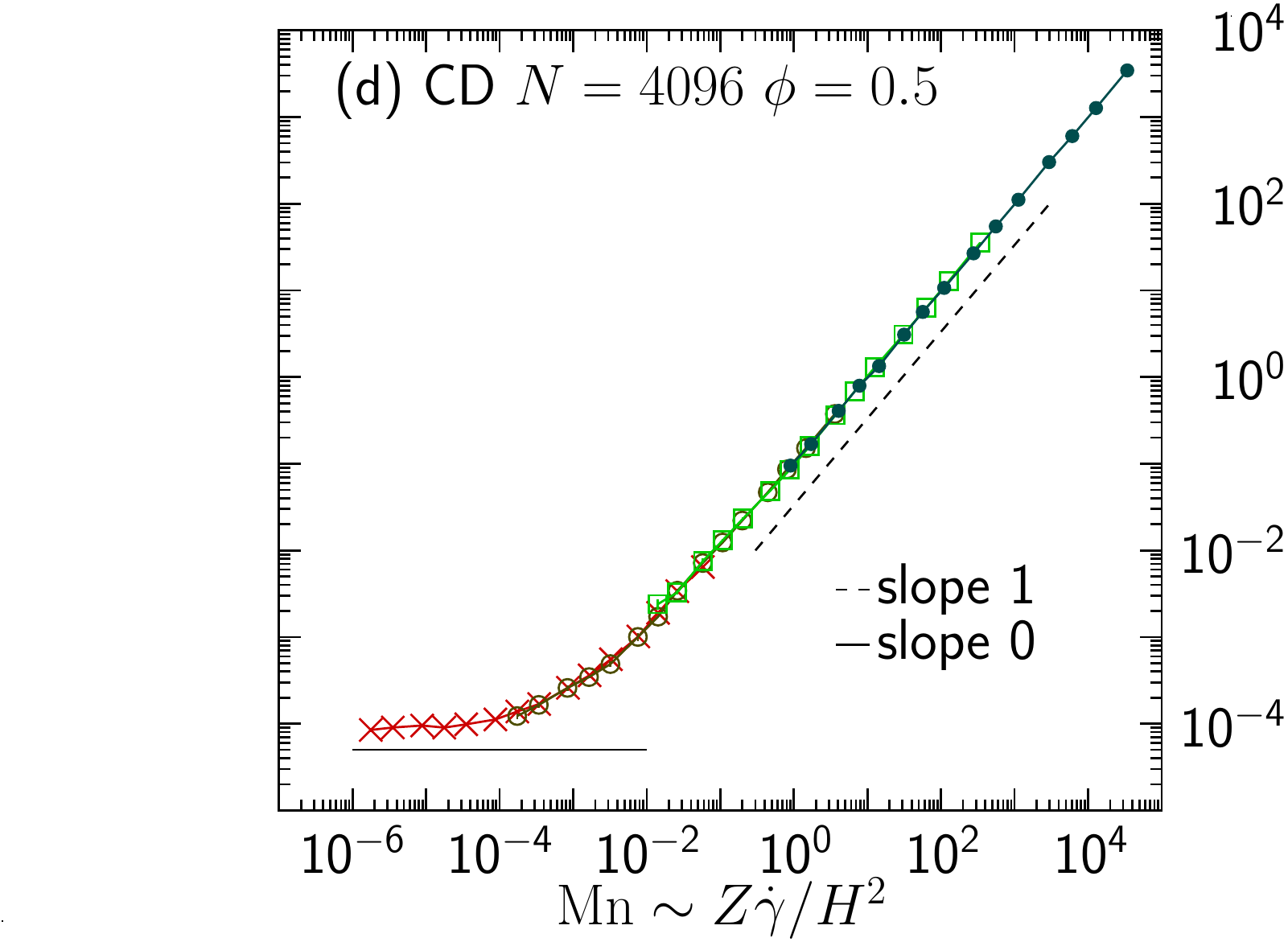}}

  \subfigure{\hspace{-0.2cm}\includegraphics[height=3.7cm] {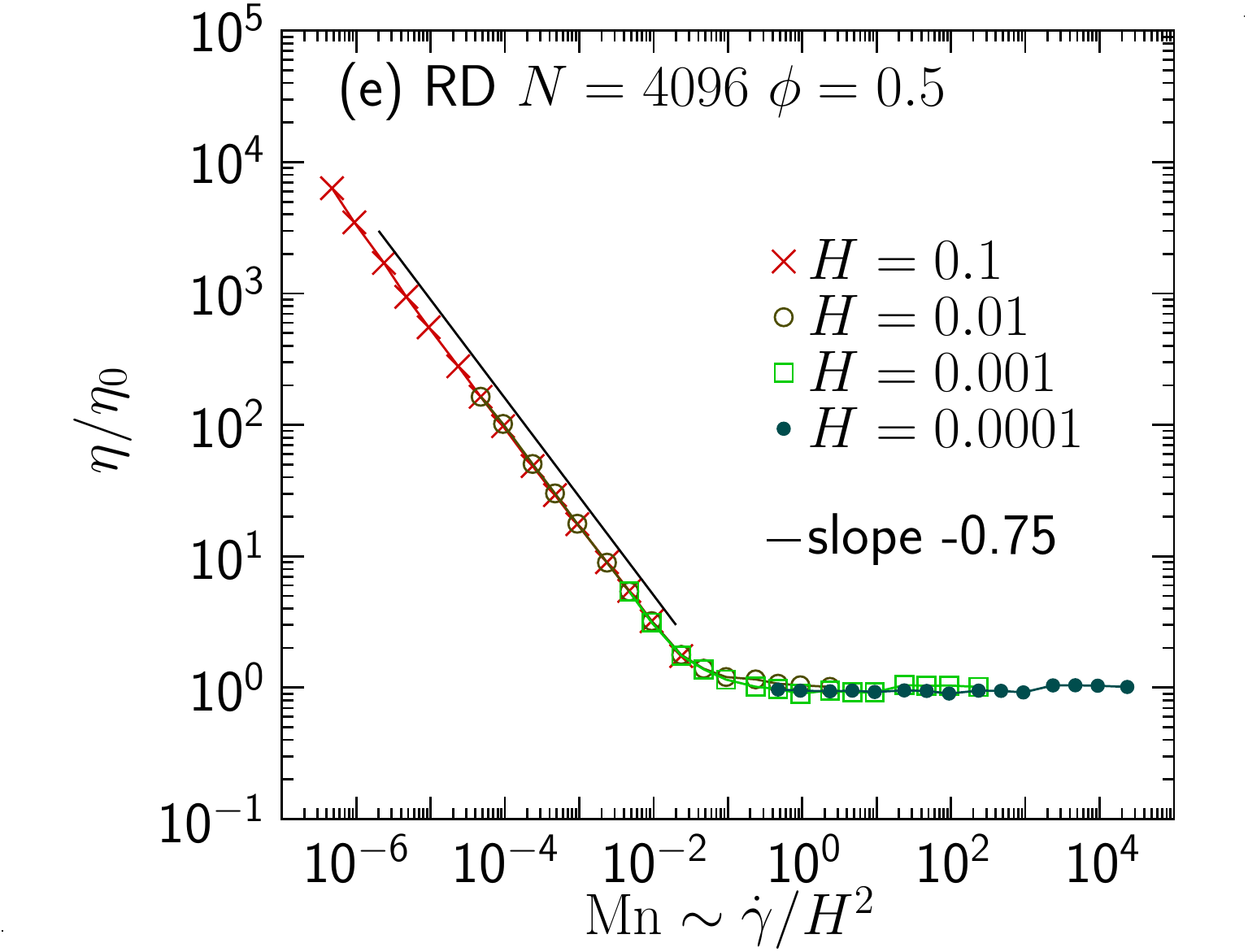}
             \hspace{-1.48cm}\includegraphics[height=3.7cm] {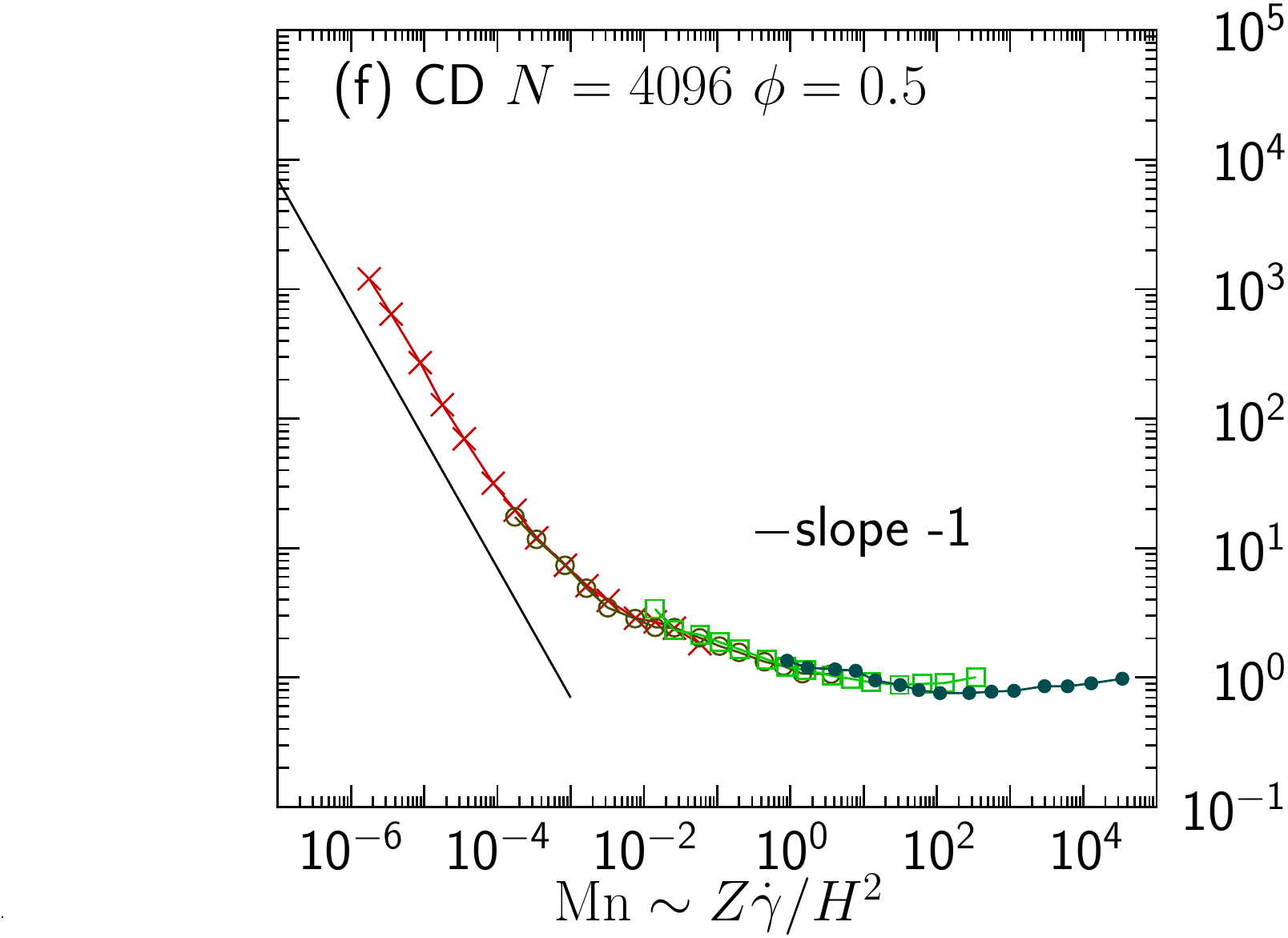}}
 \end{center}
  \caption{Flow curves for the reservoir damping (RD, left column) and contact damping (CD, right column) models. Top row: Shear stress $\sigma$ versus shear rate $\dot\gamma$ for varying field strength $\mathbf{H}$ at fixed packing fraction $\phi$.  Middle row: Data from the top row rescaled using dimensionless shear stress $\tilde \sigma$ and Mason number $\Mn$. Bottom row: The same data replotted in terms of the viscosity enhancement viscosity $\eta/\eta_0$.  }
  \label{fig:s_gdot_4krdcd}
\end{figure}

\section{Bulk rheology}
We start by considering the bulk rheological properties of the RD and CD models, with emphasis on the form of their steady state flow curves.

Fig.~\ref{fig:s_gdot_4krdcd} compares the rheology of the RD and CD models and its dependence on $\dot\gamma$ and $\mathbf{H}$ at fixed $\phi=0.5$ and $N=4096$. We first consider rheology of the RD model, shown in the left column of Fig.~\ref{fig:s_gdot_4krdcd}.  From top to bottom we plot the same data set as dimensionful flow curves, dimensionless flow curves, and in terms of the viscosity enhancement, respectively.
The dimensionless data displays excellent collapse to a master curve that  exhibits two flow regimes: a Newtonian regime, $\sigma\sim \Mn$, at high Mason numbers, and a magnetically dominated regime at low Mason numbers. It is clear that the RD model does not exhibit a  yield stress over the accessible range of $\Mn$; instead we find $\Delta\approx0.75$ in the magnetically-dominated regime.
The corresponding panels for the CD model (Fig.~\ref{fig:s_gdot_4krdcd}, right column) display a striking difference. There are again two flow regimes, but in this case there is a more gradual crossover to a yield stress in the limit of low $\Mn$, hence $\Delta = 1$.

\begin{figure}[tb]
 \begin{center}
   \subfigure{\hspace{-0.02cm}\includegraphics[height=3.7cm]{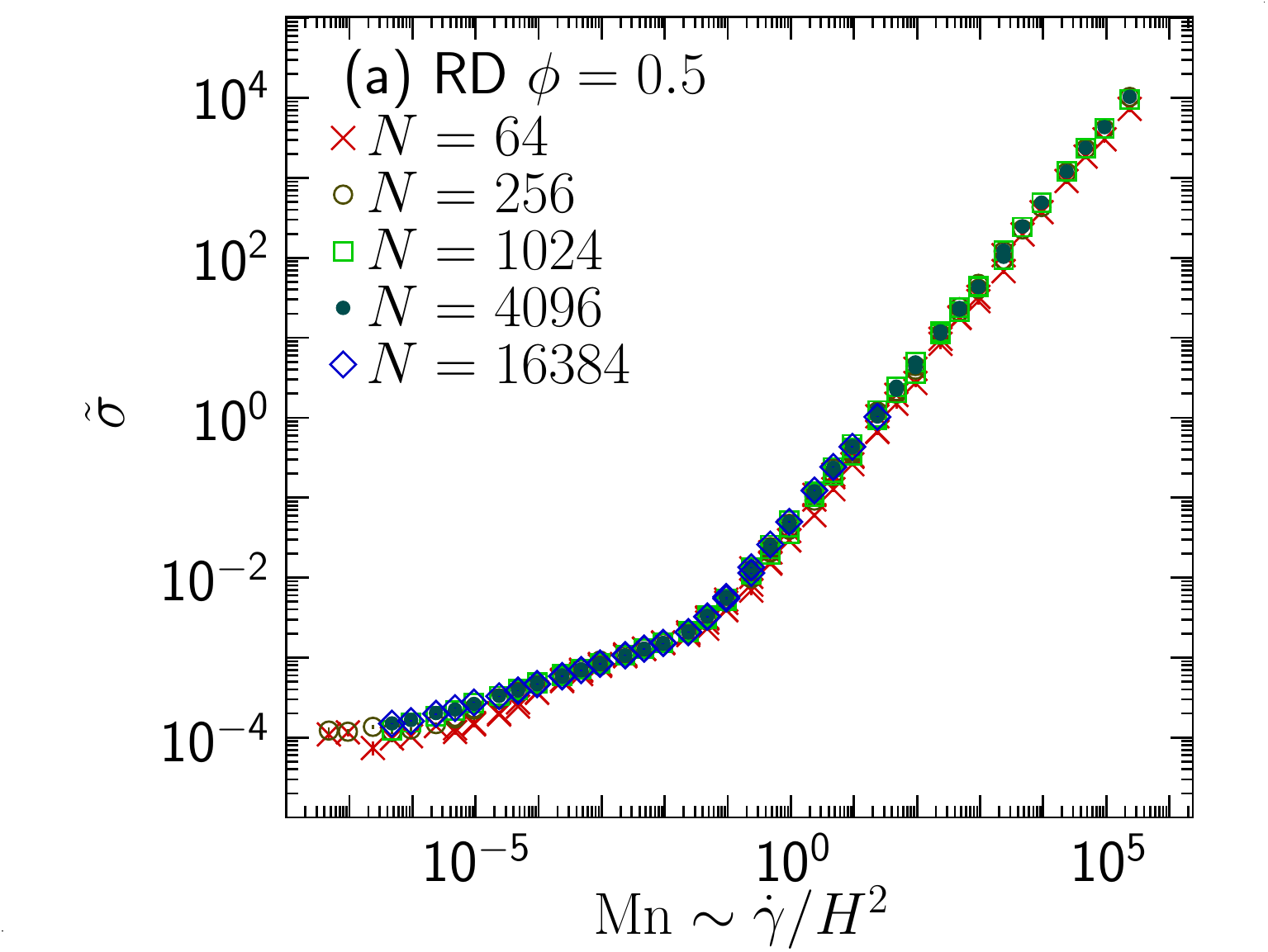}
              \hspace{-1.5cm}\includegraphics[height=3.7cm]{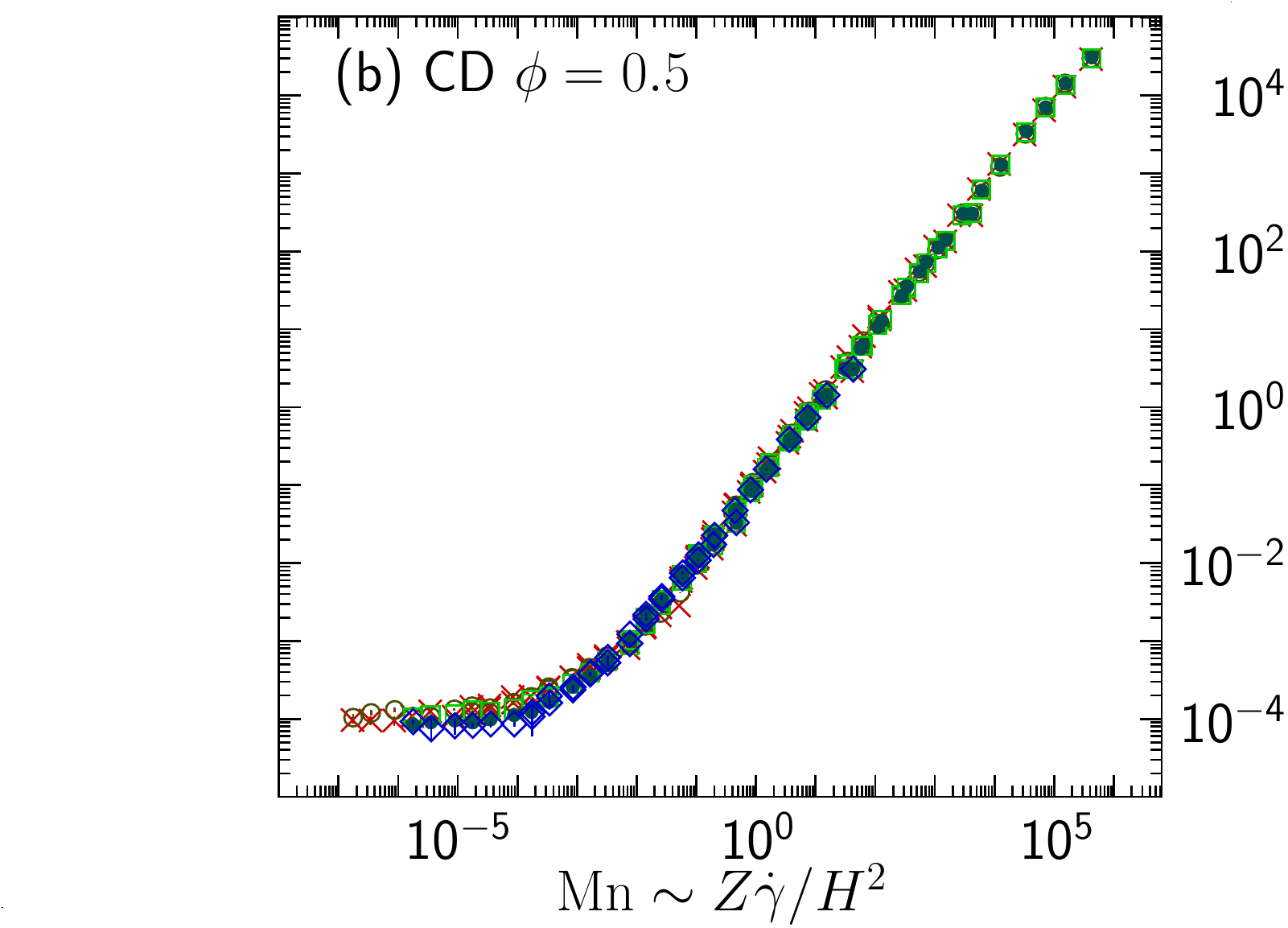}}
   
   \subfigure{\hspace{-0.02cm}\includegraphics[height=3.7cm]{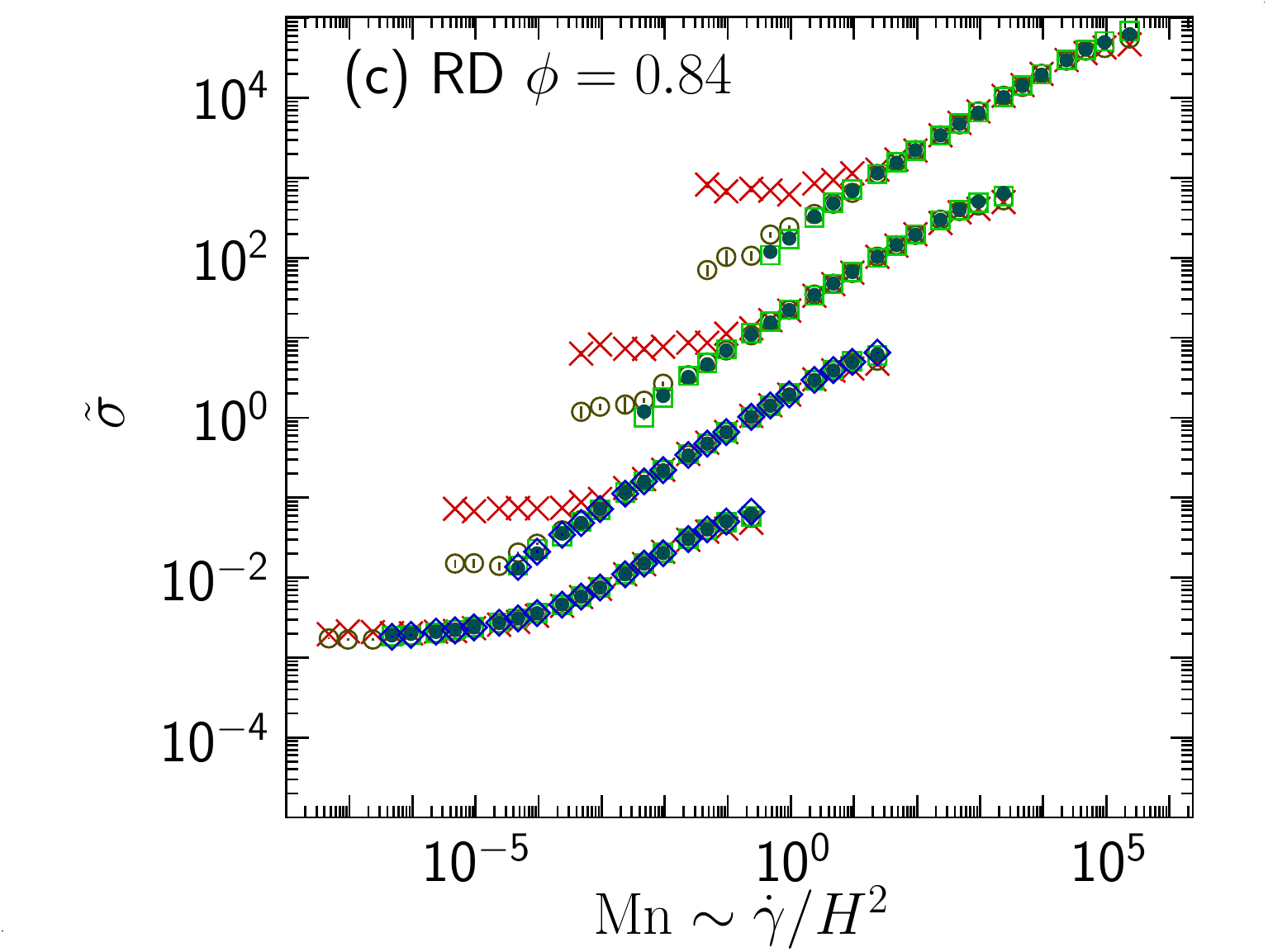}
              \hspace{-1.5cm}\includegraphics[height=3.7cm]{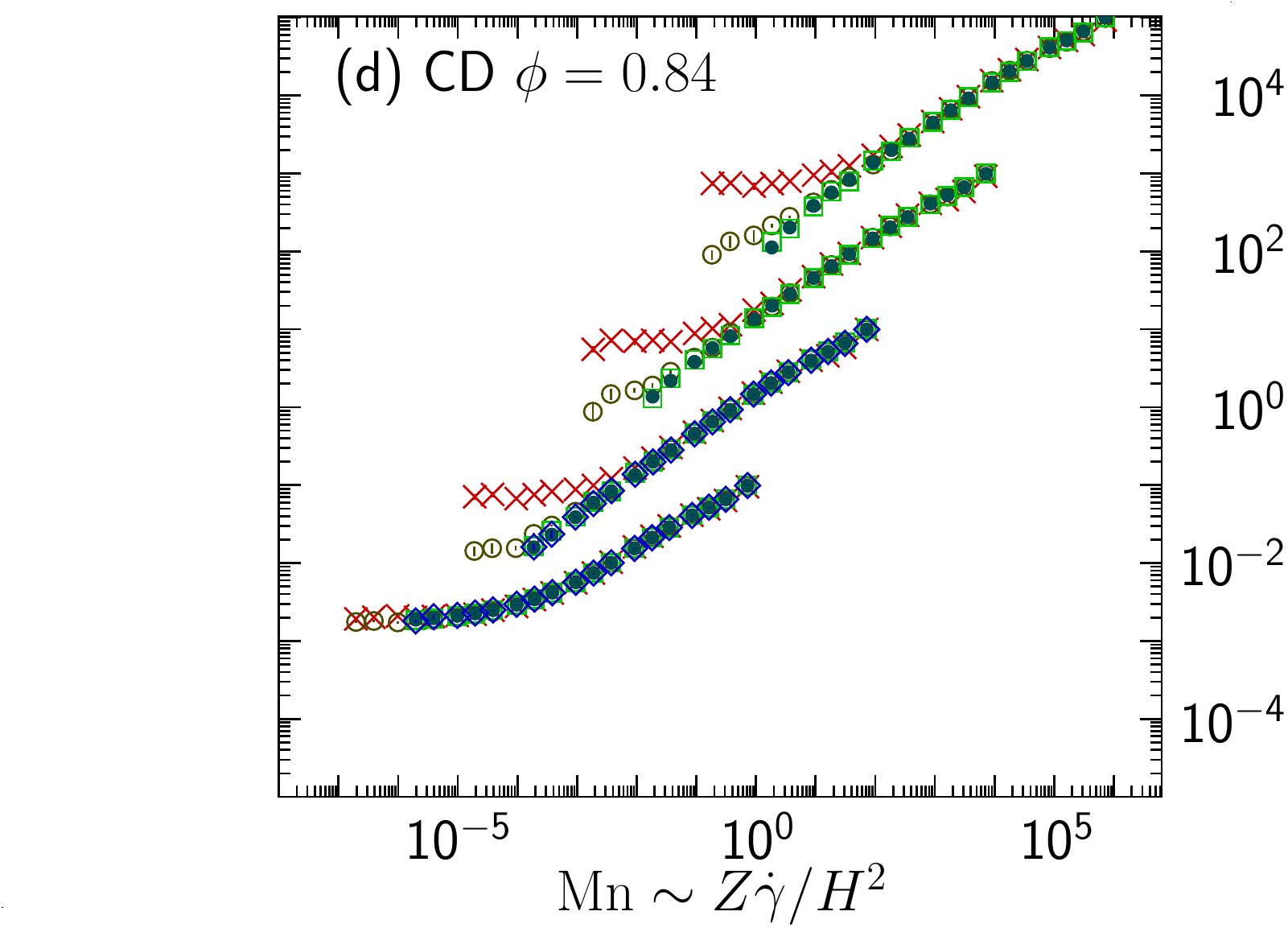}}

 \end{center}
  \caption{Finite size effects in the steady state shear stress $\tilde \sigma$ for varying Mason number $\Mn$ and particle number $N$.  Data for the RD and CD models (left/right column) at packing fraction $\phi = 0.5$ (a, b) and $\phi=0.84$ (c,d).  Data for $N = 4096$ is identical to Fig.~\ref{fig:s_gdot_4krdcd}.}
  \label{fig:s_finite_size}
\end{figure}

It is natural to ask if the qualitative differences in the flow curves of Fig.~\ref{fig:s_gdot_4krdcd} are due to finite size effects. To answer this question, we simulate steady state shear flow for a range of system sizes $N = 64 \ldots 16384$; the corresponding flow curves for the RD and CD models are plotted in Fig.~\ref{fig:s_finite_size}a and b, respectively. In both cases, we obtain good data collapse over the entire sampled range of $\Mn$, independent of $N$. We therefore conclude that differences between the RD and CD flow curves are not due to finite size effects.

\begin{figure}[tb]
 \begin{center}
   \subfigure{\hspace{-0.2cm}\includegraphics[height=3.7cm] {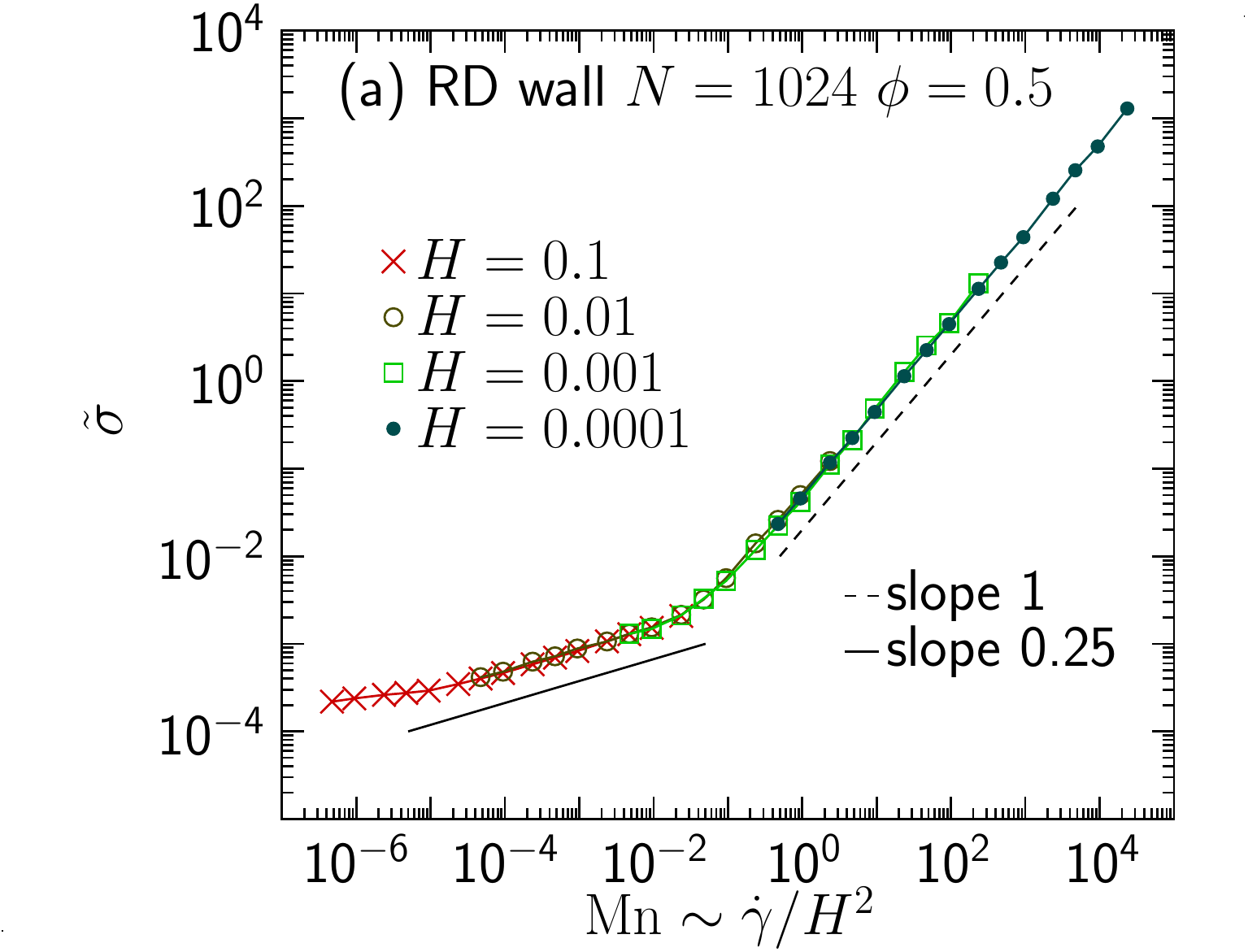}
              \hspace{-1.48cm}\includegraphics[height=3.7cm] {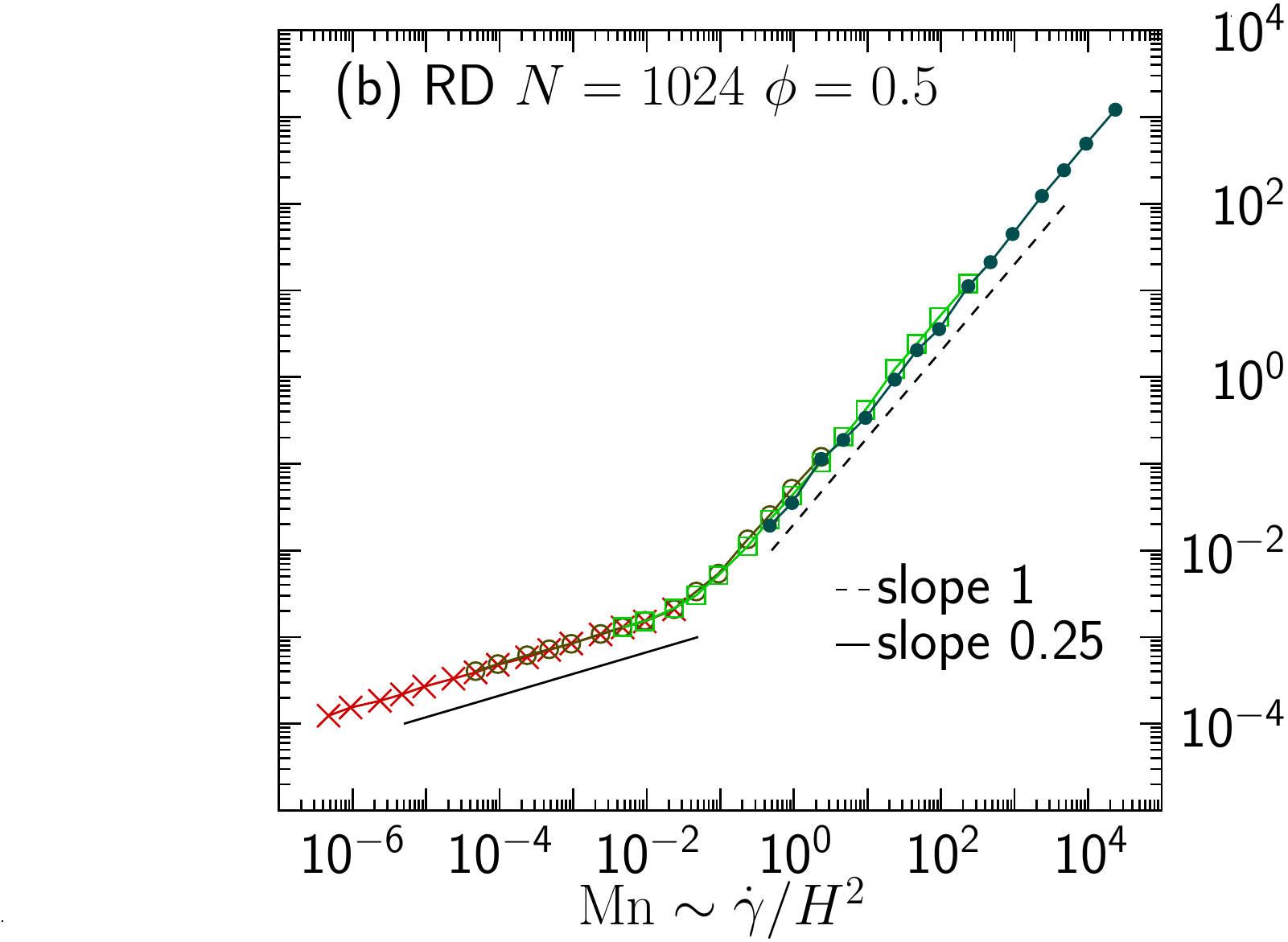}}
 \end{center}
  \caption{
  Flow curves with (a) and without (b) a wall.    }
  \label{fig:wall_effect}
\end{figure}

Boundary effects are closely related to finite size effects. They are also particularly relevant to experiments, of course, as shearing surfaces are necessary to sustain flow. To probe the influence of boundaries on the flow curve, we introduce a wall by fixing the positions of a thin layer of particles intersecting the line $y = 0$ (in the center of the cell). The resulting RD flow curves are plotted in Fig.~\ref{fig:wall_effect}. One clearly sees that the system with a wall develops a plateau at low $\Mn$ that is absent in the wall-free case for the same system size. This effect is clearly not a material property, but should be borne in mind when interpreting experimental data.

\begin{figure}[tb]
 \begin{center}

   \subfigure{\hspace{-0.2cm}\includegraphics[height=3.7cm]{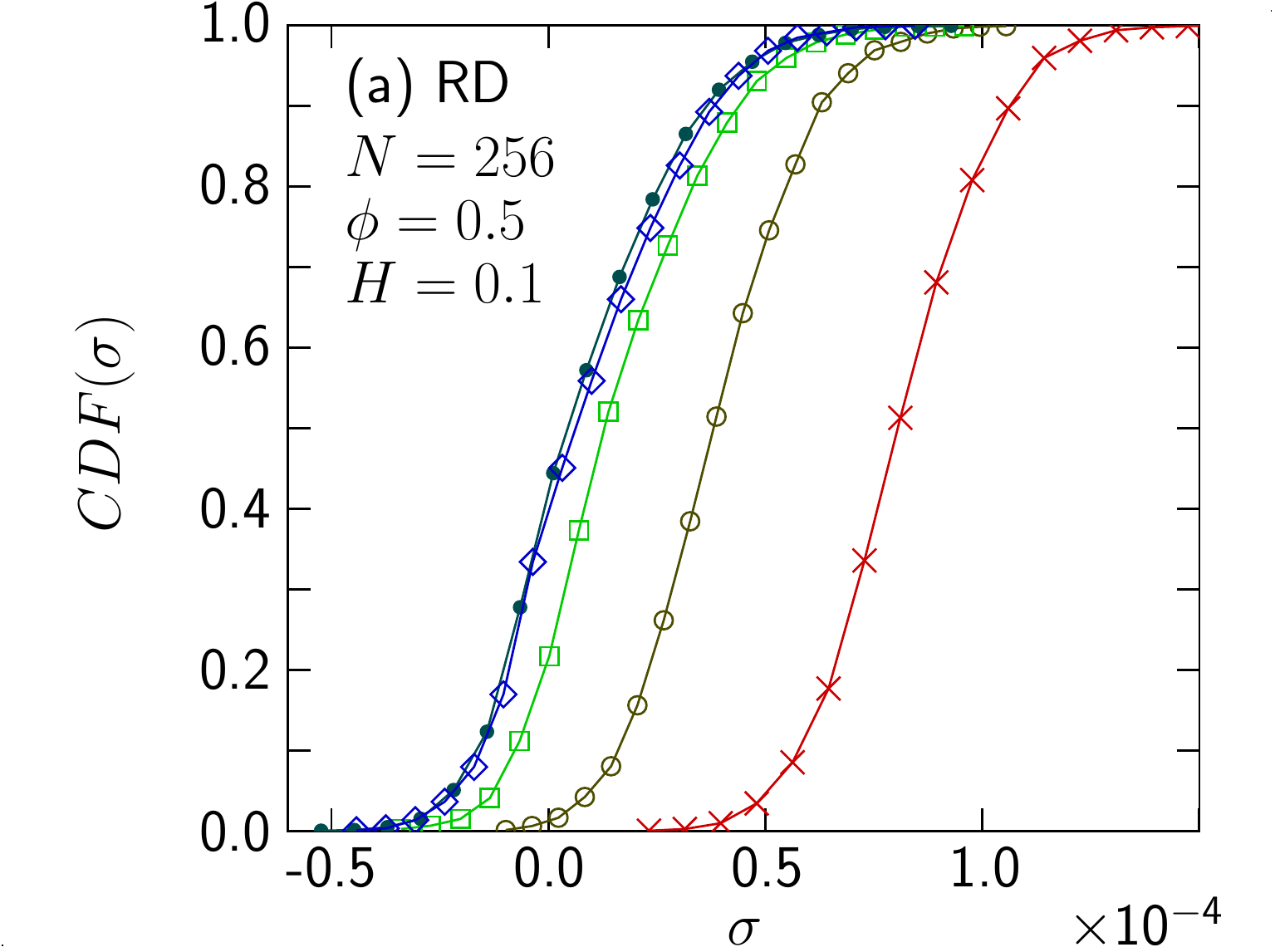}
              \hspace{-1.505cm}\includegraphics[height=3.7cm]{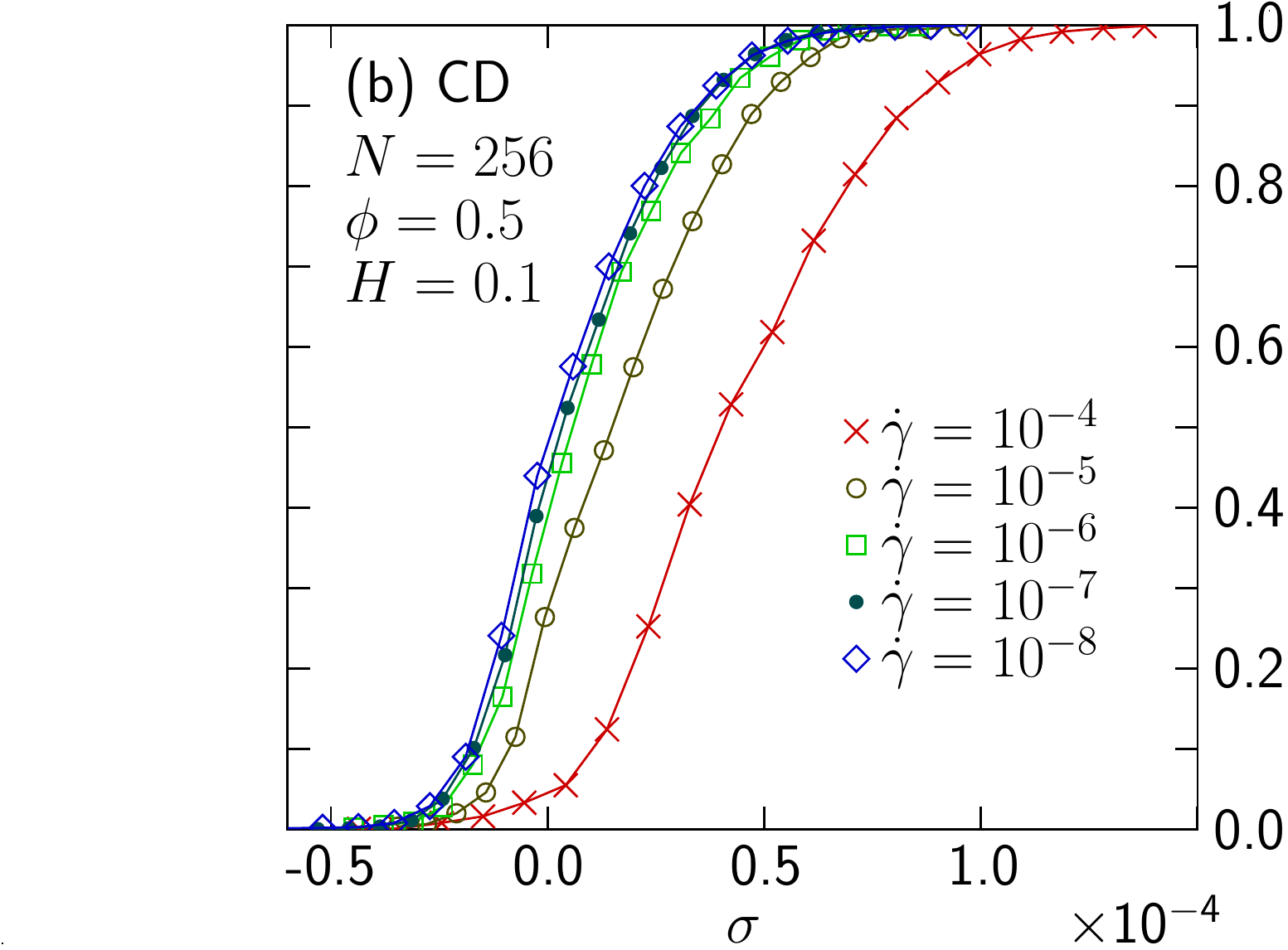}}

   \subfigure{\hspace{-0.2cm}\includegraphics[height=3.7cm]{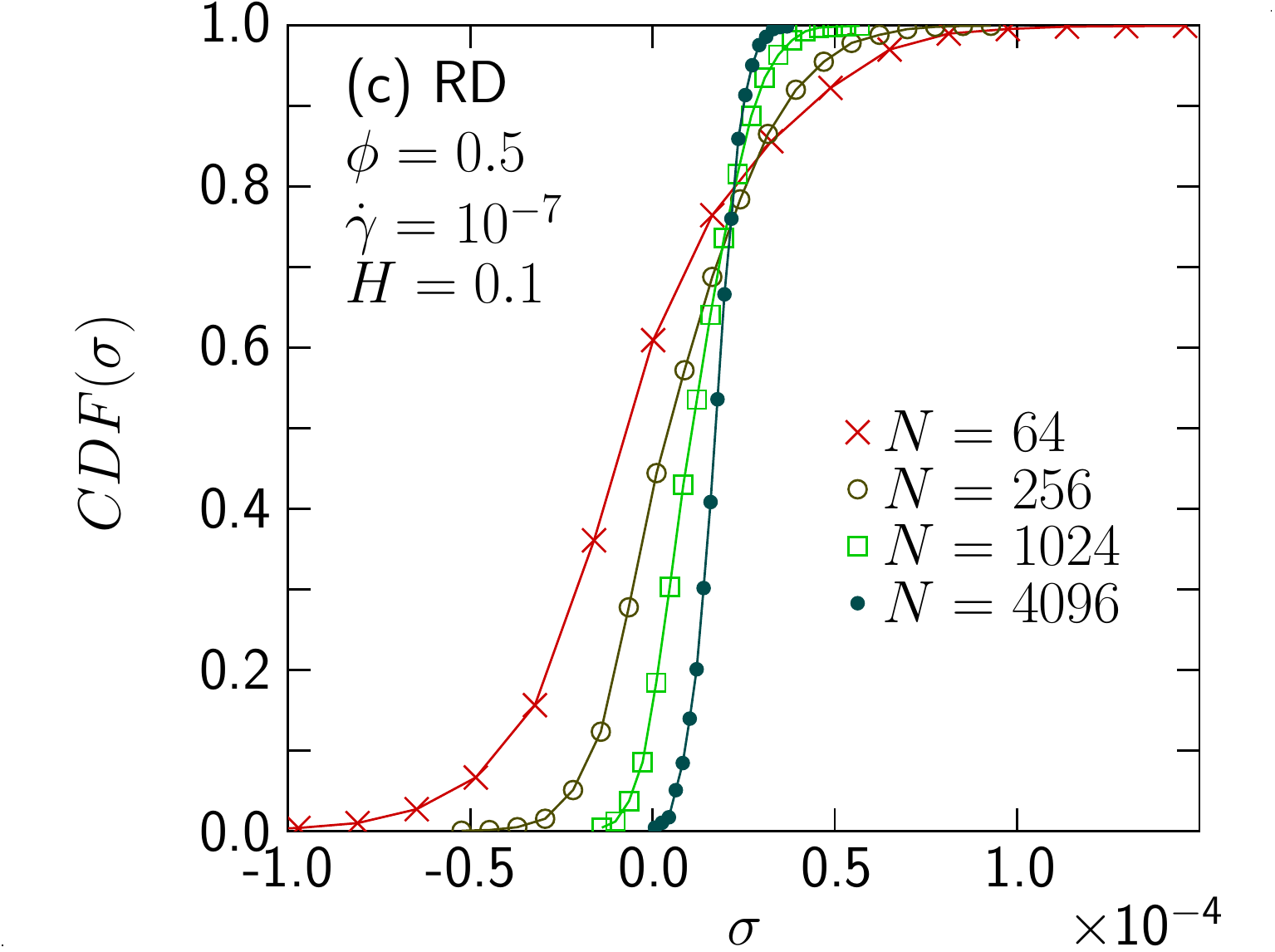}
              \hspace{-1.505cm}\includegraphics[height=3.7cm]{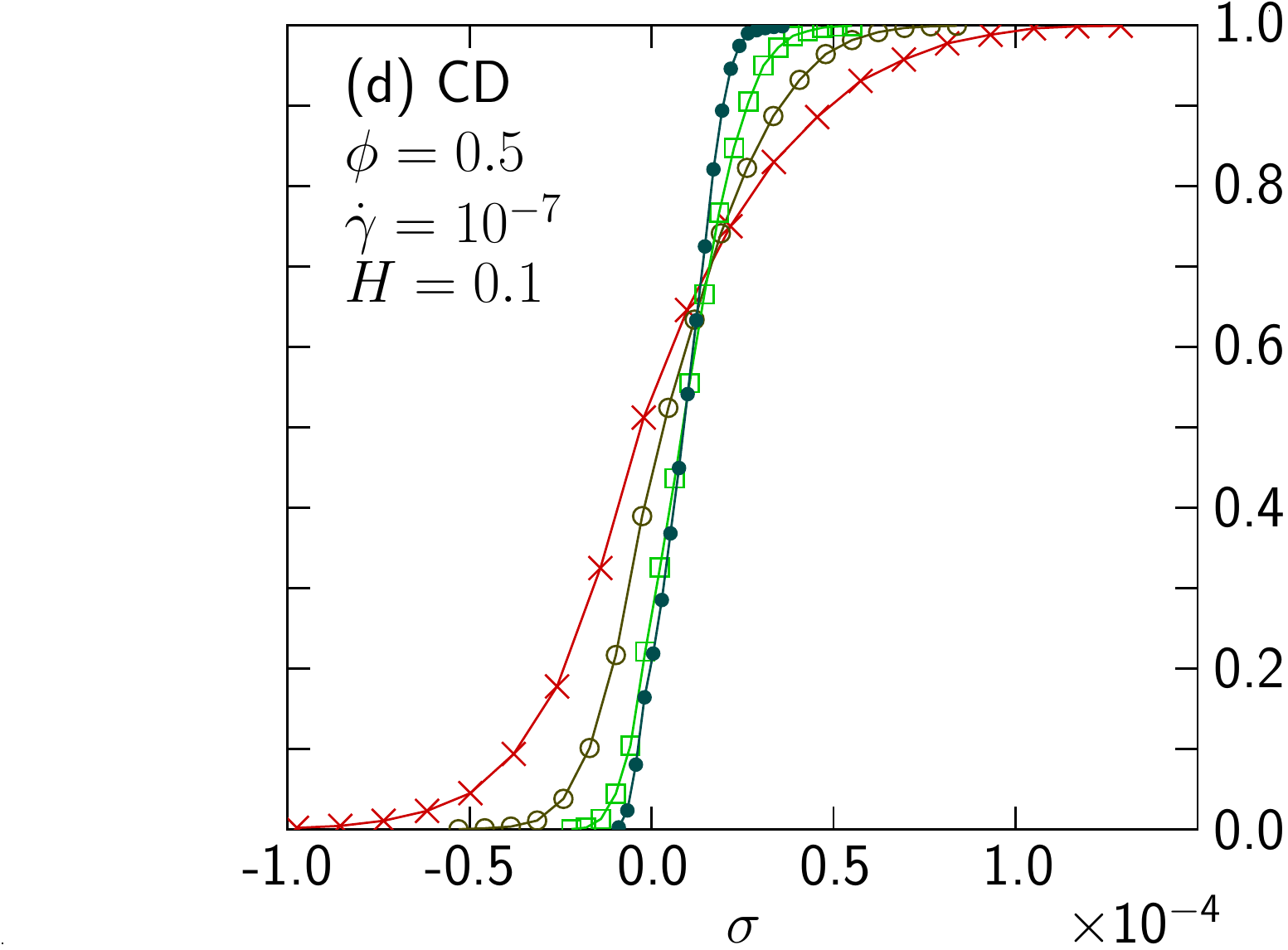}}
 \end{center}
  \caption{Stress statistics of the RD and CD models (left/right column). Cumulative distribution functions (CDF's) of $\tilde \sigma$ at $H = 0.1$, $N = 256$, and varying strain rate $\dot \gamma$ (a,b). CDF's of the steady state stress histogram for $\dot \gamma = 10^{-7}$, $H = 0.1$, and varying $N$ (b,c). Data for $N = 4096$ is identical to Fig.~\ref{fig:s_gdot_4krdcd}. }
  \label{fig:s_cdf}
\end{figure}

In order to quantify stress fluctuations in flow, we have also sampled the cumulative distribution function (CDF) of shear stress in steady state. In Fig.~\ref{fig:s_cdf}a and b we plot the CDF for a range of strain rates, with the highest chosen to correspond roughly to the ``elbow'' in the RD flow curve. While the curves shift left with decreasing $\dot \gamma$ (as already apparent from the flow curve), their overall shape changes little, indicating that stress fluctuations are insensitive to the strain rate. In Fig.~\ref{fig:s_cdf}c and d we plot the CDF for low $\Mn$ and a range of system sizes $N$. There is a slow systematic increase of the median stress (${\rm CDF} = 0.5$) with $N$, which is too weak to be seen on the log-log plots of Fig.~\ref{fig:s_finite_size}a and b. For small system sizes the flow regularly samples states with negative shear stress; however increasing the system size causes the CDF's to sharpen, reducing the fraction of negative stress states. For $N = 4096$ and $\Mn = 10^{-6}$ (Figs.~\ref{fig:s_cdf}c and d), the fraction sampled by the RD model is negligible, while in the CD model it is less than 0.1.

Based on the above results, we conclude that the bulk flow curve of the RD model at $\phi = 0.5$ has no apparent yield stress over the experimentally (and numerically) accessible range of Mason numbers. The CD model, by contrast, does have an apparent yield stress.

\begin{figure*}[tb]
 \begin{center}
  \subfigure{
        \hspace{-1.0cm}\includegraphics[height=3.755cm]{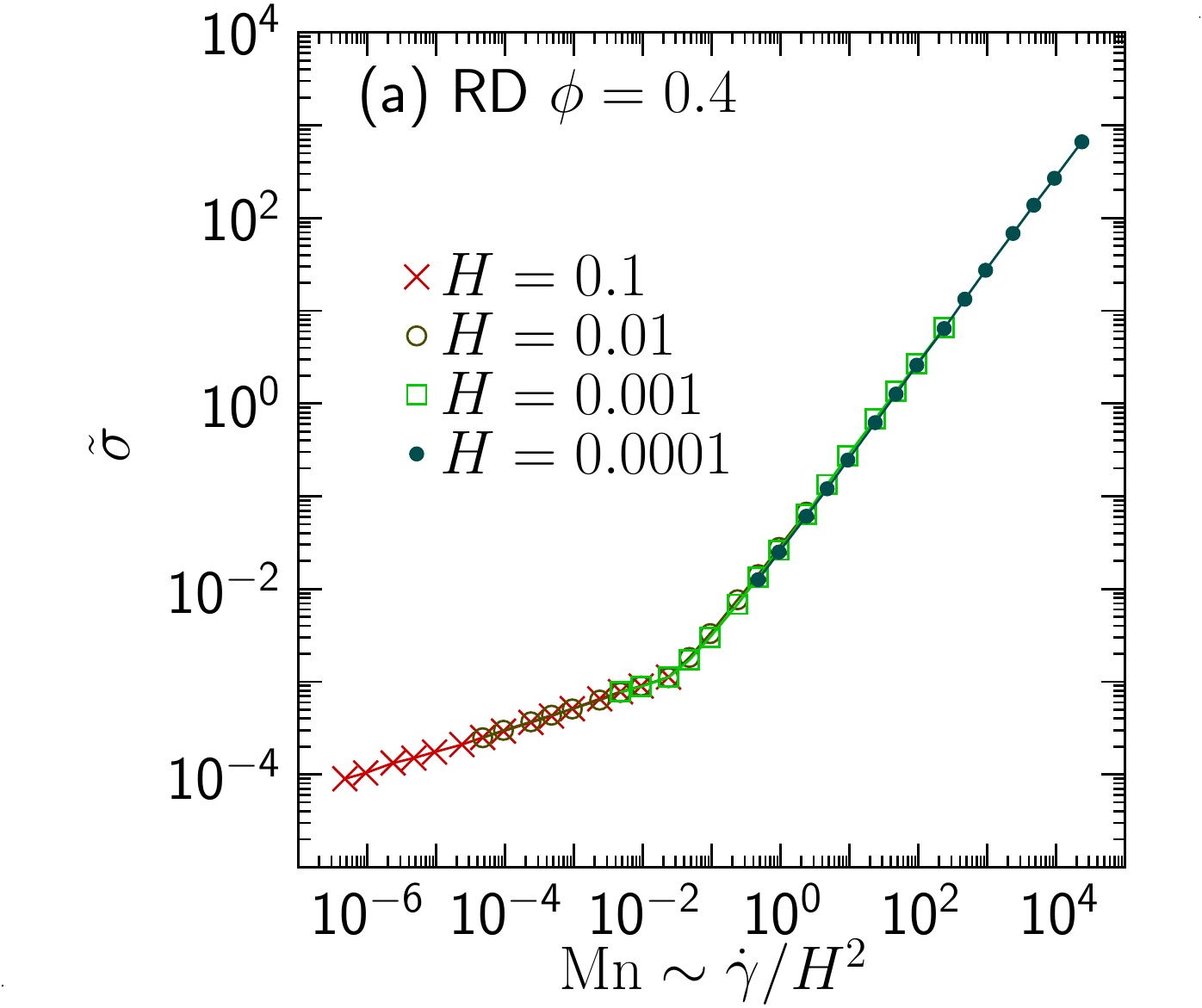}
        \hspace{-1.5cm}\includegraphics[height=3.7cm]{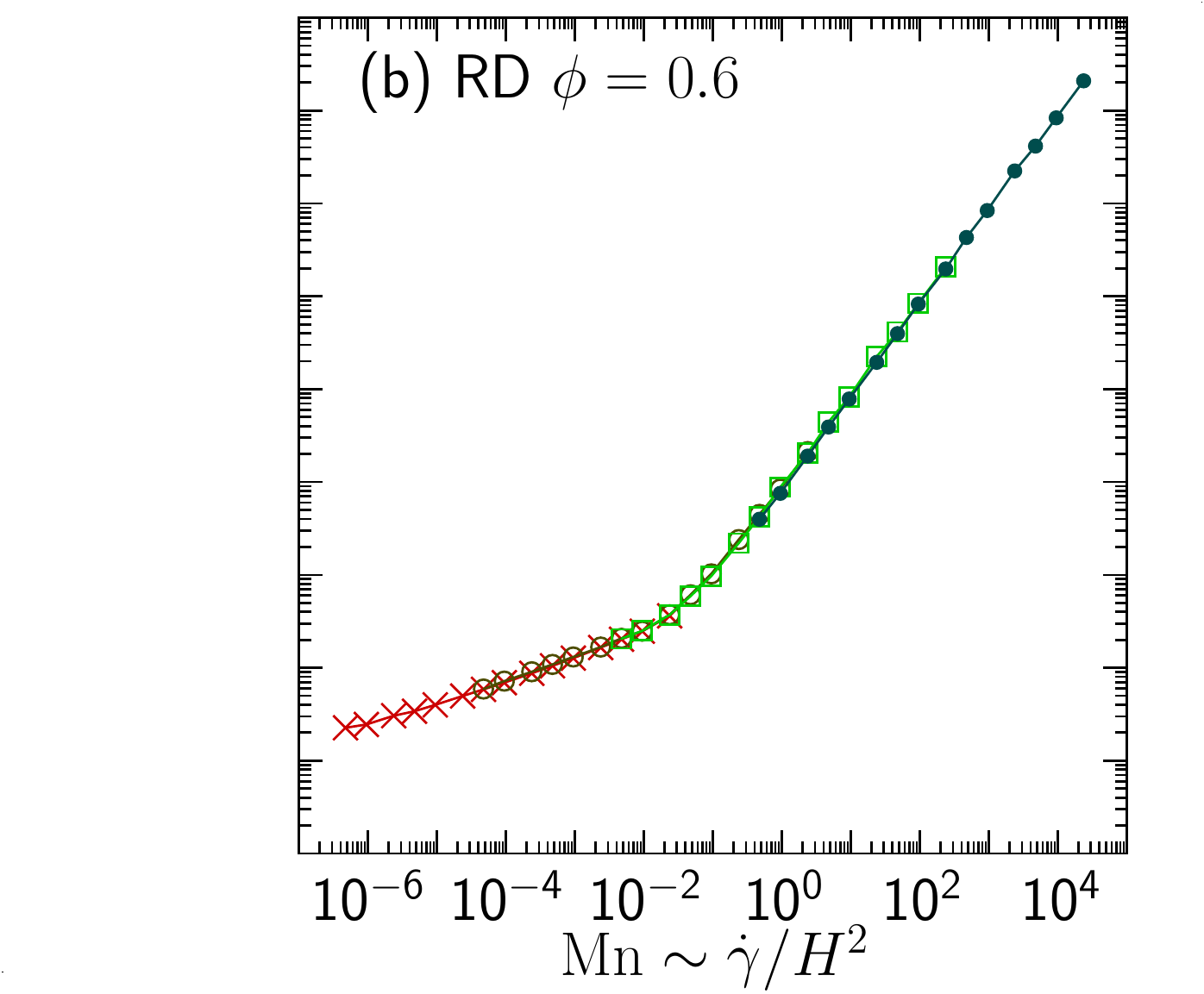}
        \hspace{-1.5cm}\includegraphics[height=3.7cm]{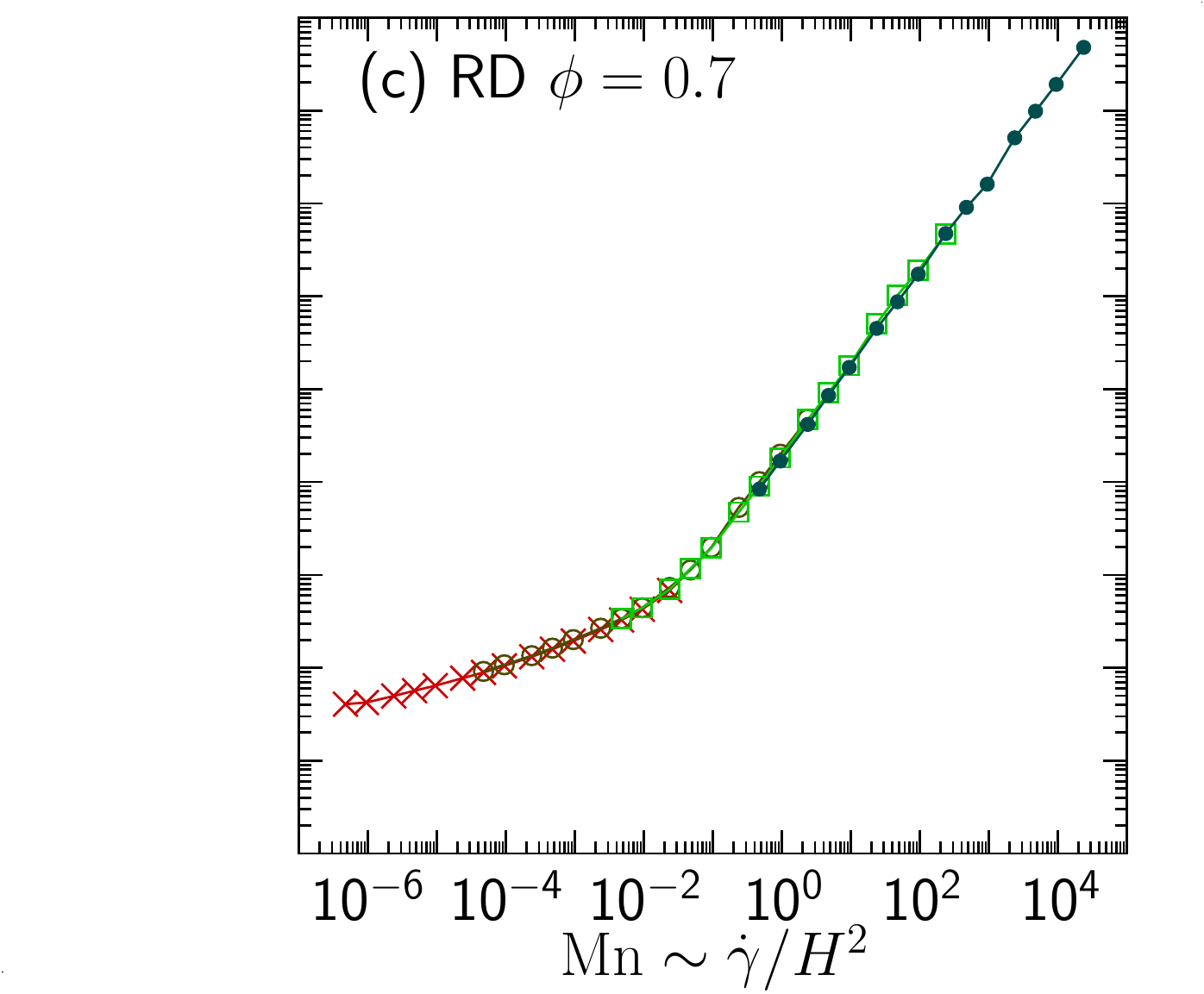}
        \hspace{-1.5cm}\includegraphics[height=3.7cm]{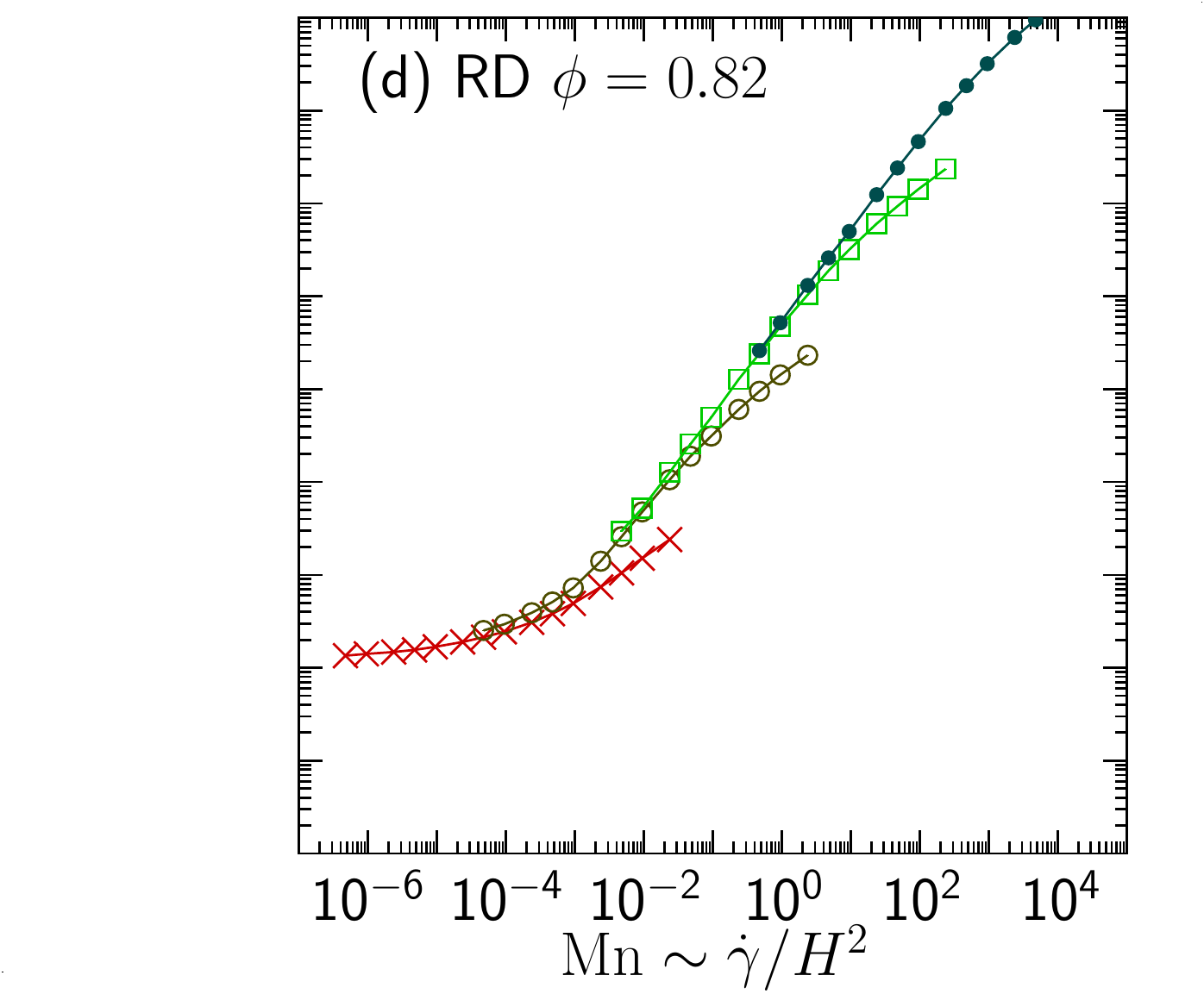}
        \hspace{-1.5cm}\includegraphics[height=3.755cm]{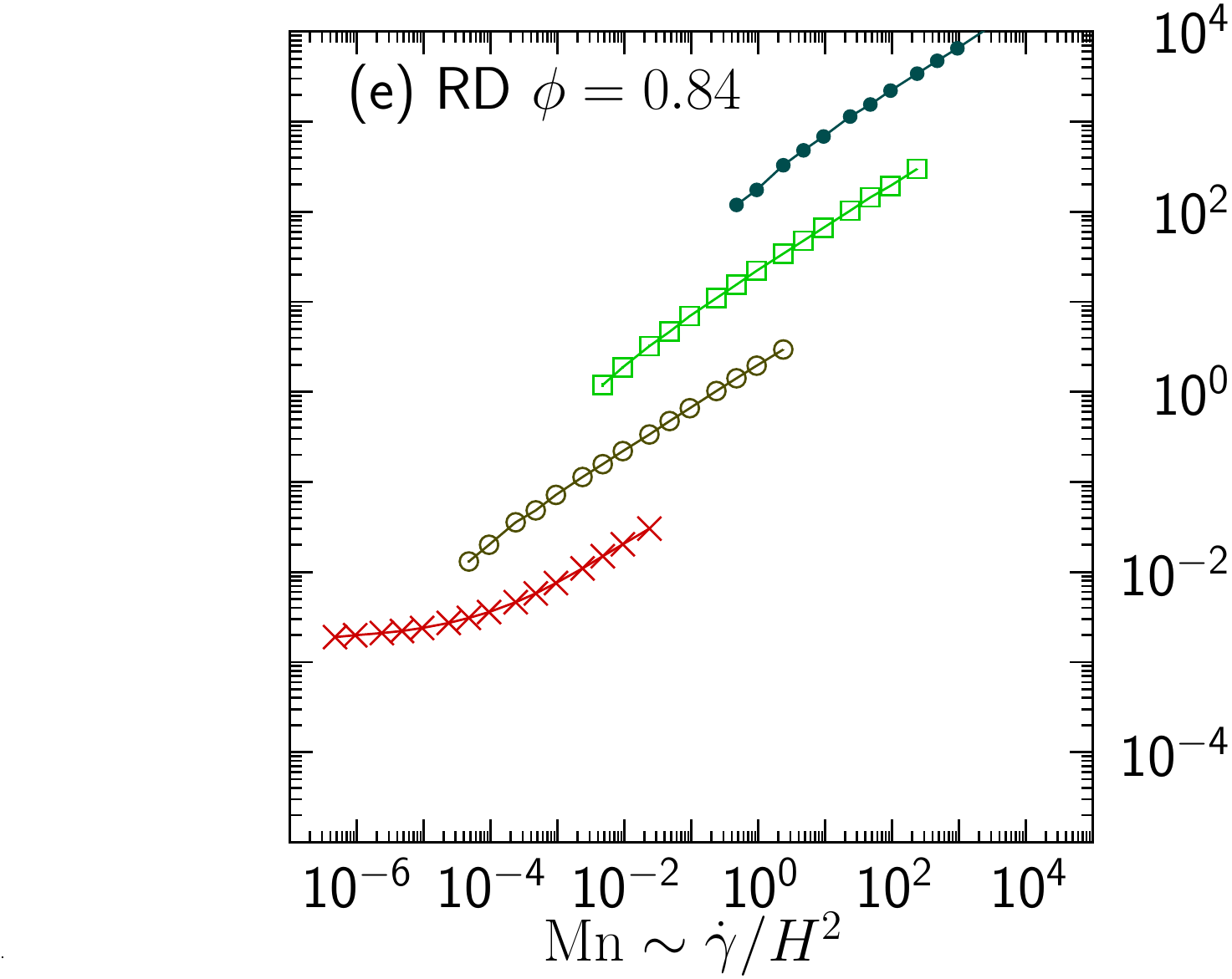}}

  \subfigure{
        \hspace{-1.0cm}\includegraphics[height=3.755cm]{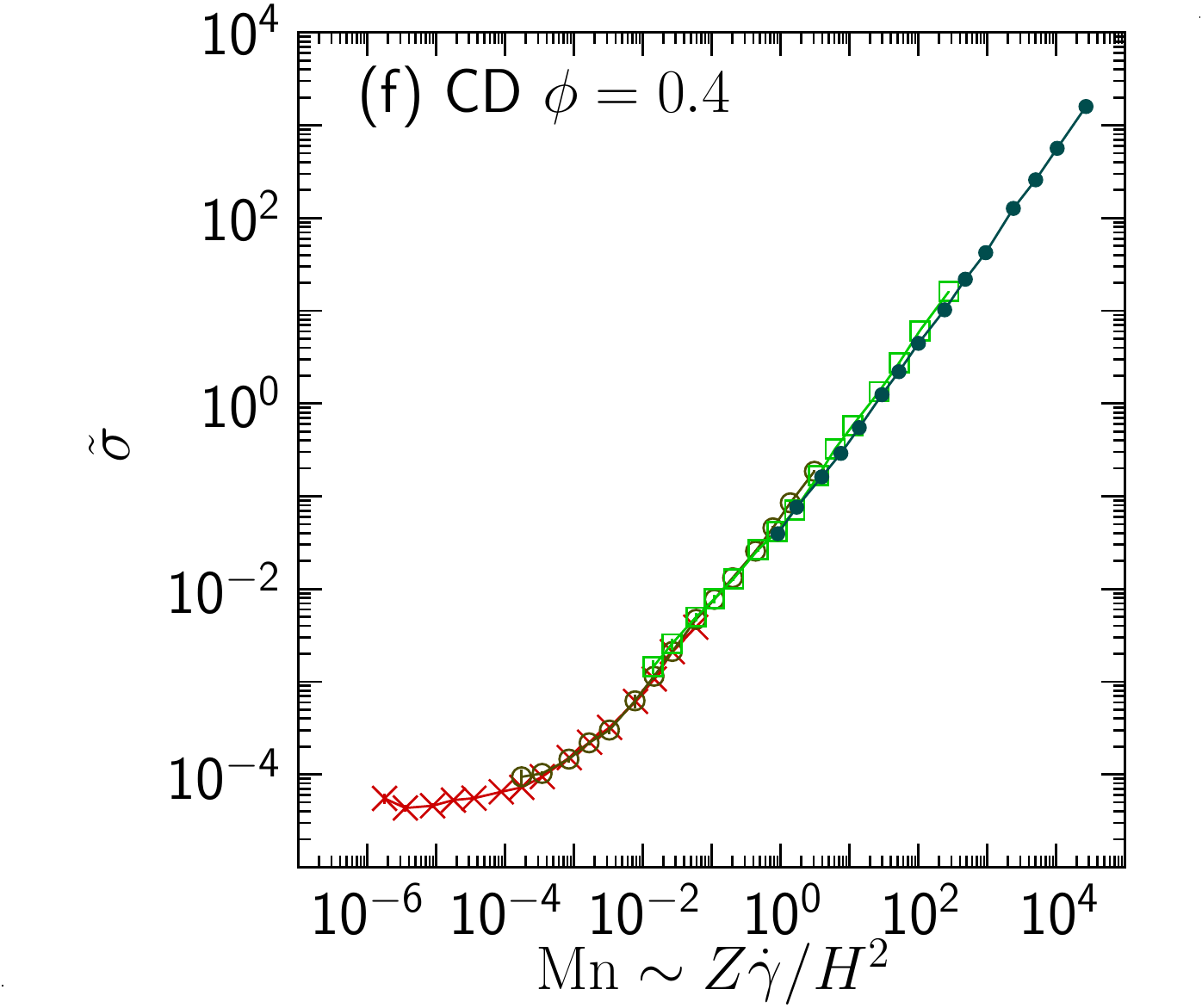}
        \hspace{-1.5cm}\includegraphics[height=3.7cm]{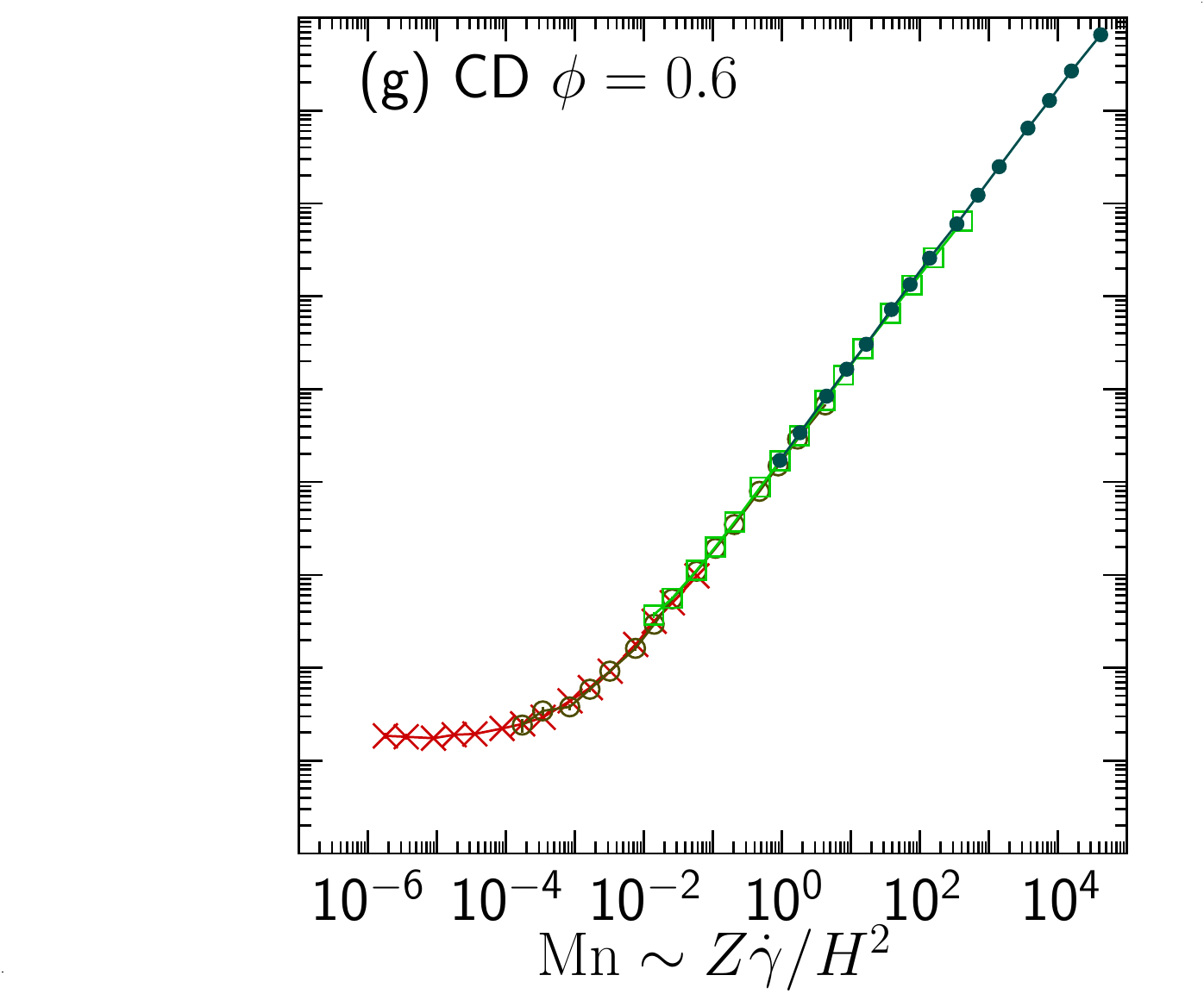}
        \hspace{-1.5cm}\includegraphics[height=3.7cm]{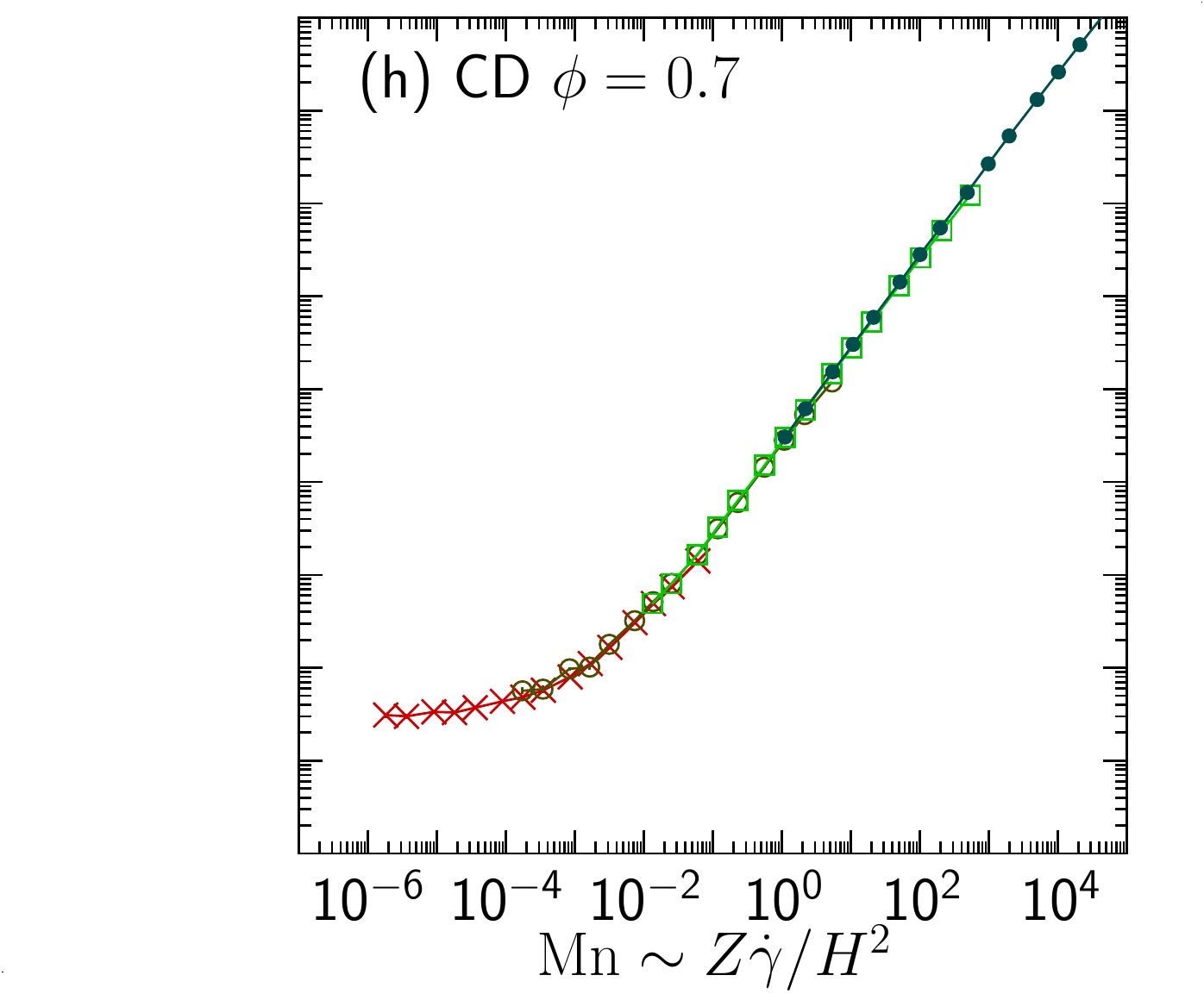}
        \hspace{-1.5cm}\includegraphics[height=3.7cm]{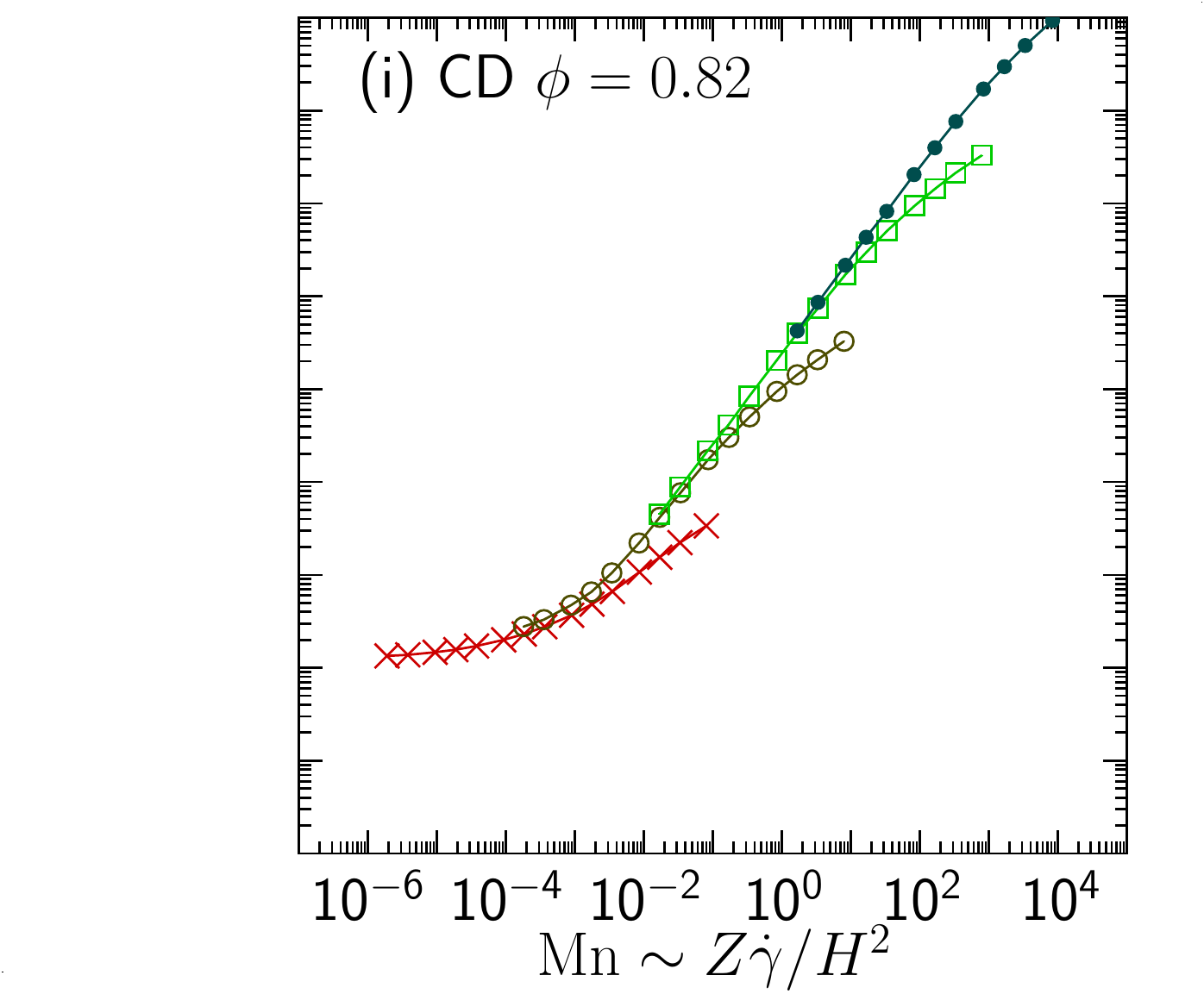}
        \hspace{-1.5cm}\includegraphics[height=3.755cm]{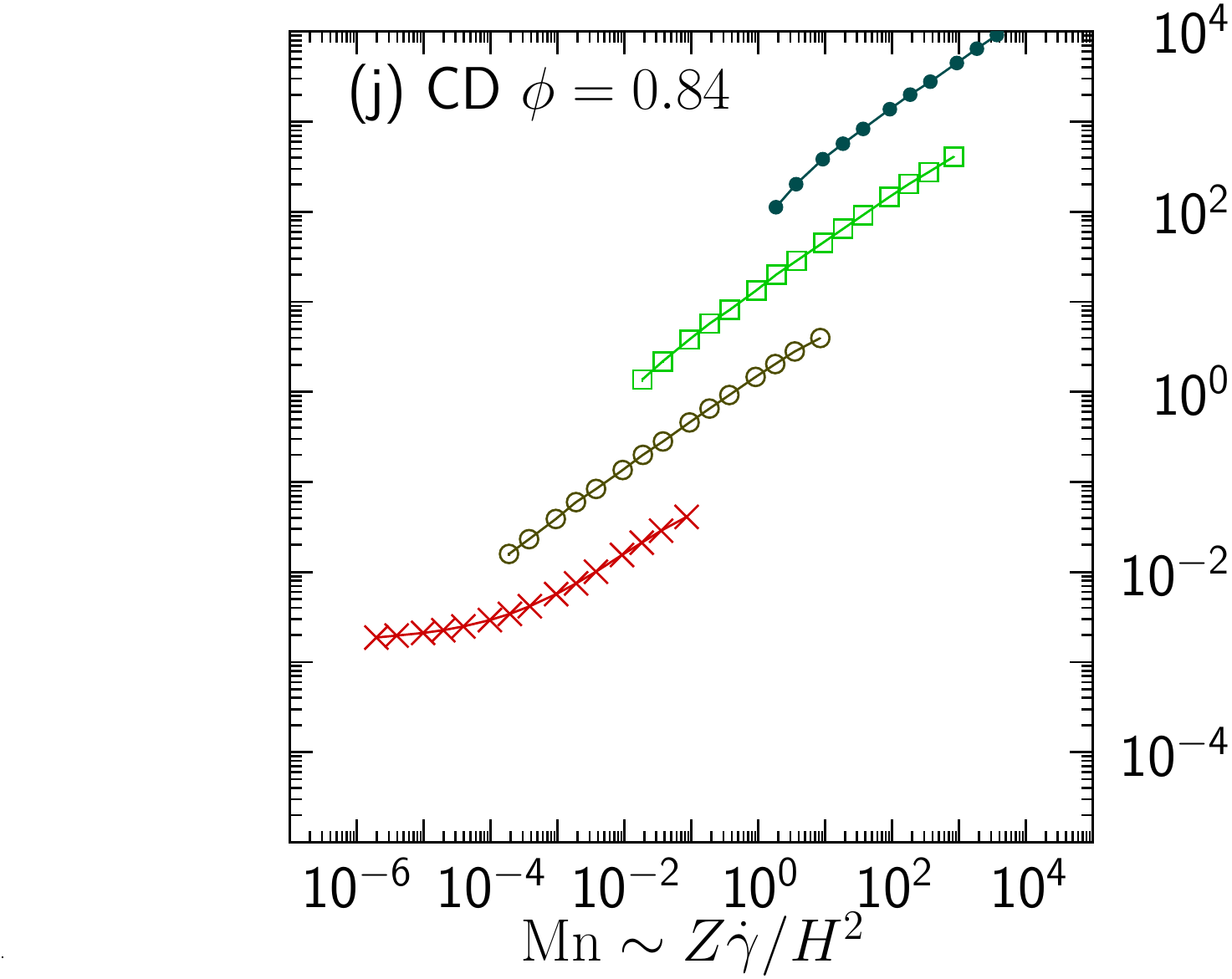}}

 \end{center}
  \caption{(top row) Flow curves at varying field strengths $H$ for five packing fractions $\phi$ for system size $N=4096$ in the RD model. (bottom row) Corresponding flow curves in the CD model.}
  \label{fig:fse_s}
\end{figure*}

\subsection{Towards Jamming}

We now consider the role of packing fraction $\phi$ in the bulk rheology. Intuitively, one expects  the stress required to sustain steady flow to increase with $\phi$. Moreover, soft sphere packings in the absence of an applied field (i.e.~$H = 0$) are known to develop a yield stress at a critical volume fraction $\phi_c$ (the jamming point)\cite{liu1998,ohern2003,olsson2007}. The precise value of $\phi_c$ depends and particle size distribution \cite{koeze2016} as well as the protocol used to generate the packings \cite{vaagberg2011glassiness}. For sheared systems in the quasistic limit and $H=0$ both the CD and RD model have been shown to jam at the same packing fraction $\phi_c\approx0.8433$ \cite{olsson2012hb,vagberg2014universality}.

It is therefore reasonable to ask what happens when the volume fraction is increased towards $\phi_c$ in the presence of a magnetic field $H>0$. 
We  start by looking at the RD model.

In the top row of Fig.~\ref{fig:fse_s}, panels (a-e), we plot RD flow curves for $\phi = 0.4$, 0.6, 0.7, 0.82, and 0.84 at varying strain rate and field strength. 
For $\phi \le 0.7$ we do not observe a plateau, although fitting a power law $\tilde \sigma \sim \Mn^{1-\Delta}$ to the low $\Mn$ data reveals an effective exponent $\Delta$ approaching 1 as $\phi$ increases.
For $\phi = 0.82$ there is an unambiguous plateau at low $\Mn$. Data above the plateau no longer collapse with $\Mn$, which is an indication that critical effects near jamming have begun to play a significant role; at the same time, flow curves at $\phi = 0.82$ and $H = 0$ do not show a yield stress \cite{vagberg2014dissipation}.  For $\phi = 0.84$ the dynamics is completely dominated by the proximity to the jamming transition and data collapse with $\Mn$ is wholly absent. There are also strong finite size effects (as expected near a critical point), as seen in Fig.~\ref{fig:s_finite_size}c. The flow curves at high $\Mn$ are no longer Newtonian but shear thinning -- also a signature of the approach to jamming. For comparison, in the bottom row of Fig.~\ref{fig:fse_s}, panels (f-j), we plot flow curves for the CD model for the same volume fractions; in all cases there is a plateau at low $\Mn$, and we observe identical trends regarding data collapse with $\Mn$. 

\begin{figure}
 \begin{center}
   \subfigure{\hspace{-1.0cm}\includegraphics[height=3.7cm] {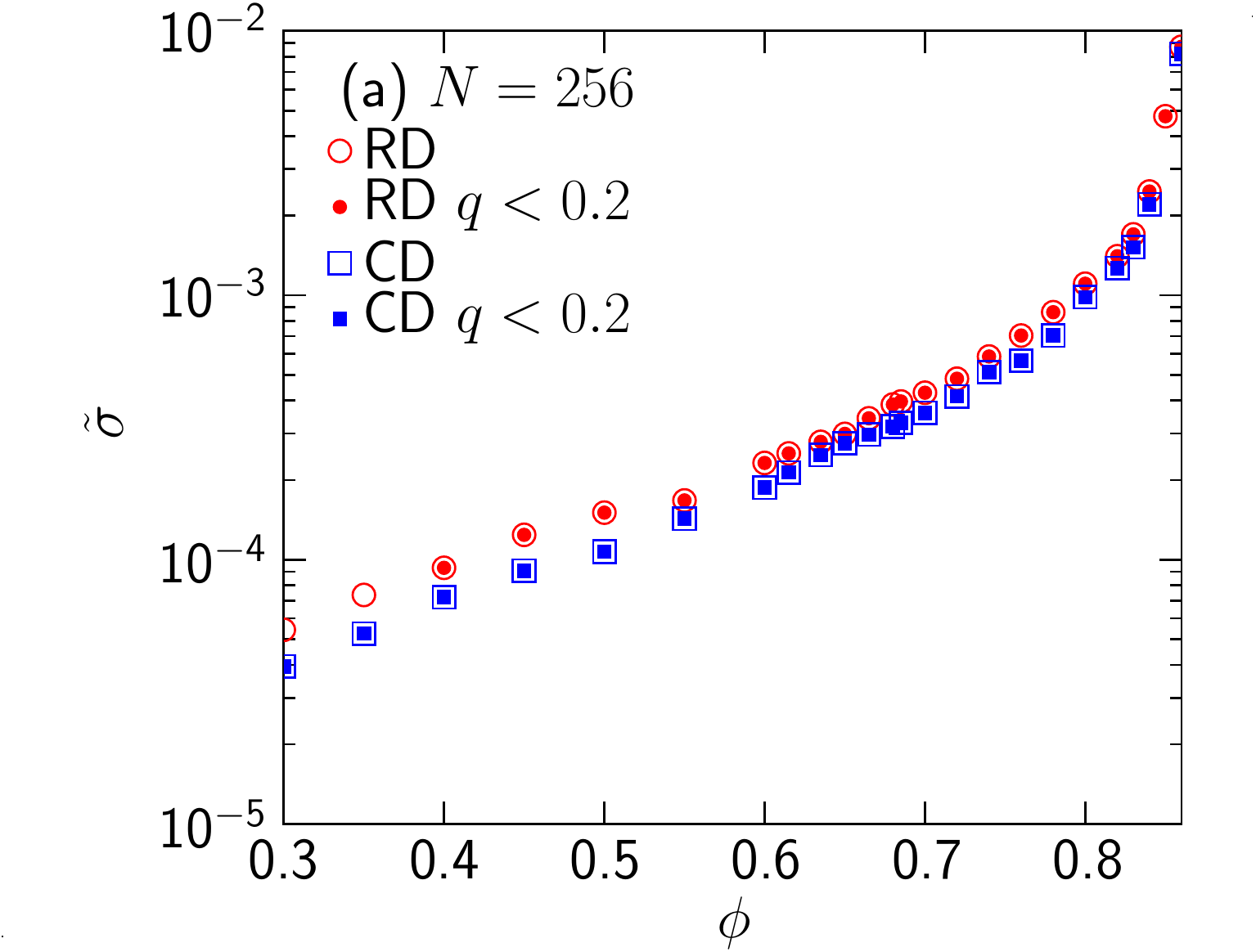}
              \hspace{-1.5cm}\includegraphics[height=3.7cm] {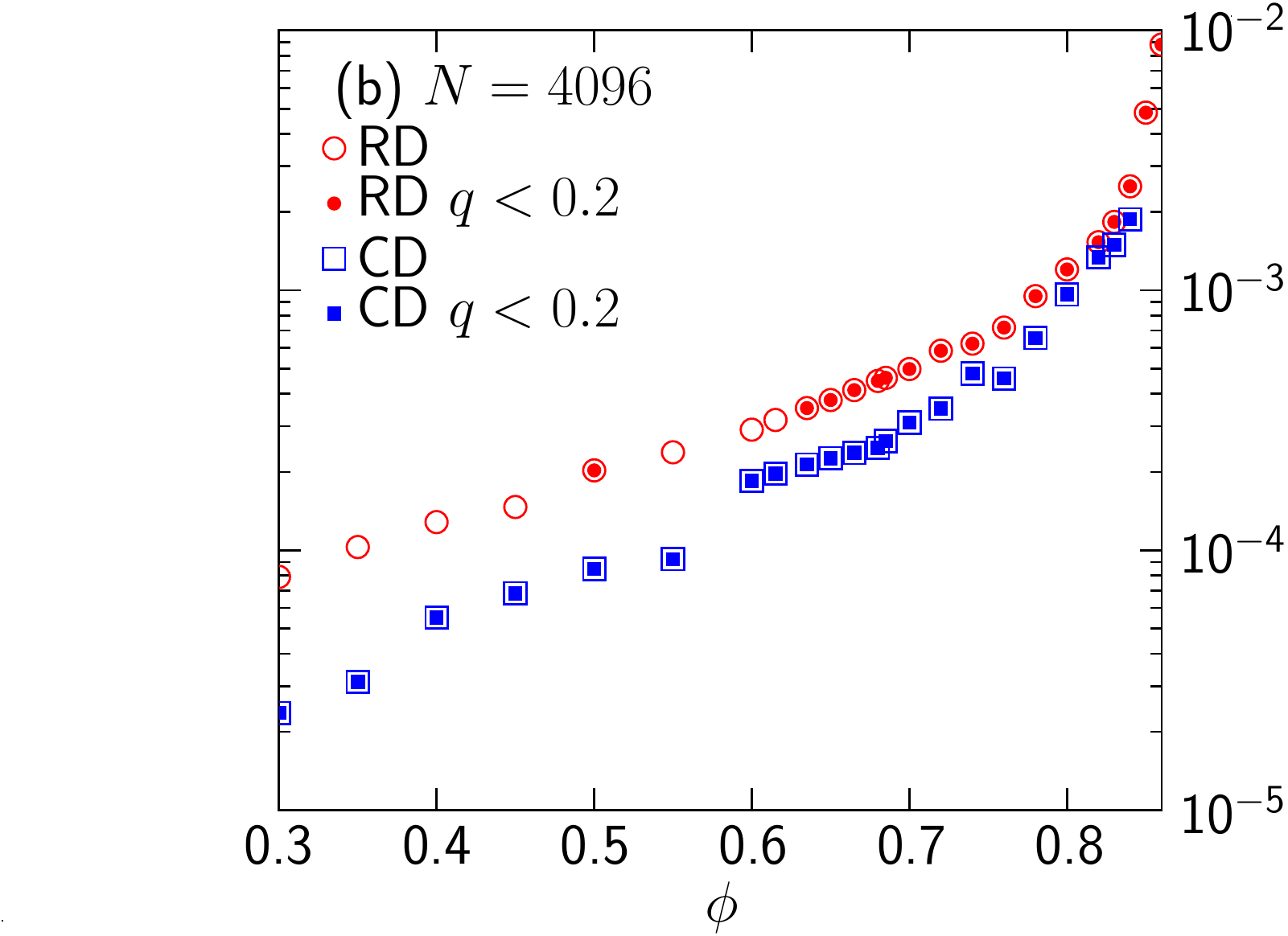}}
 \end{center}
 
\caption{$\tilde\sigma$ vs $\phi$ for $N=256$ and $N=4096$. Open symbols shows the average stress calculated over a narrow interval in $\Mn$ centered at $\Mn=2\times10^{-6}$ for a given $\phi$.  The filled symbols indicates for each point if $q<0.2$ (indicating the onset of the plateau in $\tilde\sigma$ vs $\Mn$) based on linear fitting of $q$ over the same range of $\Mn$ used to calculate the averages.} 
\label{fig:s_vs_phi}

\end{figure}

In order to compare stresses at low $\Mn$ directly, we plot the stress over a range of volume fractions for constant $\Mn = 2 \times 10^{-6}$ in both drag models -- see Fig.~\ref{fig:s_vs_phi}. The stresses display an approximately exponential growth with $\phi$ over a wide range of volume fractions, before increasing more rapidly close to jamming. To test whether the flow curve has approached a plateau, we numerically evaluate the logarithmic derivative $q \equiv {\rm d} \, \ln {\tilde\sigma}/{\rm d}\, \ln{\dot \gamma}$ and plot the stress only when $q < 0.2$ (filled symbols). For comparison we also plot the unfiltered stress (open symbols). It is apparent that the CD model always reaches a plateau (apart from a small number of outliers), while the RD model only shows a clear plateau at sufficiently high volume fractions. The particular value of $\phi$ where the plateau appears has some dependence on system size (compare panels (a) and (b)). While  the stress in the RD model always exceeds that in the CD model, the two curves grow closer with decreasing Mason number. This is suggestive of convergence to a common asymptote, and therefore indirect evidence that the RD flow curves display a plateau at asymptotically low strain rates.

The data of Fig.~\ref{fig:fse_s} demonstrates that a plateau in the flow curve (i.e.~an apparent yield stress) emerges in the RD model at sufficiently high volume fractions. We speculate that the plateau is present for all $\phi$ where the particles form a percolating cluster, which at lower $\phi$ values occurs for smaller Mason numbers  than those accessed here. This hypothesis cannot be tested directly using present methods, but in the following Section we provide supporting evidence based on the evolution of microstructural measures with $\Mn$ and $\phi$.

\begin{figure*}[tb]
 \begin{center}
  \subfigure{\parbox{1cm}{\vspace{-3.5cm}RD \break (a)-(d)}} 
  \subfigure{\includegraphics[scale=0.2] {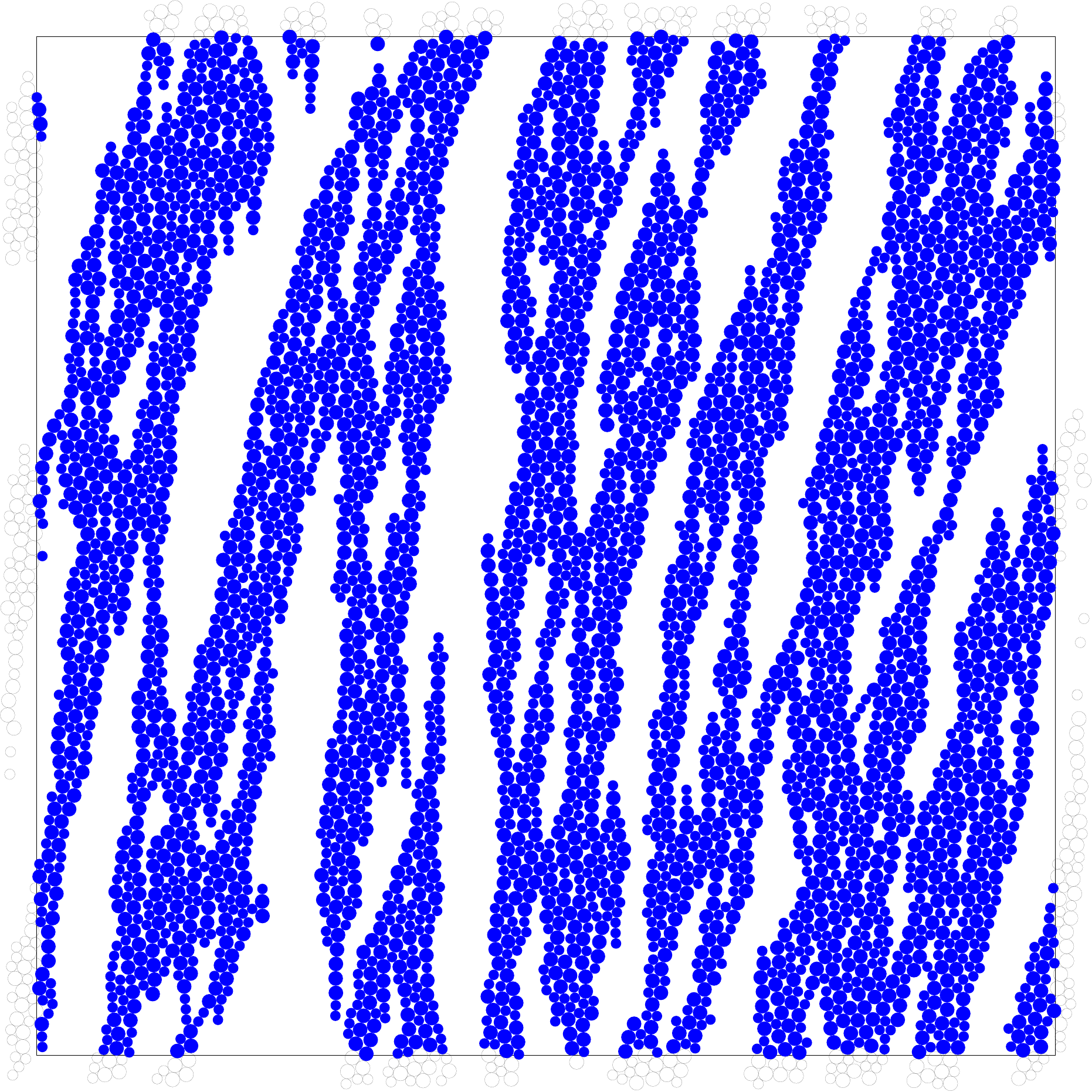}
             \includegraphics[scale=0.2] {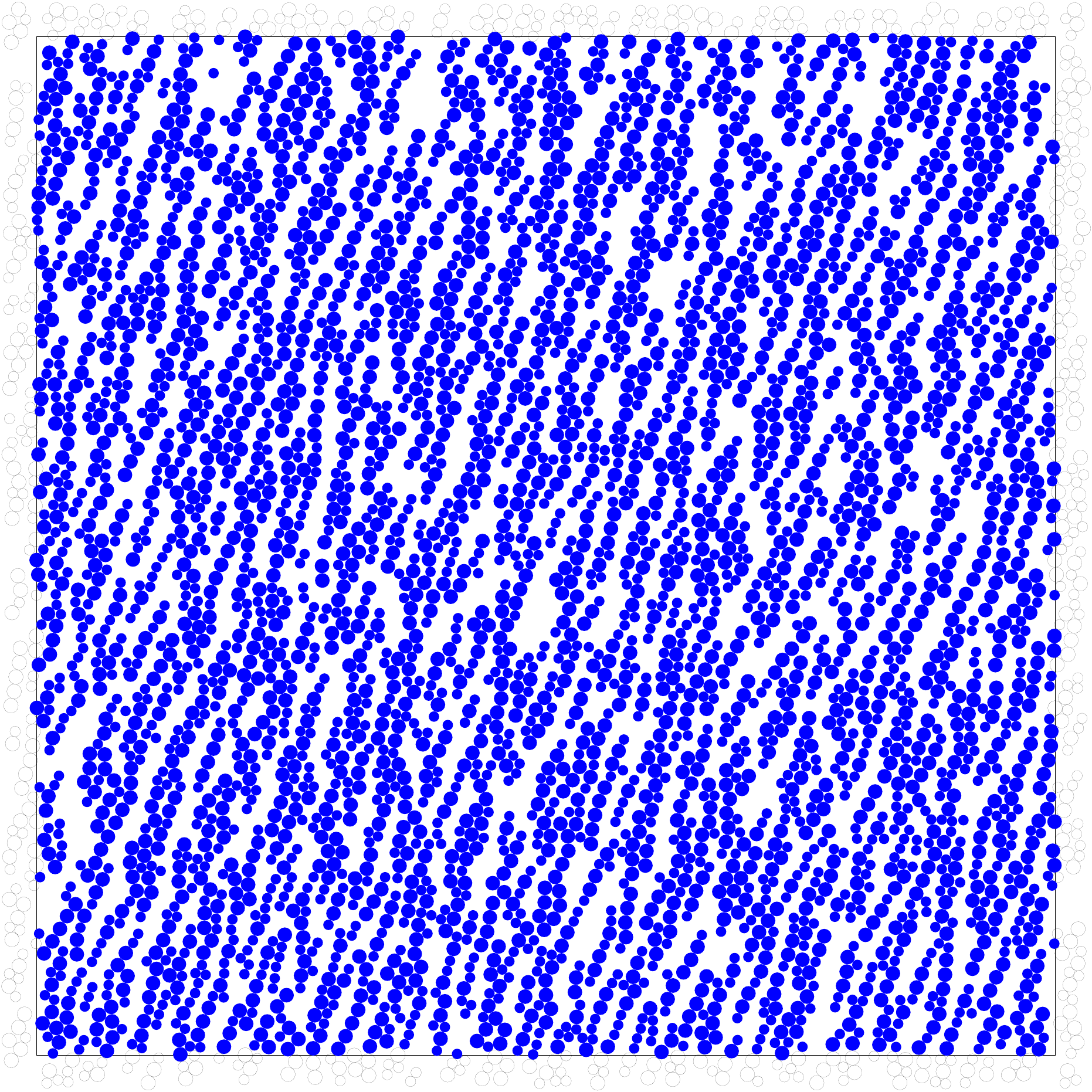}
             \includegraphics[scale=0.2] {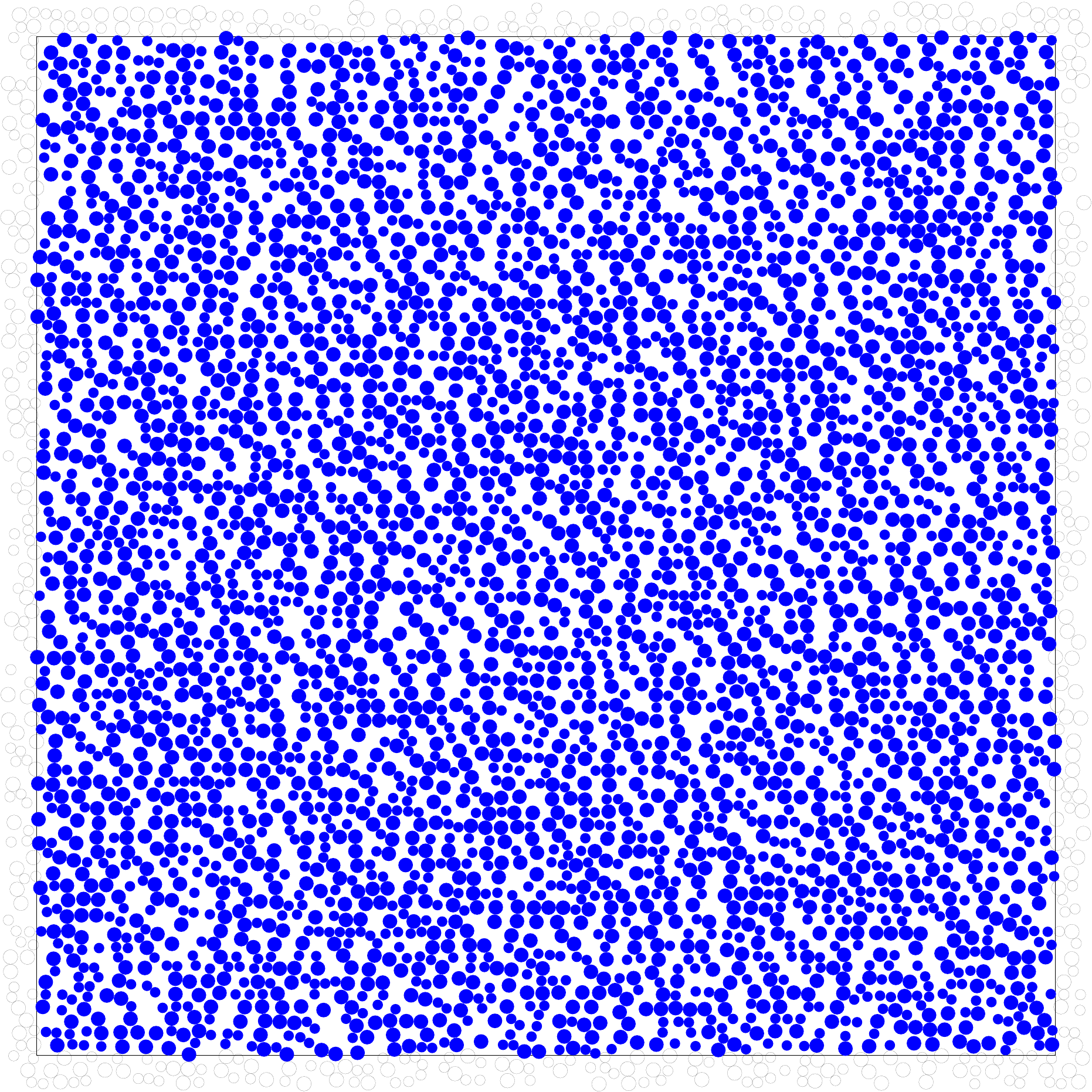}
             \includegraphics[scale=0.2] {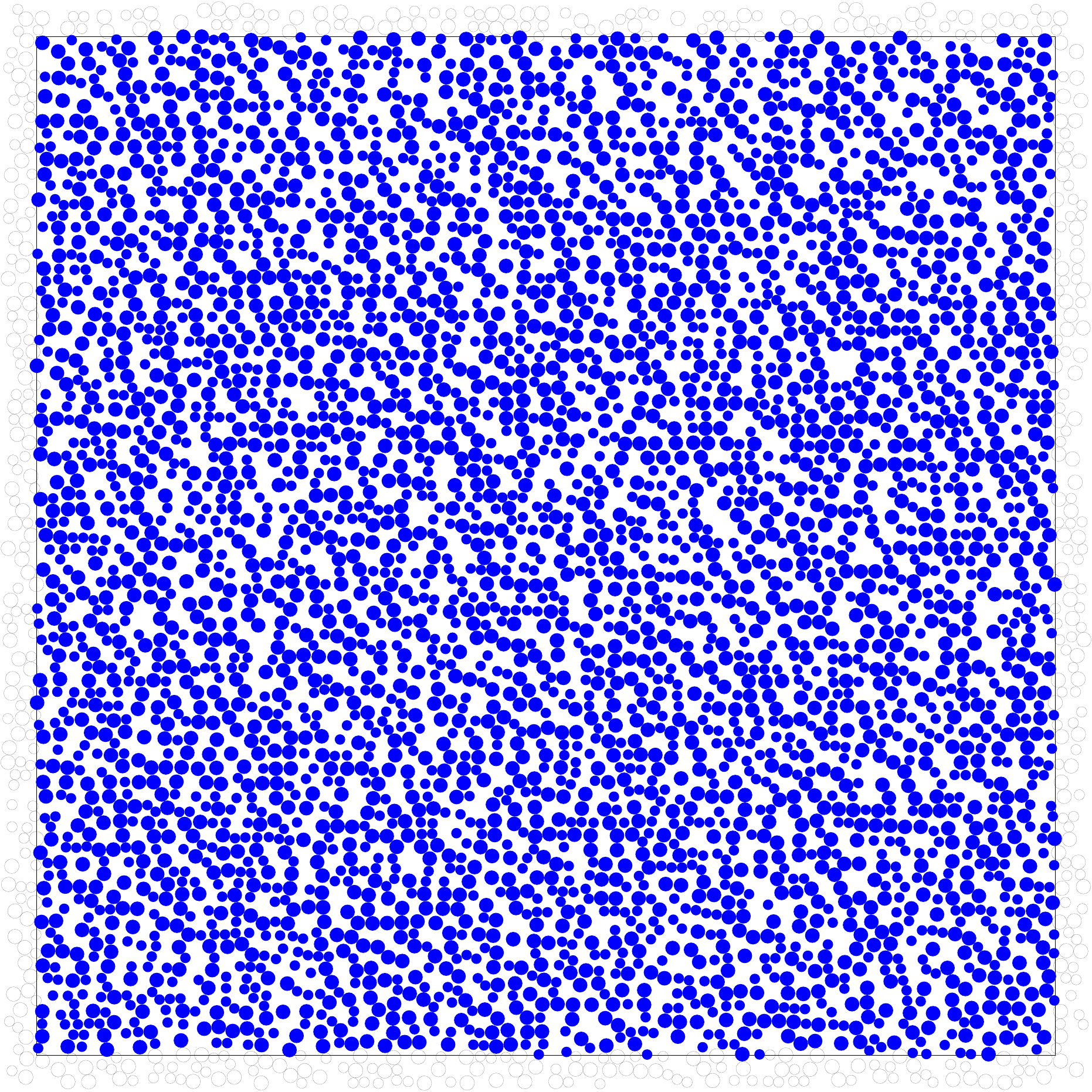}}

  \subfigure{\parbox{1cm}{\vspace{-3.5cm}CD \break (e)-(h)}} 
  \subfigure{\includegraphics[scale=0.2] {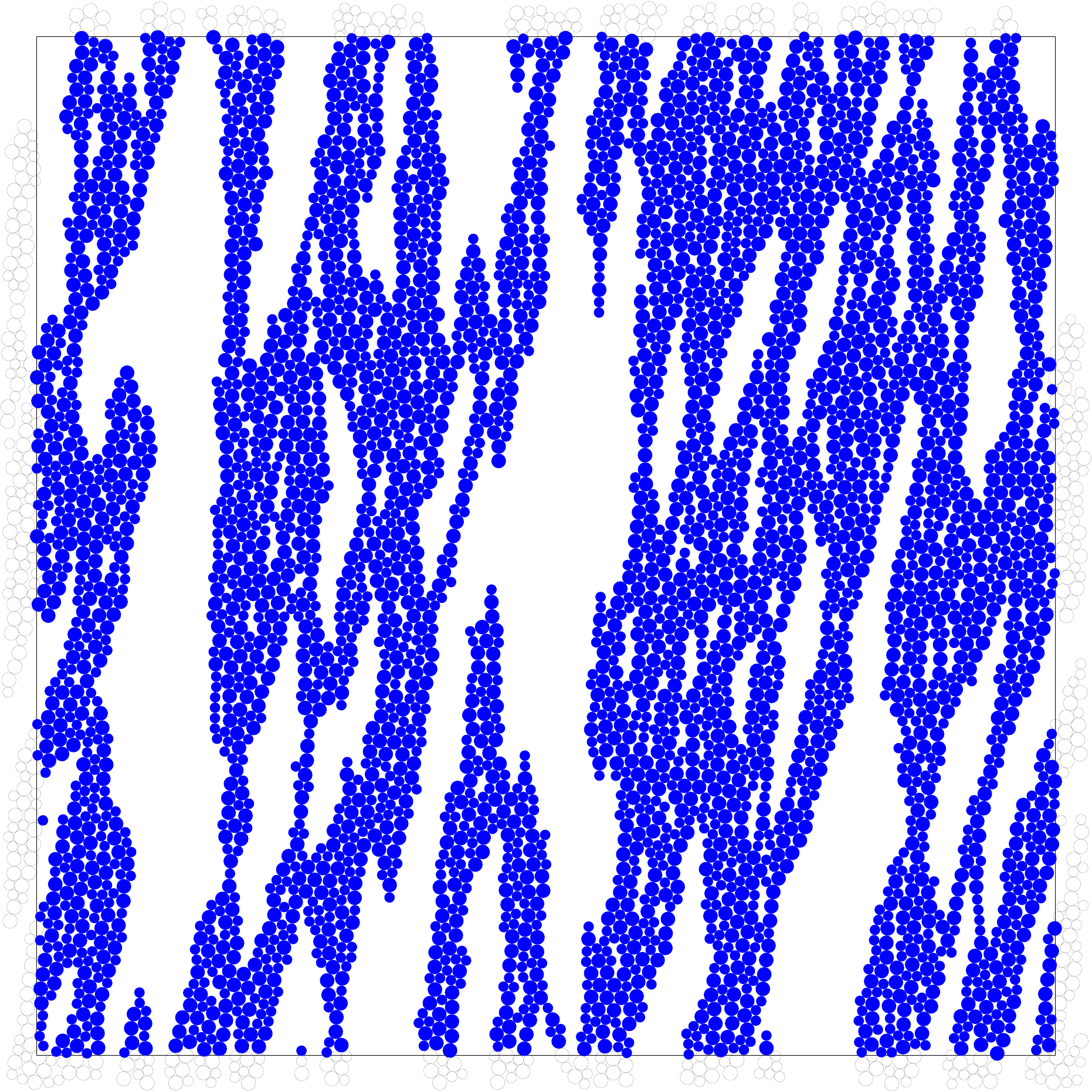}}
  \subfigure{\includegraphics[scale=0.2] {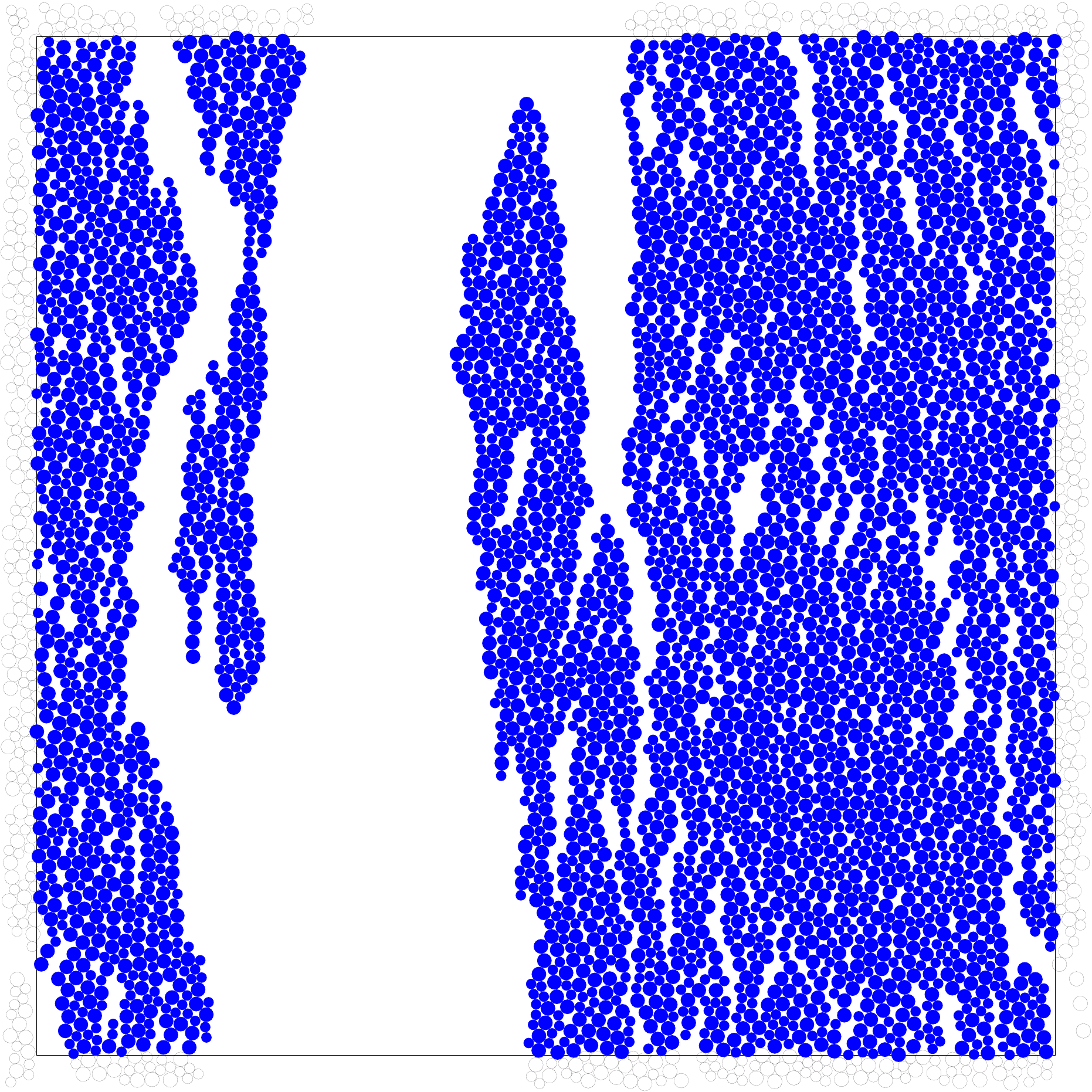}}
  \subfigure{\includegraphics[scale=0.2] {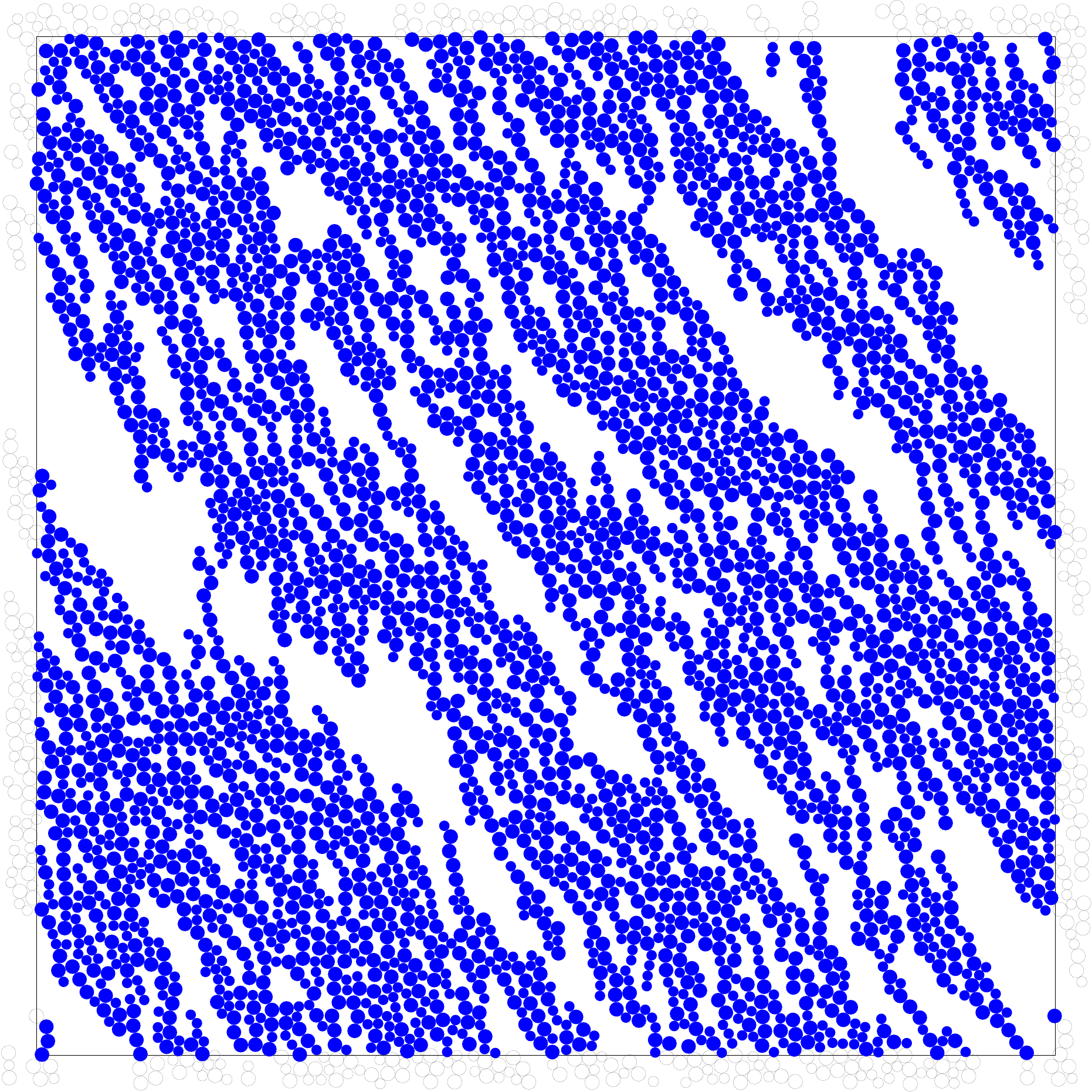}}
  \subfigure{\includegraphics[scale=0.2] {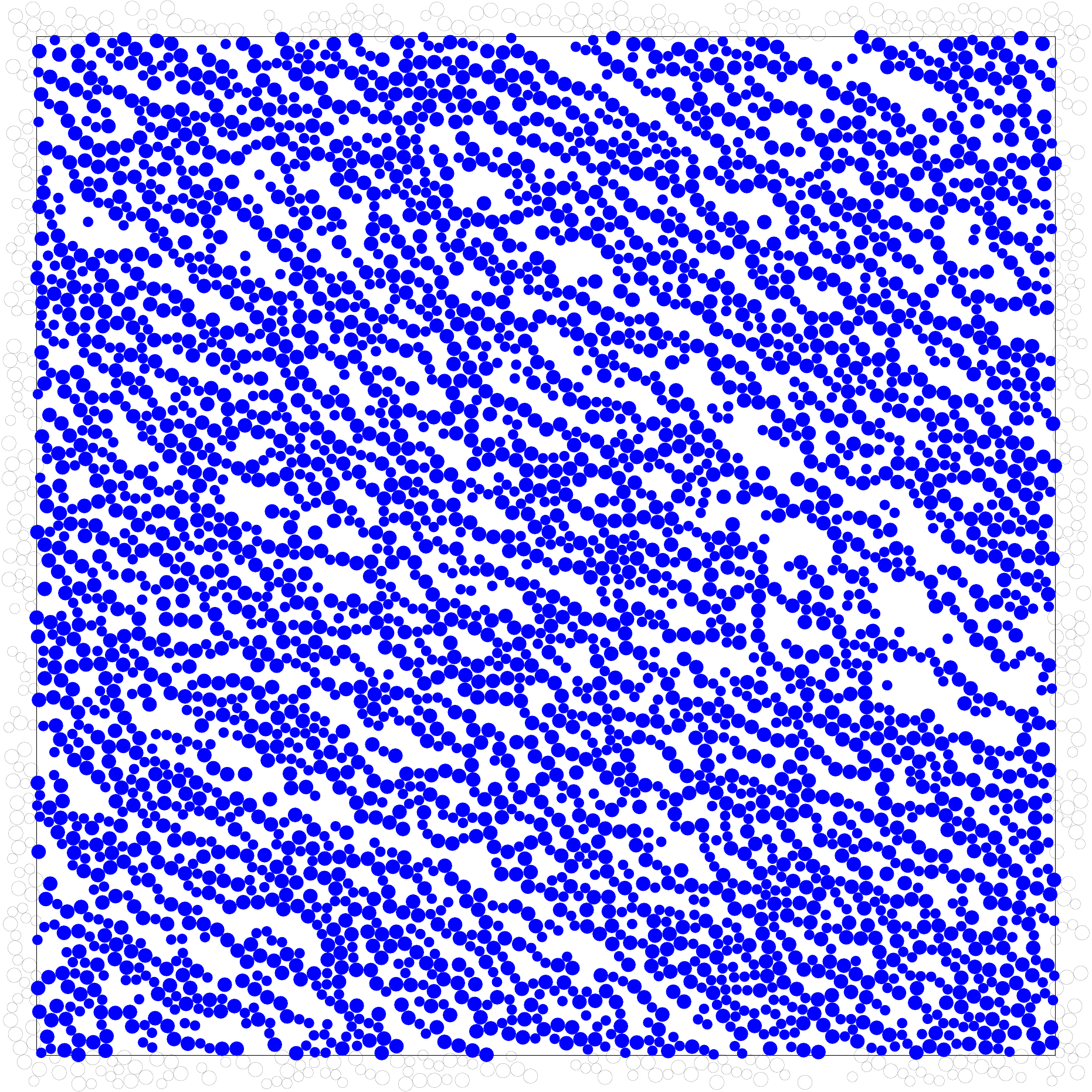}}


 \end{center}
  \caption{Particle configurations at varying Mason number $\Mn \propto \dot \gamma/H^2$ in the RD and CD models (top/bottom row).  From left to right: $\dot\gamma/H^2=10^{-5}$, $10^{-1}$, $10$, $10^4$. }
  \label{fig:conf}
\end{figure*}

 \section{Microscopic structure}
What microscopic features of the system correlate with (changes in) the bulk rheology? To gain insight into the qualitative differences between the RD and CD models apparent in the flow curves, we now seek to characterize the microstructural evolution of MR fluids in steady shear flow as a function of the Mason number and volume fraction. Our goals are twofold. First, at sufficiently low Mason numbers one expects MR fluids to quasistatically sample states that minimize the sum of the elastic and magnetic potential energies, with viscous forces playing a negligible role. Hence we will seek evidence that our simulations are approaching, if not definitively reaching, this asymptotic regime. Second, the qualitatively different flow curves in the RD and CD models should be reflected in their microstructure. Therefore we seek evidence of qualitative differences, in general, and competing clustering mechanisms in particular.

Snapshots of the system are presented in Fig.~\ref{fig:conf} for both the RD (top row) and CD (bottom row) model and several values of $\Mn$ (increasing from left to right). In the RD model there is an apparently homogeneous and isotropic microstructure in the Newtonian regime at high $\Mn$. Chains gradually emerge as the Mason number is lowered and magnetic interactions increasingly dominate. The microstructural evolution in the CD model is comparatively complex. There is anisotropy even in the Newtonian regime. More strikingly, large clusters appear at intermediate $\Mn$. These clusters are more compact than the chains that eventually form at low $\Mn$, and which resemble those seen in the RD model.
In the remainder of this Section we quantify the above observations.

\begin{figure}[tb!]
 \begin{center}
\subfigure{\hspace{-0.2cm}\includegraphics[height=3.7cm] {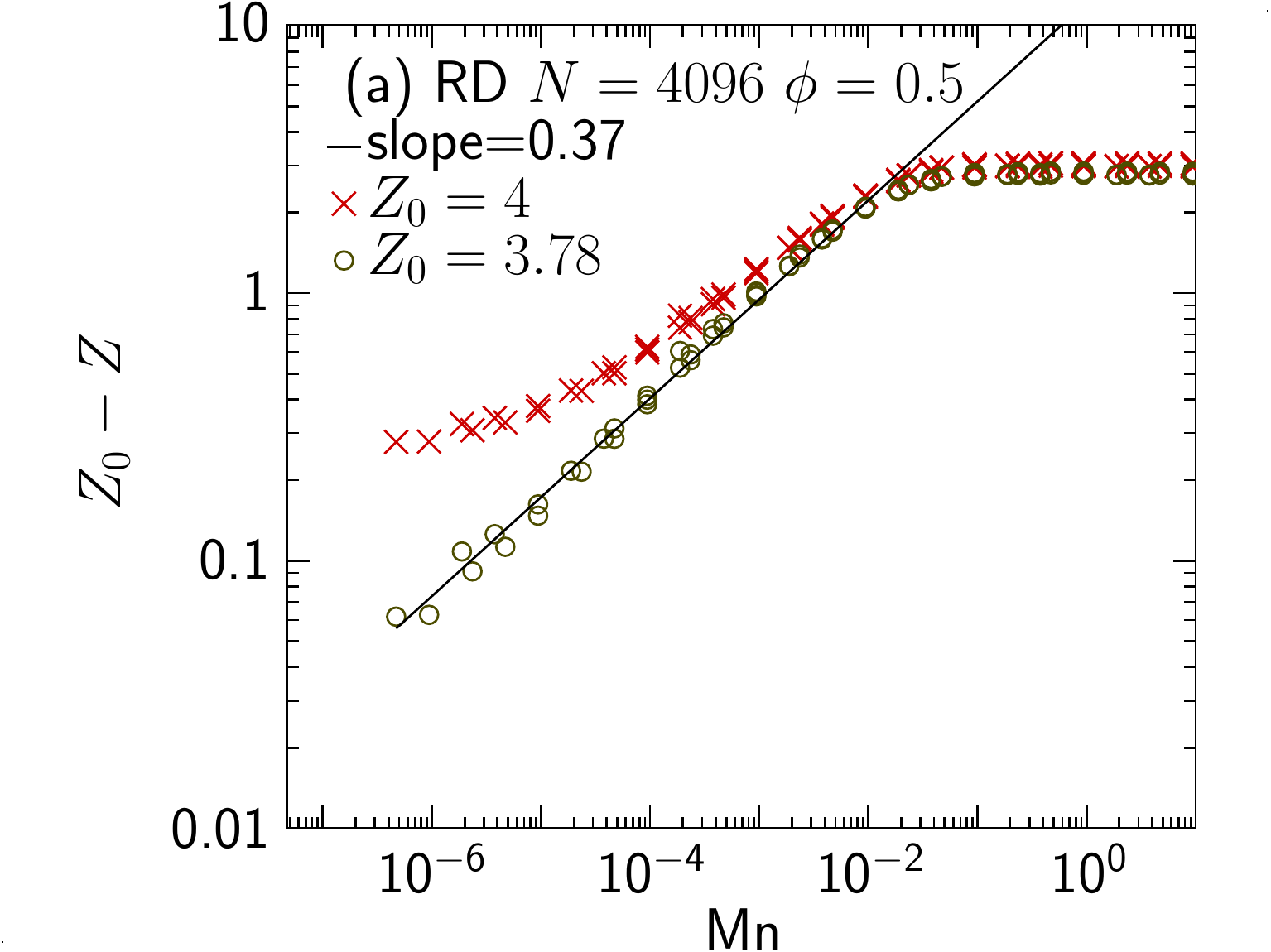}   
           \hspace{-1.5cm}\includegraphics[height=3.7cm] {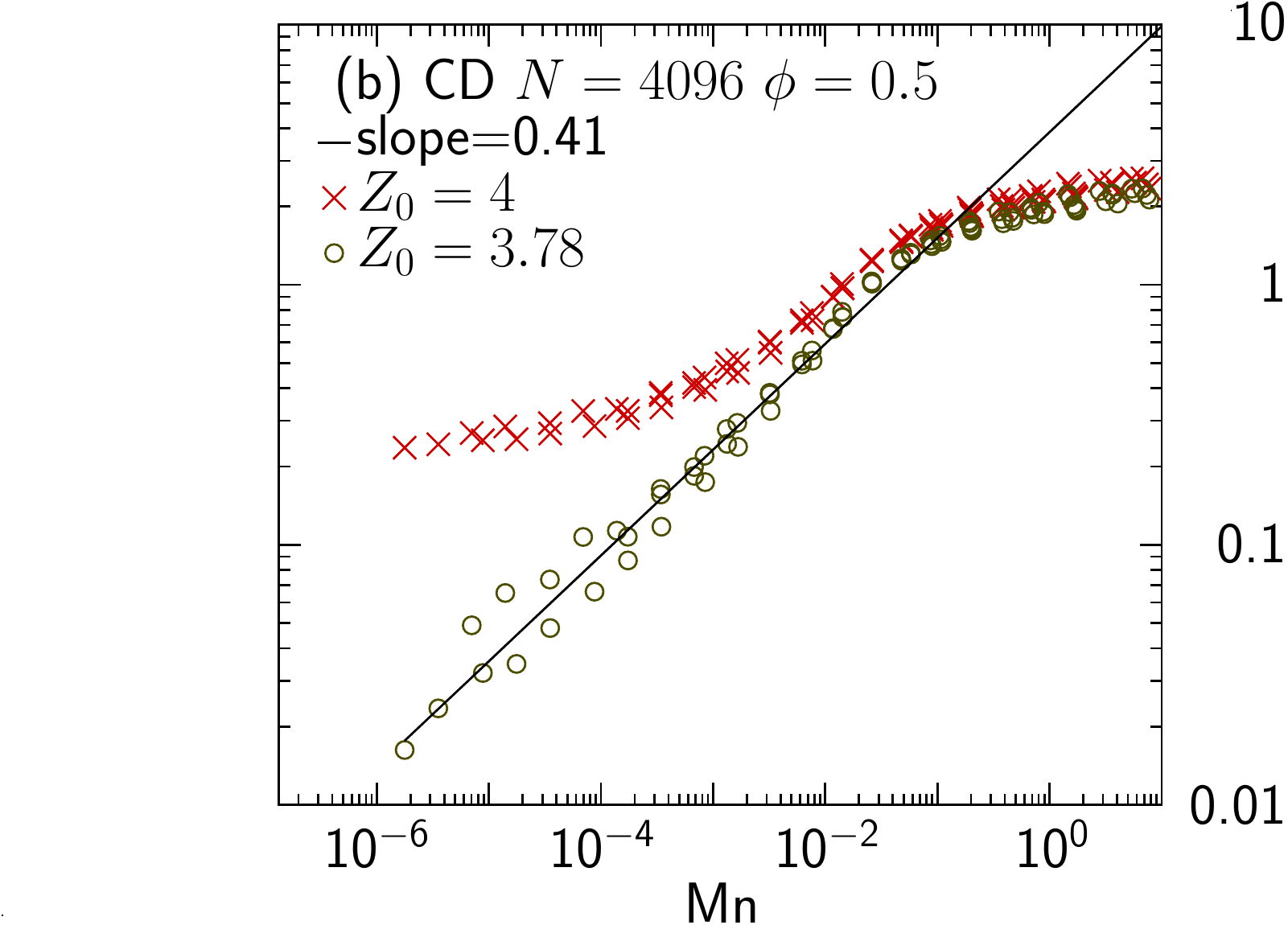}  }

\subfigure{\includegraphics[height=3.7cm] {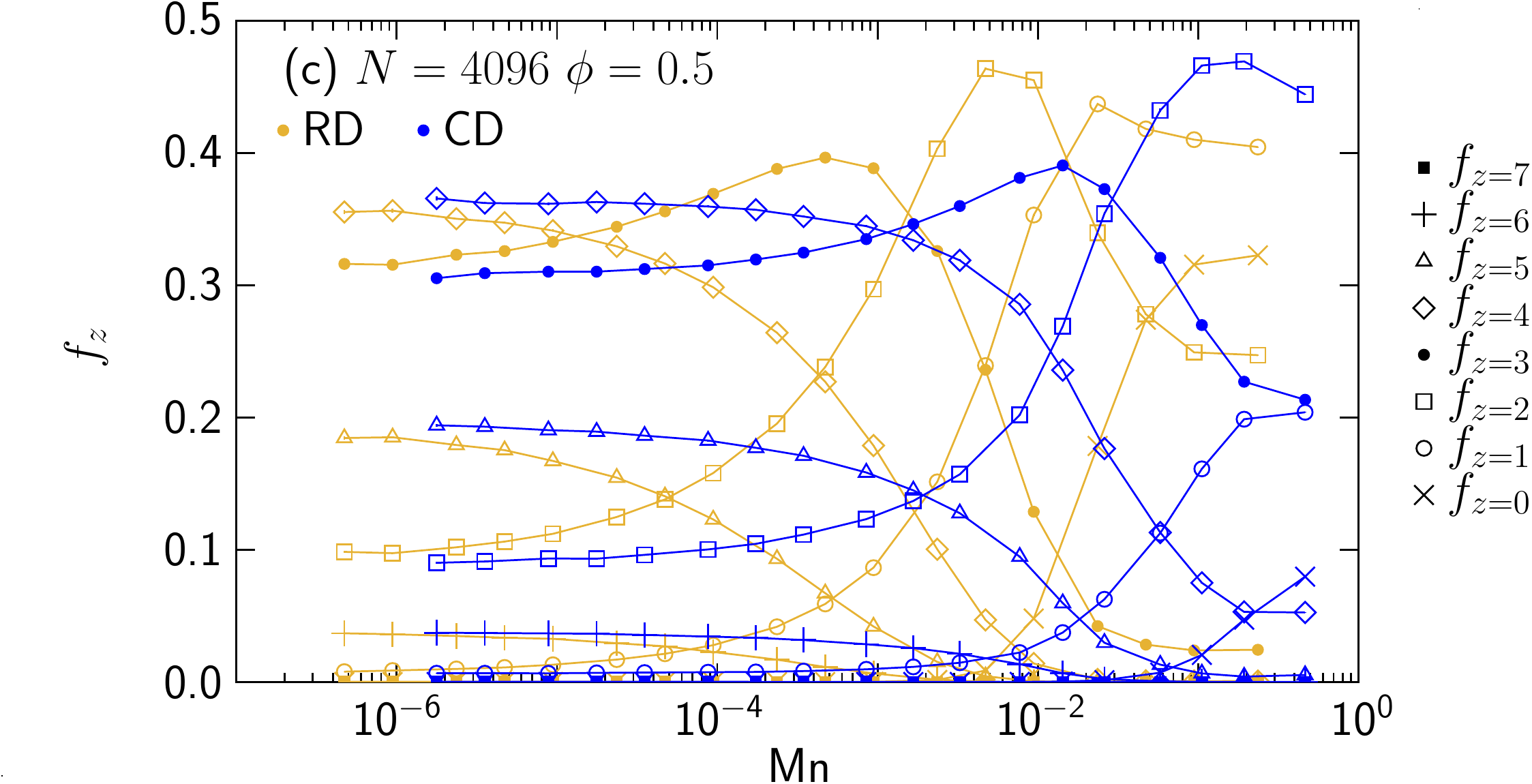}}
  \end{center}
   \caption{ (a) Evolution of the mean contact number $Z$ as a function of the Mason number in (a) the RD model and (b) the CD model. (c) The fractions of particles with $0, 1,  \ldots 7$ contacts approach asymptotic values at low $\Mn$ that appear to be the same in the RD and CD models.}
  \label{fig:z0_phi}
\end{figure}

\subsection{Coordination}

At asymptotically low Mason numbers, particles must follow quasistatic trajectories that track minima of the (magnetic and elastic) potential energy as parameterized by the strain coordinate; viscous dissipation can only play a subdominant role. Therefore the $\dot \gamma \rightarrow 0$ (and hence $\Mn \rightarrow 0$ at fixed $H$ and $\phi$) limit of the flow curve $\sigma(\dot \gamma; H, \phi)$, i.e.~the ``true'' yield stress, must be the same in both the RD and CD models. To obtain evidence of the approach to this asymptotic limit, we now study the evolution of the mean contact number $Z$ at low $\Mn$.
$Z$ plays an important role in determining whether a network (e.g.~the contact network of a soft sphere packing) can elastically support a load. Here we present evidence that microstructure is indeed independent of the damping mechanism in the limit of vanishing strain rate.

In the absence of a magnetic field, a packing jams (develops a shear modulus and yield stress) when it satisfies Maxwell's \cite{maxwell1864} counting argument $Z \ge Z_{\rm iso} = 2D + {\cal O}(1/L^{D-1})$, where $Z$ is the mean number of contacts per particle calculated after removing non-load bearing ``rattlers'' and $D$ is the spatial dimension. The correction term accounts for boundary effects. For several reasons, one expects magnetic interactions to generate elastically rigid states with mean contact numbers $Z < Z_{\rm iso}$. First, magnetic interactions enhance boundary effects due to clusters' anisotropic shape \cite{tighe2011b,goodrich2012,dagois-bohy2012,goodrich2014}. They also introduce long range, potentially tensile forces between particles. The connectivity of the contact network still provides a useful characterization of the flow, however, because the tail of the magnetic interaction potential falls off rapidly with distance, so that the strongest magnetic forces are between nearest neighbors. Finally, when chains are present at low Mason numbers, to minimize the potential energy the particles will arrange such that nearest neighbor magnetic forces are nearly always tensile. Tensile forces increase the likelihood of a structure containing states of self stress, which reduce the number of contacts needed to render a structure rigid. Maxwell's original counting argument can be extended to correctly count states of self stress as described by Calladine \cite{calladine1978}, a procedure which has also been adopted for studying dense sphere packings \cite{lois2008,wyart2005,head2007}.

We now empirically determine the scaling of $Z(\Mn)$ at low Mason number, including its asymptote $Z_0$ as $\Mn$ tends to zero. The contact number is a ``bare'' $Z$ with no correction for rattlers.  
Recalling that  $Z_{\rm iso} \approx 4$ in large systems with no magnetic interactions, in Fig.~\ref{fig:z0_phi}a  (crosses) we plot $4 - Z$ as a function of Mason number the RD model with $\phi = 0.5$ and $N = 4096$. While in the Newtonian regime at high Mason number the contact number is insensitive to $\Mn$, the quantity $4-Z$ decreases ($Z$ increases) as chains form in the magnetically dominated regime. There is an apparent leveling off at the lowest simulated values of $\Mn$, suggesting that $Z$ asymptotes to a value below 4. In order to estimate this value, we plot $Z_0 - Z$ (Fig.~\ref{fig:z0_phi}b, circles) and adjust the value of $Z_0$ to find the cleanest power law at low $\Mn$. For $Z_0 = 3.78$ we find a  power law $Z_0 - Z \sim \Mn^a$ with exponent $a \approx 0.37$. Interestingly, a similar scaling relation $Z_{\rm iso} - Z \sim \dot \gamma^{0.38}$ has been observed in hard sphere suspensions with no magnetic interactions \cite{lerner2012}.  
 In Fig.~\ref{fig:z0_phi}b we plot the same quantities for the CD model, finding nearly identical values for the extrapolated asymptote $Z_0 = 3.78$ and exponent $a \approx 0.41$. We note that the small difference in the exponent $a$ seems to be entirely due to the $Z$ factor in the definitions of $\Mn$, which differs between the RD and CD model. If we fit both data sets using the same definition of $\Mn$ the exponent $a$ is the same for both models within statistical error.
We have verified that both $a$ and $Z_0$ are independent of $N$ for sufficiently large system sizes, and that their values vary little over a wide range of volume fractions (not shown). Between $\phi = 0.3$ and $0.7$ the value of $Z_0$ trends from $Z_0 \approx 3.78$ to $3.85$ and eventually approaches $Z_0\approx4$ as $\phi\rightarrow\phi_c$ for both RD and CD models.

To further verify that the microstructure in both models is statistically indistinguishable in the zero Mason number limit, we now investigate the distribution of local contact numbers. In Fig.~\ref{fig:z0_phi}c we plot the fraction of contacts $f_z$ having $z$ contacts, for $z = 0 \ldots 7$, in both the RD and CD models. At large Mason numbers, $f_z$ differs strongly between the two models, both in its magnitude and its trend with $\Mn$. However, at low $\Mn$ each fraction $f_z$ approaches a constant value. To within the accuracy of our measurements, the asymptotes of each $f_z$ are equal in the RD and CD models. 

To summarize our results on contact number, we have seen that for two types of damping, the flow samples states with the same mean value $Z_0$ of the contact number, as well as the same contact number frequencies $\lbrace f_z \rbrace_{z = 0 \ldots 7}$. This provides strong evidence that steady shear flows in the RD and CD models sample the same ensemble of states as $\Mn$ tends to zero. However it is also clear that the asymptotically low-$\Mn$ regime is at the limit of the lowest Mason numbers we can practically access numerically.

\begin{figure}[tb]
 \begin{center}
   \includegraphics[scale=0.2] {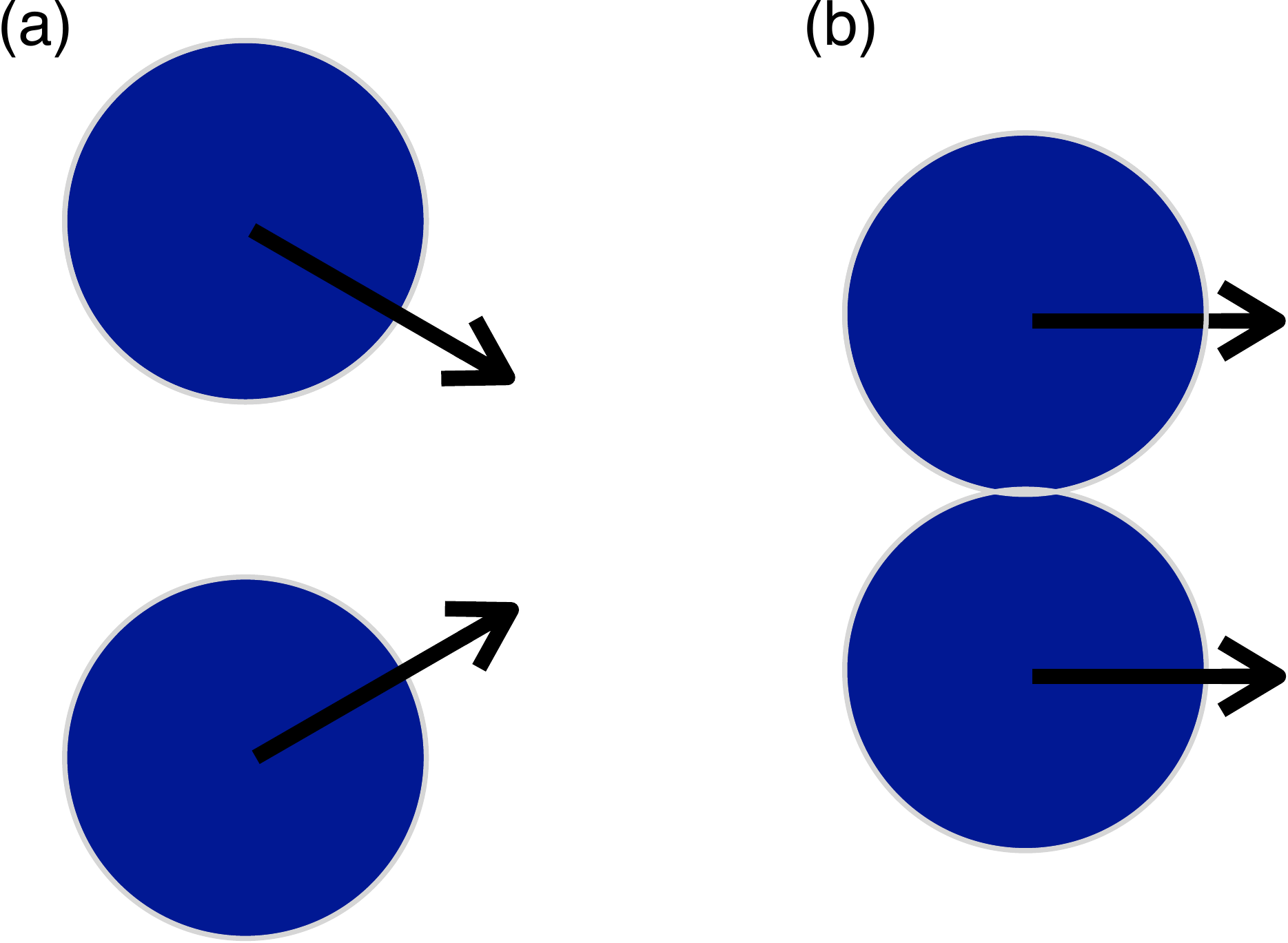}
 \end{center}
 \caption{Particles (a) before and (b) after colliding in the contact damping (CD) model.}
 \label{fig:collision}
\end{figure}

\begin{figure}[tb]
 \begin{center}

   \subfigure{\hspace{-0.2cm}\includegraphics[height=3.7cm] {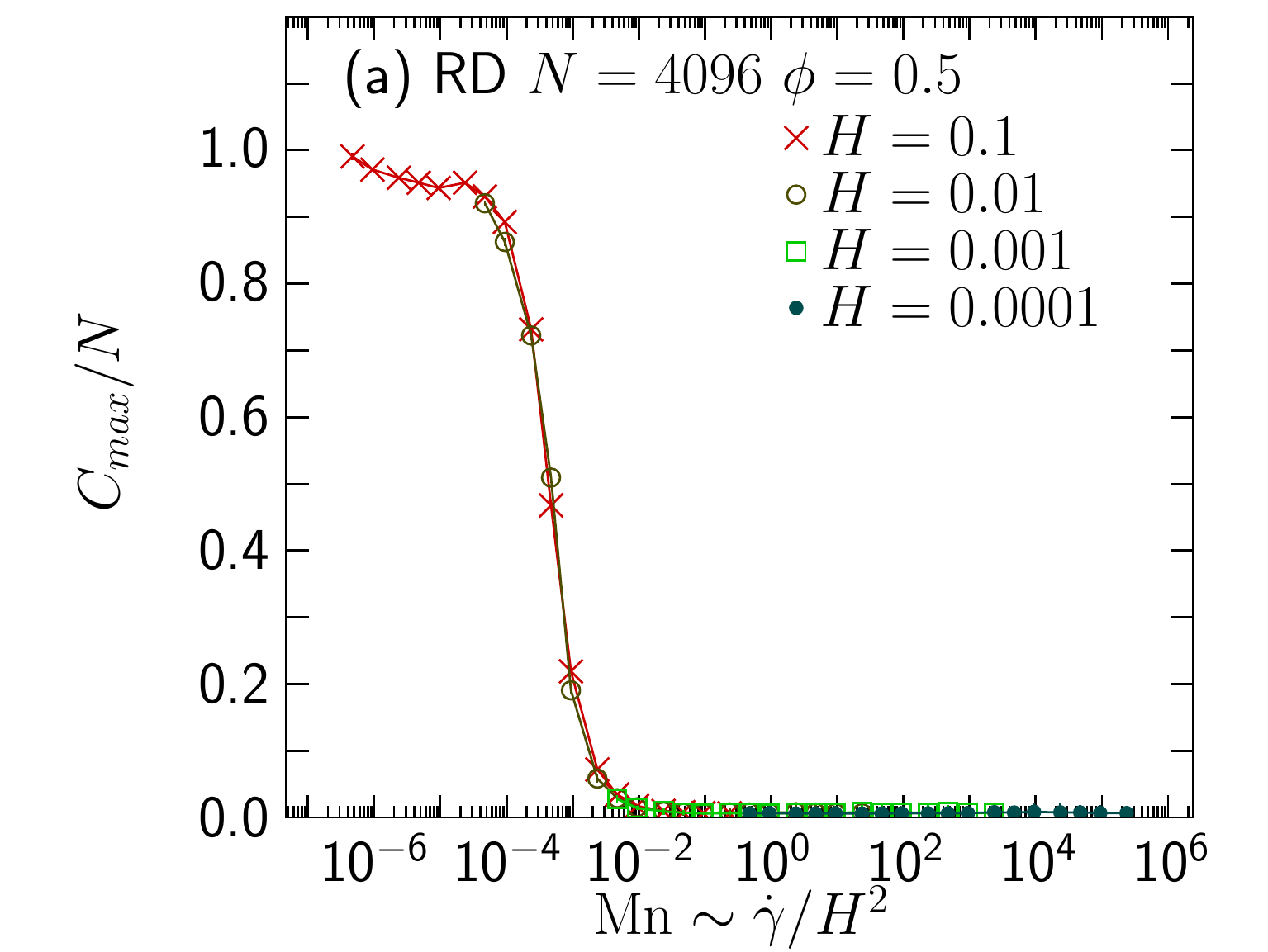}
             \hspace{-1.5cm}\includegraphics[height=3.7cm] {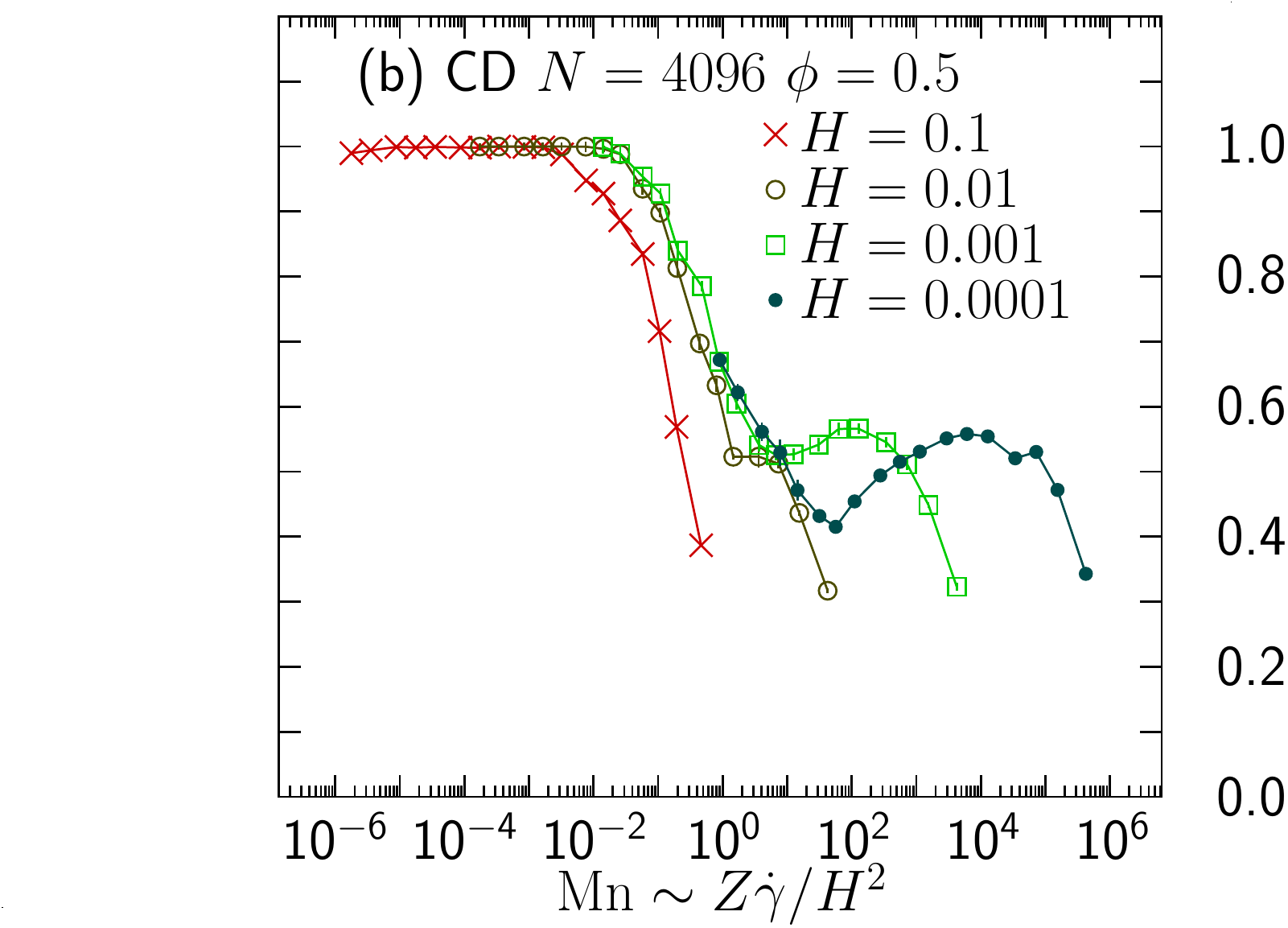}}

 \subfigure{\hspace{-0.2cm}\includegraphics[height=3.7cm] {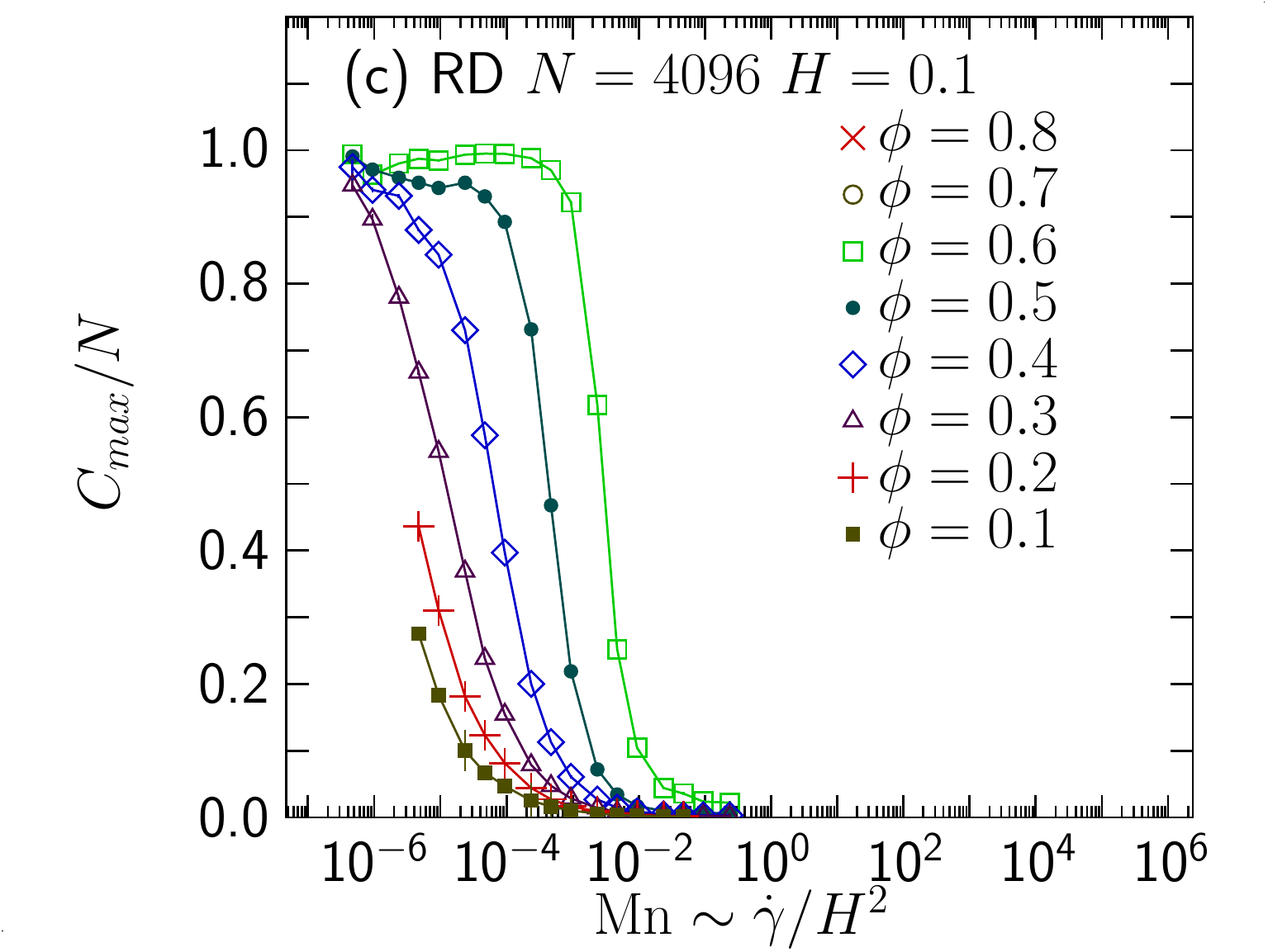}
             \hspace{-1.5cm}\includegraphics[height=3.7cm] {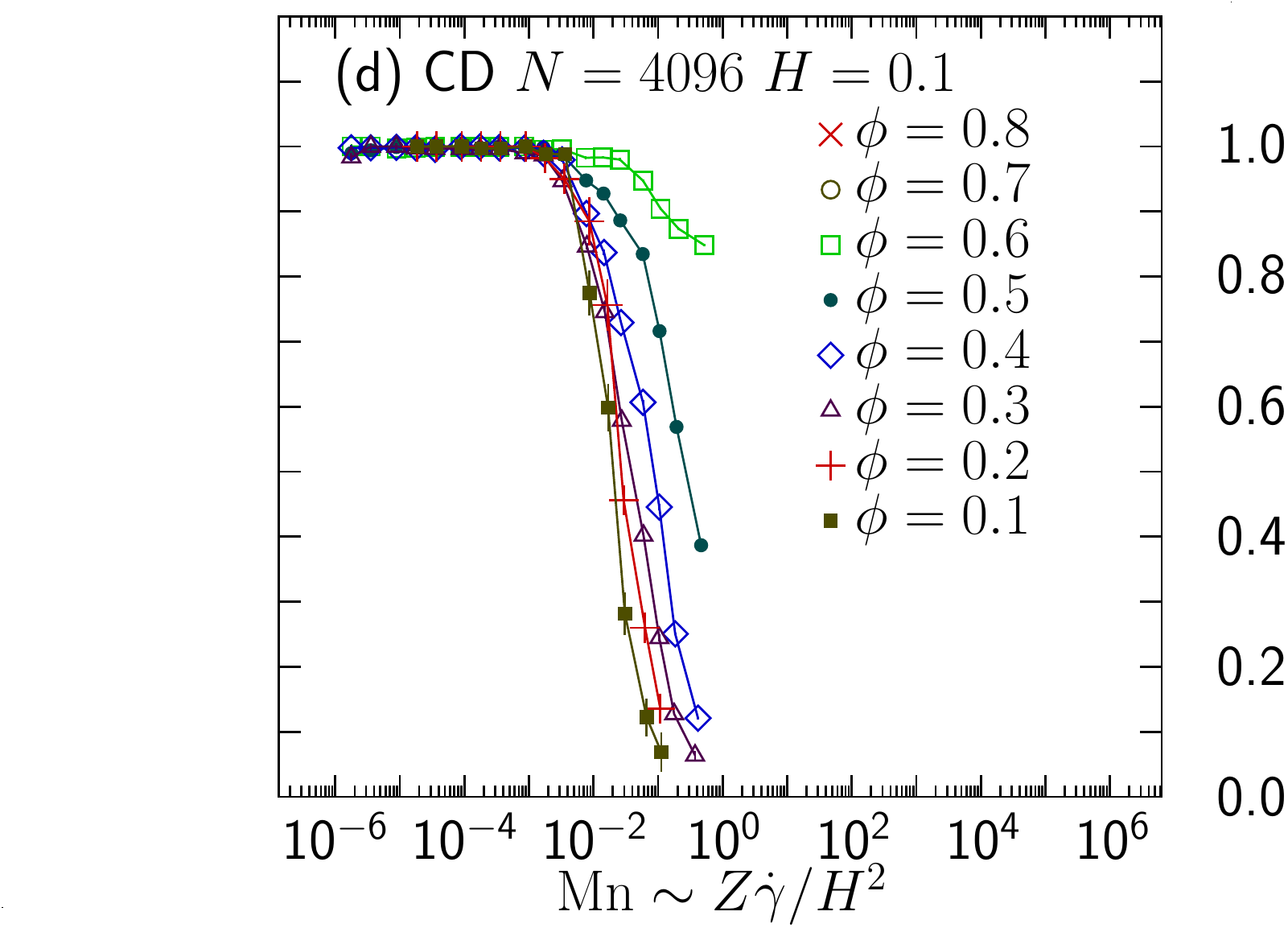}}

 \end{center}
  \caption{
  Top: Average size  $C_{\rm max}$ of the largest cluster in the system for varying field strength ${H}$ at fixed packing fraction $\phi$ (top) and varying $\phi$ at fixed $H$ (bottom) for the RD (left column) and CD (right column) model.
  }
  \label{fig:cluster}
\end{figure}

\begin{figure*}[tbh!]
 \begin{center}
   \subfigure{\includegraphics[height=6cm] {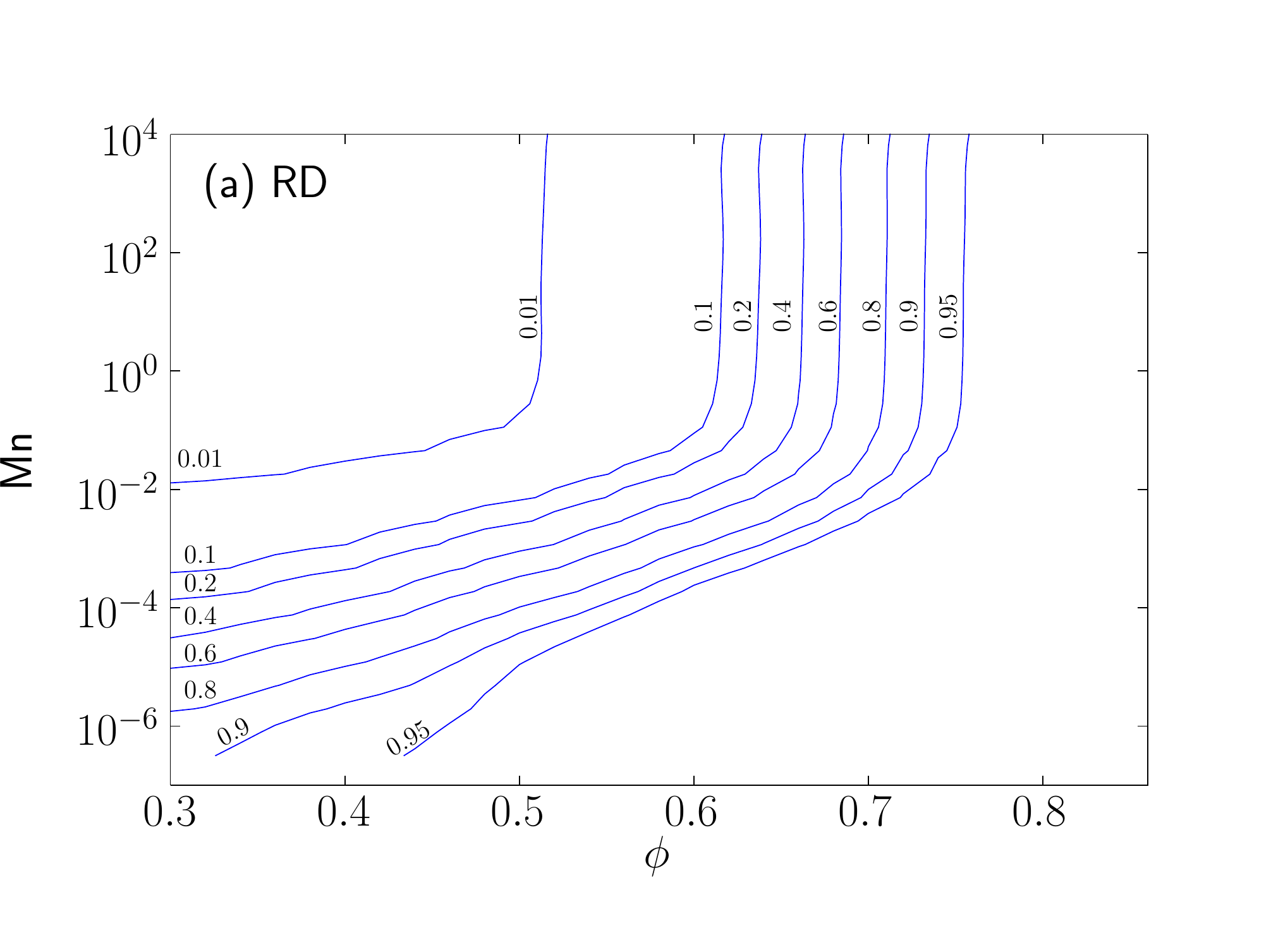}}
   \subfigure{\includegraphics[height=6cm] {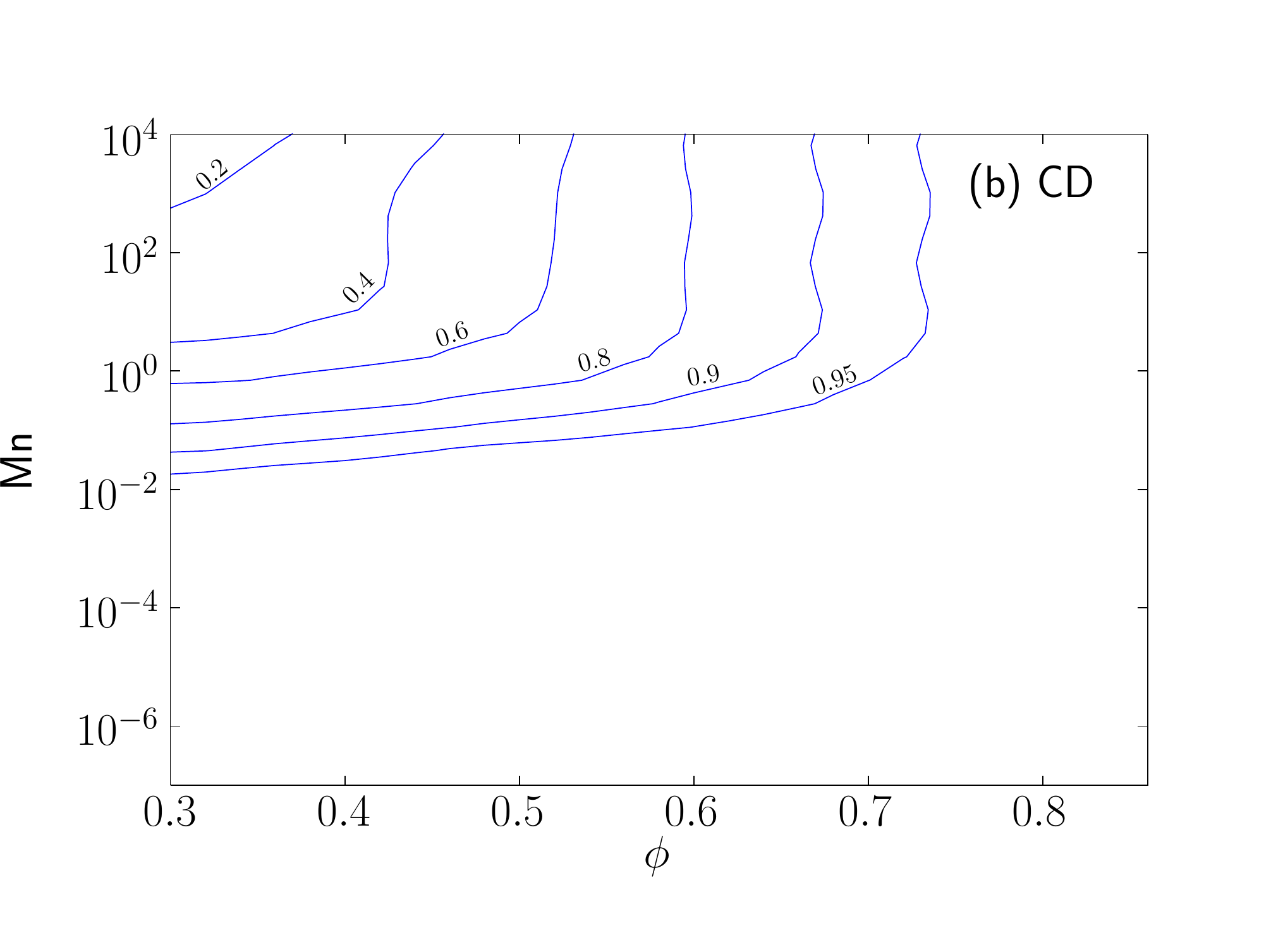}}
 \end{center}
  \caption{Contour plot showing the largest cluster size $C_\text{max}/N$ for the RD (left) and CD (right) models.  Since the same $\Mn$ can correspond to several different combinations of $H$ and $\dot\gamma$, the highest obtained value for any given combination of $\Mn$ and $\phi$ were used to generate the contours. All data is for $N=4096$. }
  \label{fig:cluster_plane}
\end{figure*}

\subsection{Cluster statistics}
From the snapshots in Fig.~\ref{fig:conf} it is apparent that the build-up of clusters proceeds differently in the RD and CD models. Here we present evidence that, whereas clustering in the RD model is driven solely by magnetic interactions, inelastic collisions between particles provide a second, unrelated clustering mechanism in the CD model. Clustering due to inelastic collisions is well known in granular gases: particles exit a collision with a lower relative velocity, and hence tend to stay closer together\cite{goldhirsch1993,mcnamara1996}. In the CD model, and unlike the RD model, dissipation indeed occurs via collisions. Moreover, due to the model's overdamped dynamics, particles remain in contact after colliding; i.e.~their relative velocity is zero (see Fig~\ref{fig:collision}). 

We now seek to quantify the degree of clustering in the RD and CD models. If, as hypothesized above, inelastic clustering is present only in the CD model, one should find differences in, e.g., the  time-averaged size $C_{\rm max}$ of the largest cluster in the system.
We consider a particle to belong to a cluster if it has a non-zero overlap with any other particle belonging to that cluster. 
A size $C_{\rm max} = N$ indicates that every particle participates in one cluster.
In the left panel of Fig.~\ref{fig:cluster}, we plot $C_{\rm max}/N$
in the RD model as a function of Mason number. Note, first, that the data collapse with $\Mn$. Second,  there are no clusters of significant size at high values of $\Mn$, when the rheology is Newtonian; however, there is a sharp rise in cluster size 
below $\Mn \sim 10^{-3}$, coinciding with the magnetically-dominated regime in the flow curve (c.f.~Fig.~\ref{fig:s_gdot_4krdcd}). We conclude that ``clusters'' in the RD model correspond to chains supported by magnetic interactions.

As with the flow curves, the clustering data for the CD model (Fig.~\ref{fig:cluster}b) are comparatively complex. First, there is a degree of clustering even in the Newtonian regime. Second, the data do not collapse with Mason number. This clearly indicates the presence of a clustering mechanism independent of magnetic interactions, which we identify with inelastic collisions. Finally, for sufficiently low $\Mn$ all particles participate in a single cluster, as in the RD model.

$C_{\rm max}$ also shows qualitatively different  dependence on the volume fraction $\phi$ in the two models. In Fig.~\ref{fig:cluster}c and d we plot $C_{\rm max}/N$ as a function of $\phi$ at high field strength $H=0.1$. It is clear that the clustering in the RD model shows a much stronger $\phi$ dependence than in the CD model. This $\phi$ dependence is consistent with our previous observations that the $\Mn$ needed to reach the plateau in $\sigma$ decreases as $\phi$ is lowered, and that this shift is stronger in the RD model. Here we also include  data for $\phi=0.1$ and $0.2$. At these low values of $\phi$, the Mason number needed to reach the yield stress plateau is currently inaccessible in simulation. However there is an increase in $C_{\rm max}$ at low $\Mn$, suggesting that a plateau does emerge at lower $\Mn$. 
Another way of visualizing the $\phi$ dependence over a wider range of $\Mn$ is shown in Fig.~\ref{fig:cluster_plane}a and b, where we plot contours of $C_{\rm max}/N$ over the same range of $\phi$ and $\Mn$ for the RD and CD model, respectively. Differences are most easily seen by considering, e.g., the $C_{\rm max}/N = 0.9$ contour. In the CD model this contour is nearly independent of $\phi$, up to some maximum $\phi$ close to $\phi_c$. This suggests that large clusters appear in the CD model at a characteristic Mason number that is independent of $\phi$. In the RD model, by contrast, the value of $\Mn$ where clusters appear is an increasing function of $\phi$.

\begin{figure}[tb]
 \begin{center}
  \subfigure{\includegraphics[height=3.7cm] {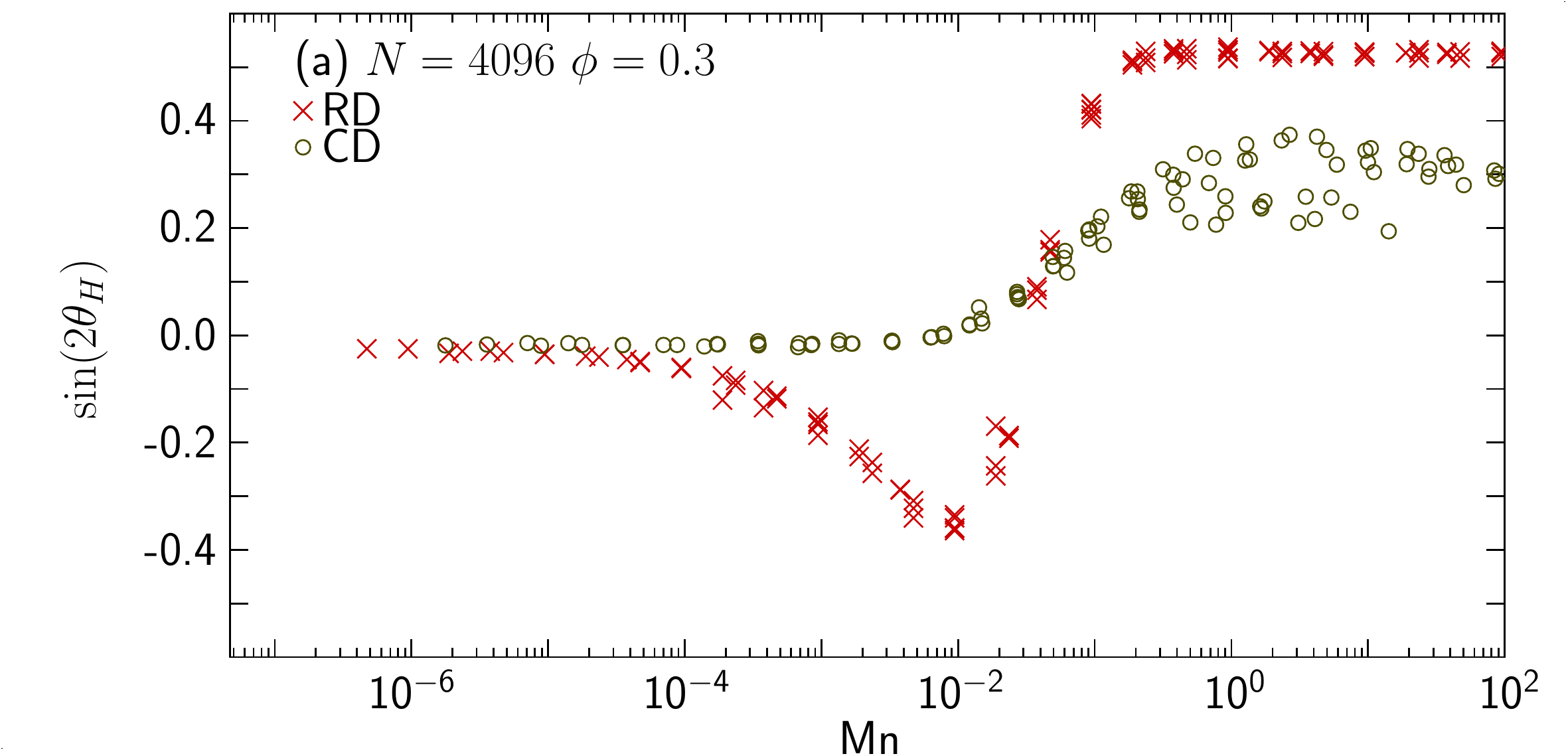}}
  \subfigure{\includegraphics[height=3.7cm] {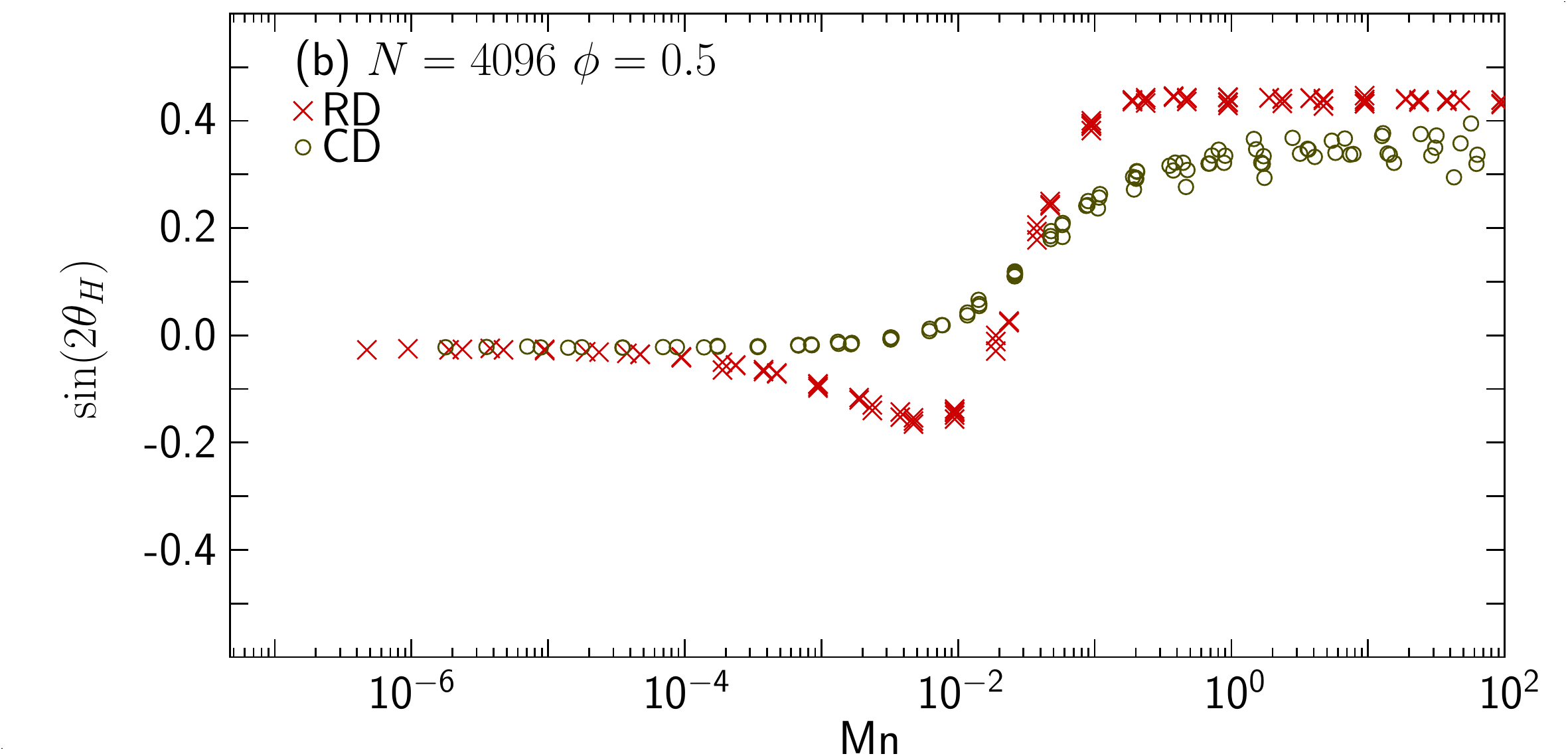}}
  \subfigure{\includegraphics[height=3.7cm] {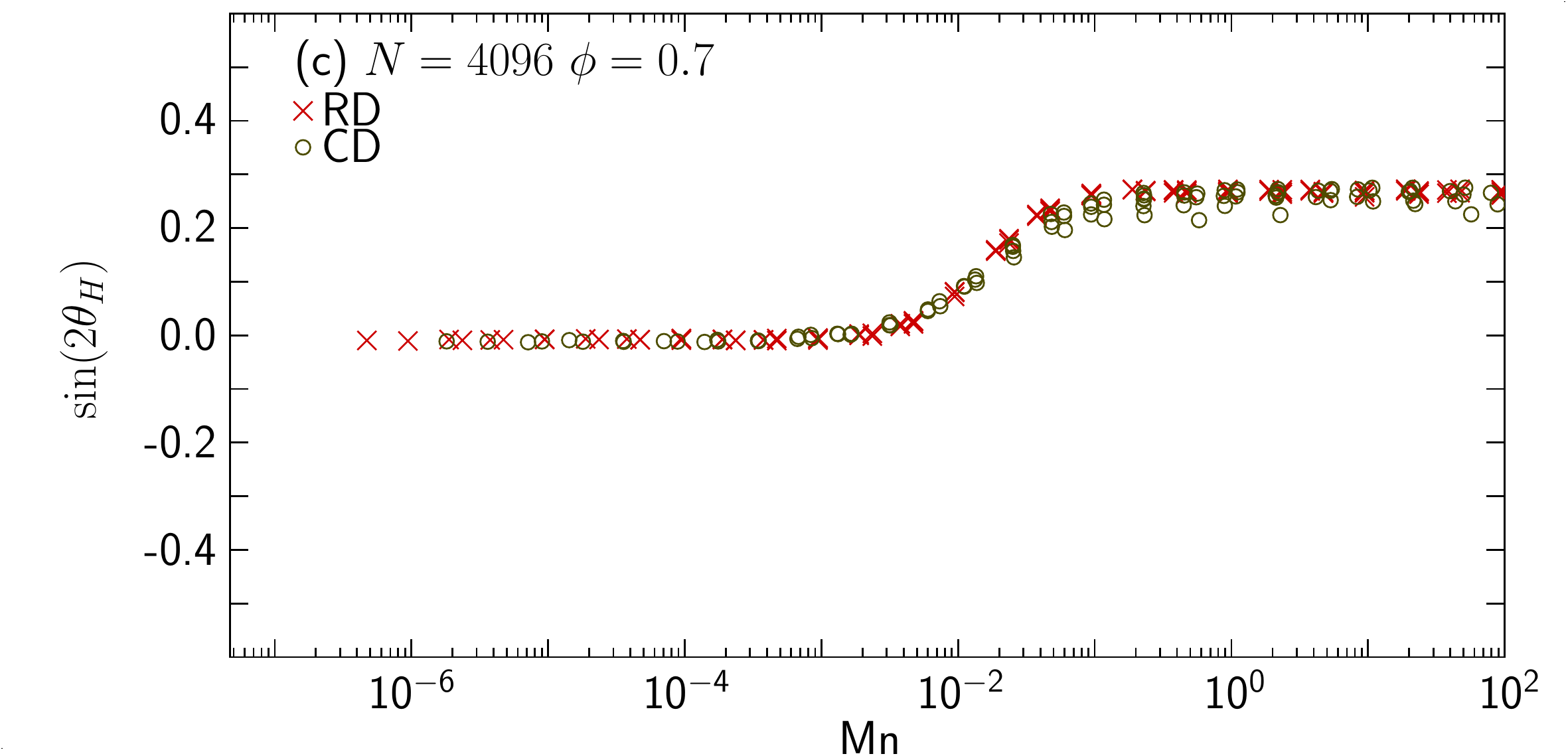}}
  \end{center}
   \caption{ Comparison of the average bond angle $\theta_H$ as a function of $\Mn$ for the RD and CD model. The panels correspond to three different values of $\phi$ }
  \label{fig:angle_comp}
\end{figure}

\begin{figure}[tb]
 \begin{center}
  \subfigure{\includegraphics[height=3.7cm] {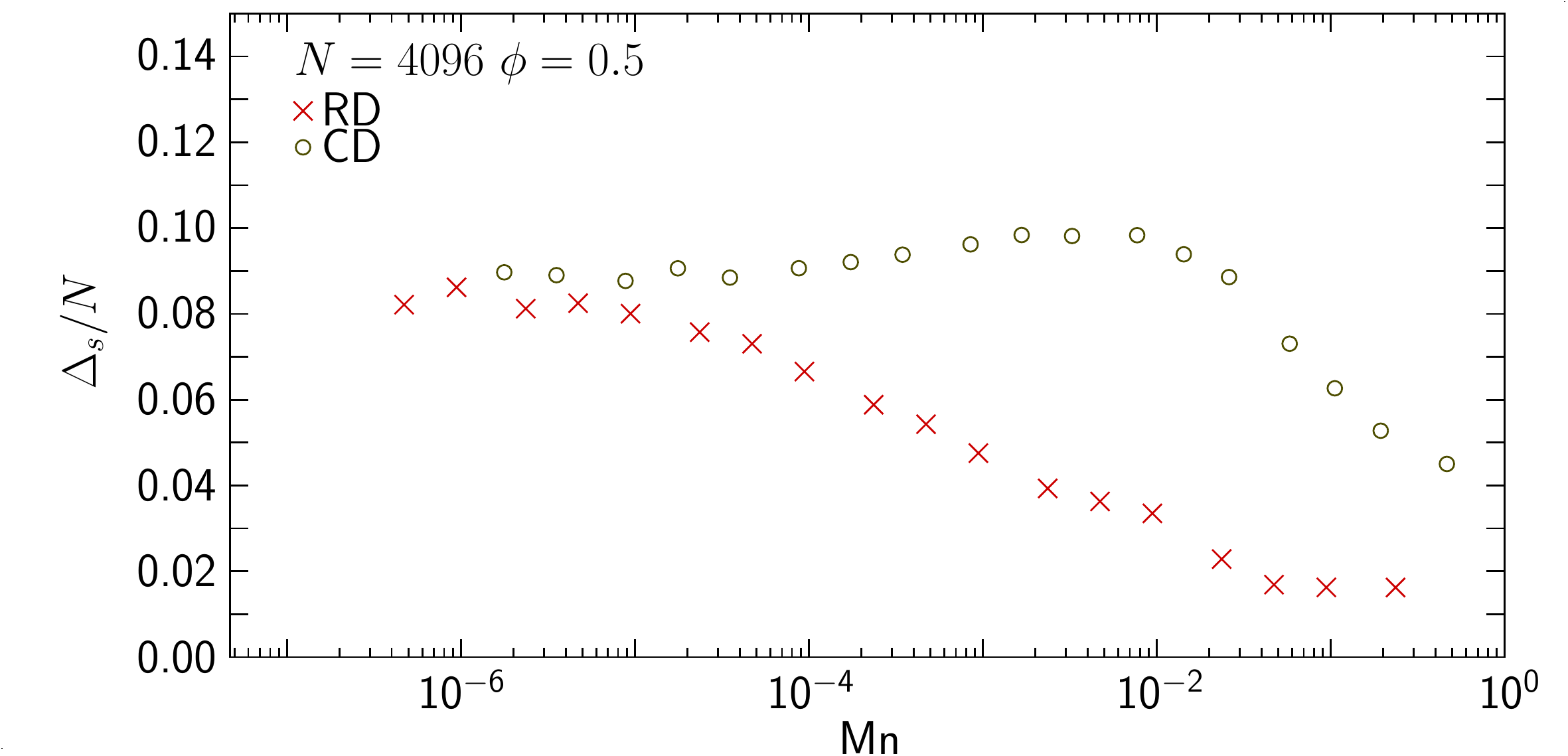}}
  \end{center}
   \caption{ The average number of small triangles per particle vs Mason number for the RD and CD model. }
  \label{fig:tris_comp}
\end{figure}

In the snapshots of Fig.~\ref{fig:conf}, it is also evident that the orientation of the emergent chains differs between RD and CD flows. To characterize chain orientation, we study $\theta_H$, defined as the average contact angle measured counter-clockwise relative the magnetic field axis (the $\hat y$-axis), 
\begin{equation}
  \theta_H= \frac{1}{NZ}\sum_{ij} \max\left(\theta(\mathbf{r}_{ij},\mathbf{H}),\theta(-\mathbf{r}_{ij} \,\mathbf{H})\right) \,.
\end{equation}
The sum runs over all bonds with a positive overlap. $\theta(\mathbf{u},\mathbf{v})$ is the angle between the vectors $\mathbf{u}$ and $\mathbf{v}$ measured counterclockwise from $\mathbf{v}$ such that $-\pi/2<\theta(\mathbf{u},\mathbf{v})<\pi/2$, giving $0<\theta_H<\pi/2$. In Fig.~\ref{fig:angle_comp} we plot $\sin{2\theta_H}$ as a function of $\Mn$ for three values of $\phi$. Chains emerge in both models for sufficiently low $\Mn$, indicated by $\sin{2\theta_H} \approx 0$. Likewise, at high $\Mn$ there is a positive bias, indicating that contacts tend to be rotated in a positive sense with respect to $\bf H$ -- as one would expect for collisions due to rapid shear flow.  The height of the plateau at high $\Mn$ shows stronger $\phi$-dependence in the RD model than in the CD model.

There is a dramatic difference in how the two models cross over between the plateaus at high and low $\Mn$. Whereas $\sin{2\theta_H}$ has a sigmoidal shape in the CD model, in the RD model the curve overshoots its low-$\Mn$ asymptote. In this intermediate range of $\Mn$, the two models approach their asymptotic values from opposite ``directions'': chains in the CD model are rotated counter-clockwise with respect to $\bf H$, while chains in the RD model have a clockwise rotation.

One expects the clusters promoted by inelastic collisions to have a different character from the chain-like structures formed due to magnetic interactions -- they should be comparatively compact and isotropic (see Fig.~\ref{fig:conf}). We find the clearest signature of this difference is found by plotting mean number of triangles $\Delta_s$ formed by small particles in contact. For a given cluster size, one expects $\Delta_s$ to be larger for a compact cluster than for an anisotropic, chain-like structure. $\Delta_s$ is plotted in Fig.~\ref{fig:tris_comp} as a function of Mason number. While $\Delta_s$ increases monotonically with decreasing $\Mn$ in the RD model, its evolution is non-monotonic in the CD model. There is a peak at intermediate $\Mn$, which we associate with the more compact collisional clusters, followed by a decrease as those clusters are converted to chains.

The data for cluster size, contact angle, and mean triangle number suggest the following picture. In the RD model chain-like clusters build up monotonically as $\Mn$ is lowered. In  the CD model, in contrast, isotropic clusters form ``earlier'' (at higher $\Mn$) due to inelastic collisions. As $\Mn$ is further lowered and magnetic interactions grow dominant, these compact clusters are reshaped into chains. All relevant observables approach the same asymptotic value in the two models, but they may do so from opposite sides (e.g.~$\theta_H$ an $\Delta_s$). This provides some insight into how the two models' flow curves can display qualitative differences even as they approach the same asymptote. It also provides indirect support of the hypothesis suggested in the previous section, namely that flow curves approach a finite yield stress plateau at inaccessible values of $\Mn$. Of course one might instead infer that the common asymptote of  the RD nor CD flow curves is at zero stress, i.e.~that neither has a true yield stress. However this interpretation is disfavored by Occam's Razor, as all simulated values of $\phi$ show a plateau in the CD flow curve.

\section{Conclusions}

We have studied the steady state rheology of MR fluids interacting via magnetic, elastic, and two distinct viscous forces. Performing numerical simulations that meet or exceed the lowest values of the Mason number accessed experimentally, we have shown that for moderate volume fractions only systems with contact damping (CD) show a clear plateau in their flow curve. Systems with reservoir damping (RD), by contrast, appear to follow a power law $\sigma \sim \dot \gamma^{1-\Delta}$ with $\Delta < 1$ -- which, if extrapolated to zero strain rate, would imply the absence of a dynamic yield stress. We have argued, instead, that viscous forces must play a subdominant role at asymptotically low $\dot \gamma$, and hence either both models possess a yield stress or neither does. The fact that both models display a plateau in their flow curves at sufficiently high volume fractions strongly suggests it is the former: both models possess a dynamic yield stress, with the plateau in the RD flow curve appearing outside the accessible window of $\Mn$ for moderate $\phi$. This interpretation is supported by statistical measures of the microstructure, which approach the same asymptote in each model -- albeit at the edge of our numerically accessible window in $\Mn$. Cluster statistics suggest that the difference in bulk rheology is related to cluster formation due to inelastic collisions in the CD model, which are absent in RD systems. Despite this conclusion, the clear qualitative difference between the RD and CD flow curves evidenced in our simulations is significant for at least two reasons. First, it persists over a wide interval in $\Mn$ including, as previously noted, the lowest values of $\Mn$ accessed experimentally. Second, the difference is clearest for moderate values of $\phi$ far below the jamming transition, which are typical of the MR fluids used in applications.

Our work raises several (computationally expensive) questions that might profitably be addressed in future work. One, of course, is whether the speculated crossover to a plateau is in fact seen in RD flow curves at volume fractions around $0.5$ or lower. We have focused on higher $\phi$ values in part to make the connection to jamming, but also because a yield stress, if present, should be more readily apparent. In practice, $\phi$ values around 0.1 are common in experiments and applications. In this dilute limit, chains form, break, and re-form slowly. Hence transients are long and it becomes necessary to simulate for comparatively (and impractically) long total strains.   

A second question concerns the role of dimensionality. Inelastic collisions are also present in the CD model (and absent in RD) in higher dimensions, which would suggest that qualitative differences persist. However simulations are needed to determine details such as the apparent value of $\Delta$ and the $\Mn$-interval over which effects are observed. 

Third, one can ask about the origins of the exponent $\Delta \approx 0.75$ in the RD flow curves. We note that the critical exponent $\beta$ in directed percolation (DP), which characterizes the mass of the percolating cluster, has a value $\beta \approx 0.276$ in 1+1 dimensions \cite{essam1988}. It is tempting to think there might be a connection to $1- \Delta$ in MR flows, with the applied field defining the time-like dimension. However such a connection is purely speculative.

Finally, one can ask about the role of Coulomb friction, which presumably plays a role in the laboratory. Insofar as Coulomb friction renders collisions between particles inelastic, we expect that shear flows in the CD model more closely resemble systems with friction.
While the results and equations in this paper are presented in the context of MR fluids, the model and the findings are more general and we expect that they can be generalized to electrorheological fluids or other similar dipolar systems.

\section{Acknowledgments}
This work was sponsored by NWO Exacte Wetenschappen (Physical Sciences) for the use of supercomputer facilities, with financial support from the Nederlandse Organisatie voor Wetenschappelijk Onderzoek (Netherlands Organization for Scientific Research, NWO). For the final stages of the work D.V. also recevied support from the European Research Council under the European Unions Seventh Framework Programme (FP7/2007-2013)/ERC Grant Agreement No. 306845. D.V. acknowledges simulating discussions with H.P. Svea.

\section{Appendix: Simulation details}

\subsection{Dipole moments}
There are many effects to consider when modeling the dipole moments $\mathbf{d}$ induced by the external field $\mathbf{H}$. For simplicity we assume our materials are ideal so that we do not need to consider saturation effects at high field strengths. 
We also assume the magnitude and direction of the induced dipole-moments are given by 
\begin{equation}
  \mathbf{B}=\mu_f\left(\mathbf{H}+\mathbf{M}\right)  \,,
\end{equation}
where $\mathbf{H}$ is  the the applied magnetic field and $\mathbf{M}$ the  magnetization. 
The dipole moment induced by the external field in a single particle is
\begin{equation} \label{m_single}
  \db_i=V_{\rm ci}\mathbf{M}=V_{\rm ci}( 3\beta \mathbf{H} ) \,,
\end{equation}
where $V_{\rm ci}$ is the core-volume of particle $i$, and 
\begin{equation}
  \beta = \frac{\mu-1}{\mu+2} \,.
\end{equation}
The relative permeability of the particles is
\begin{equation}
  \mu = \frac{\mu_i}{\mu_f} \,,
\end{equation}
and $\mu_i$ is the permeability of the core of particle $i$. The outer shell is assumed to have the same permeability as the carrier fluid.

When there are multiple particles the fields from the induced dipoles interact, giving a total dipole moment of
\begin{equation} \label{m_full}
  \db_i=3V_{ci}\beta\left[\mathbf{H}+\frac{1}{\mu_f}\sum_{i\ne j}\mathbf{B}_{ij}\right] \,,
\end{equation} 
where $\mathbf{B}_{ij}$ is given by \eqref{b_def}. This is an implicit relation, since $\mathbf{B}$ itself depends on $\db$. Eq.~\eqref{m_full} can be solved by iteratively evaluating the expression until it converges \cite{vesely1977}. 
However we find that for the parameter range investigated here, the correction due to this iterative scheme is negligible, except at the highest field strength we consider. Since our goal for this paper is to reach the lowest Mason numbers possible, and the Mason number is more sensitive to changes in the field strength than to the shear rate, we chose to ignore this effect for all values of $\mathbf{H}$. Consequently all the data presented here are generated using the much faster single particle relation for $\db$ given in equation \eqref{m_single}. A major reason why the self-interaction is so low in our system is the core-shell structure of the particles, which prevents the magnetic cores from directly touching each other and ensures the point dipoles remain separated. Note that since $V_{\rm ci}$ and $\beta$ always appear together, these parameters can be varied without changing the result as long as their product stays constant, meaning our results can be mapped to a model where $V_{ci}=V_i$ by lowering the value of $\beta$ accordingly.

\subsection{Long range interactions}
The dipole-dipole potential between two particles decays as $1/r^3$. The interaction is therefore long-ranged in 3D and decays too slowly to be easily truncated in 2D, and care must be taken to correctly include the influence of distant particles. There are several methods to do this, of which the lattice-based Ewald summation \cite{ewald1921berechnung} and cutoff-based reaction field methods \cite{barker1973monte} are the most common -- see e.g.~\cite{allen1989computer,fennell2006ewald} for comparisons of different methods. We use a cutoff-based method because it is more computationally efficient (computational complexity $O(N)$) and easier to generalize when changing the geometry of the simulation cell and applying external deformations such as shearing. 

While cutoff-based methods are commonly used to simulate MR systems, the long range correction terms used (if any) are rarely published.
We therefore include the correction terms employed here. We  consider only dipole-dipole interactions; free point charges are not treated. The expressions stated in this Appendix are for 2D systems. 

We introduce a cutoff distance $r_c$ and evaluate all pair interactions at close distances $r_{ij}<r_c$ directly. Evaluating each pair interaction at longer distances quickly becomes computationally expensive. Instead we assume the space outside the sphere given by $r_c$ is filled with a uniformly polarized continuous phase. It is then possible to analytically integrate over the continuous phase to obtain the long range correction.

For each observable $\mathcal{O}$ dependent on the dipole potential, it is necessary to calculate a correction term $\mathcal{O}_{LR} = \int_{r_c}^\infty \tilde{\mathcal{O}} dV$ by integrating the corresponding observable density function $\tilde{\mathcal{O}}$ over $r>r_c$.
The observable for a single particle $i$ is then given by 
\begin{equation}
  \mathcal{O}_i= \sum_{r_{ij}<r_c}\mathcal{O}_{ij}+\mathcal{O}_{LR} \,.
\end{equation}

We now show how this is applied to the dipole-dipole potential energy.
The magnetic flux density $\mathbf{B}_j$ from a dipole $\mathbf{m}_j$ at a distance $\mathbf{r}$ is given by  
 \begin{equation} \label{eq:b_flux}
 \mathbf{B}_j(\mathbf{r})=\frac{\mu_0}{4\pi}\left(\frac{3\mathbf{r}(\mathbf{m}_j\cdot\mathbf{r})}{r^5}-\frac{\mathbf{m}_j}{r^3}\right) \,.
 \end{equation}
At short distances $\mathbf{r}$  the local field can be calculated by summing over all particles $j$ located within a sphere of radius $r_c$. At longer distances we integrate over the uniformly polarized continuous phase 
to obtain the long range contribution to the magnetic flux density.
In order to perform the integration the discrete particle dipole moment $\mathbf{m}_i$  is replaced with the an average dipole moment density $\tilde{\mathbf{m}}$. 
There are several ways to approximate $\tilde{\mathbf{m}}$; we use  
\begin{equation}
   \tilde{\mathbf{m}}_i=\frac{1}{\pi r_c^2}\sum_{|r_{ij}|<r_c} w(r_{ij})\mathbf{m}_j \,,
\end{equation}
where we estimate the density of the whole space using the local density. An alternative would be to use the system average or the asymptotic value at infinity (if known) to estimate $\tilde{\mathbf{m}}$. Here we have introduced a weight factor $w(r_i)$ used to taper the interaction as the cut off distance $r_c$ is approached. This prevents discontinuous jumps in measured quantities when particles move in or out of the cutoff sphere. We use a simple linear taper function
\begin{equation}
w(r)=
 \begin{cases}
   1   \text{ for }  r<0.95r_c \\
   1-\frac{r-0.95r_c}{0.05r_c}   \text{ for }  0.95r_c<r<r_c \\
   0   \text{ for }  r>r_c \,.
 \end{cases}
\end{equation}
Inserting $\tilde{\mathbf{m}}_i$ into \eqref{eq:b_flux} and integrating over all $r>r_c$ yields the correction term 
\begin{equation}
  \mathbf{B}_\LR=\frac{\mu_0}{4\pi r_c^3}\sum_{|r_{ij}|<r_c}{w(r_{ij})\mathbf{m}_{j}} \,.
\end{equation}
The correction to the magnetic potential energy for a given particle $i$ then follows as 
\begin{equation}
  U_\LR= -\mathbf{m}_i \cdot \mathbf{B}_\LR \,.
\end{equation}
We note that this is an approximation. For a more careful calculation the correction term should be integrated over all space where $w(r)\ne1$ including the weight function $\mathcal{O}_{LR} = \int_{0}^\infty (1-w(r))\tilde{\mathcal{O}} dV$.

It is straight forward to repeat the above procedure for other observables. 
For the force one obtains
\begin{equation}
  \mathbf{f}_\LR=\mathbf{0} \,,
\end{equation}
as expected from symmetry. 
For the pressure one finds
\begin{equation}
p_\LR=-\frac{3\mu_0}{4\pi r_c^3}\sum_{|r_{ij}|<r_c} w(r_{ij})\mathbf{m}_i\cdot\mathbf{m_j}  \,,
\end{equation}
and correspondingly for the stress
\begin{equation}
  {\sigma_{xy}}_\LR'=
 \frac{3\mu_0}{16 \pi r_c^3}  
 \sum_{r_{ij} < r_c} w(r_{ij})\left[ m_{ix}m_{jy}+m_{iy}m_{jx}\right] \,.
\end{equation}
While this expression works for isotropic distributions of dipole moments, in our specific case all the dipoles are aligned with the $y$-axis and the correction term is identically zero. We solve this by introducing a second correction term  
\begin{equation}
  {\sigma_{xy}}_\LR=-cp_\LR
\end{equation}
where the coefficient 
\begin{equation}
 c=\left|\frac{\sum_{r_{ij} < r_c} w(r_{ij})r_{ijx}r_{ijy}}{\sum_{r_{ij} < r_c} w(r_{ij})r_{ij}^2}\right|
\end{equation}
is  a measure of the anisotropy of the packing. This correction term approximates the $\phi$ and $\Mn$ dependence over the parameter range we study. However it still assumes that all the dipoles are aligned with the $y$-axis, and it becomes increasingly inaccurate at $\phi<0.3$.

\begin{figure}[tb]
 \begin{center}
  \subfigure{\includegraphics[height=3.7cm] {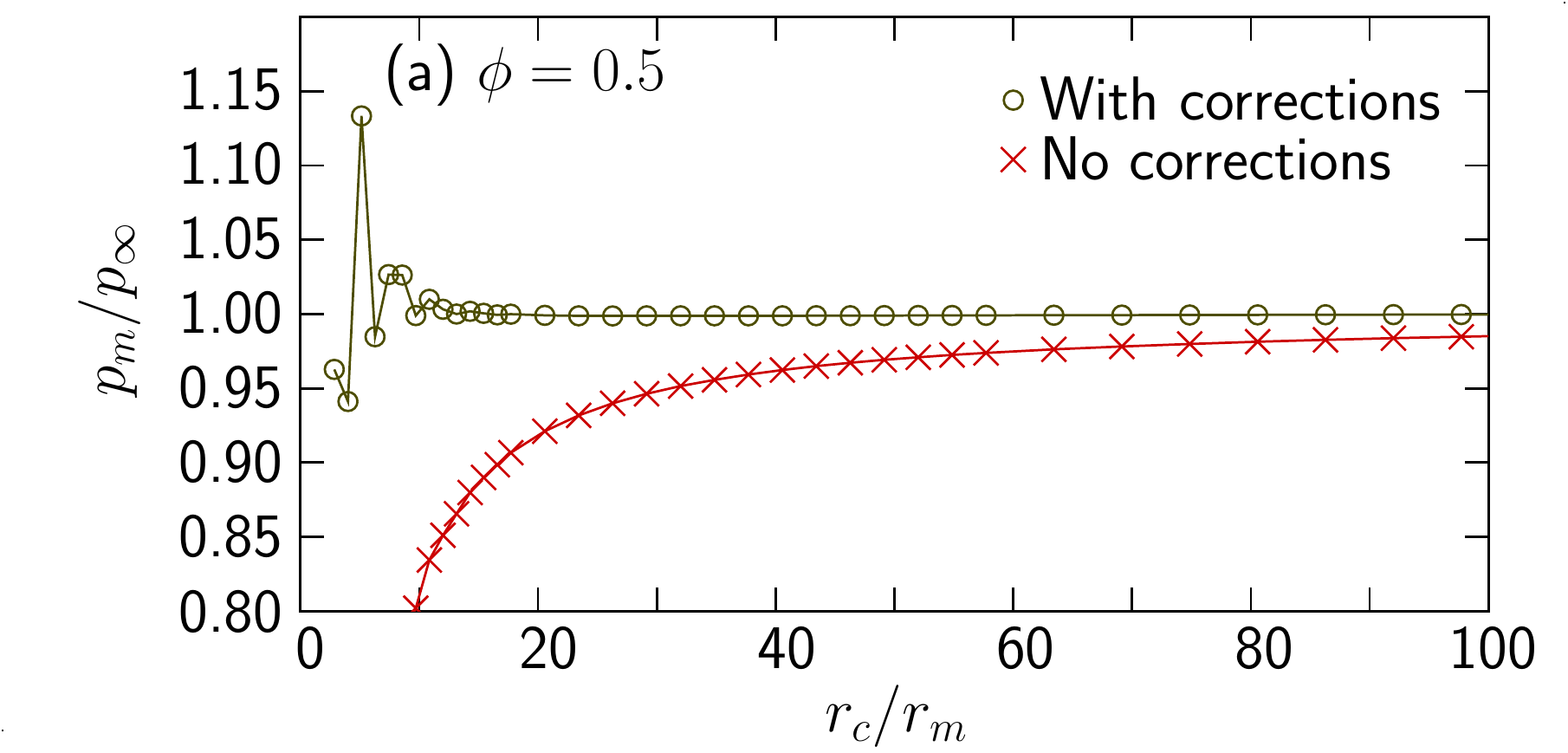}}
  \subfigure{\includegraphics[height=3.7cm] {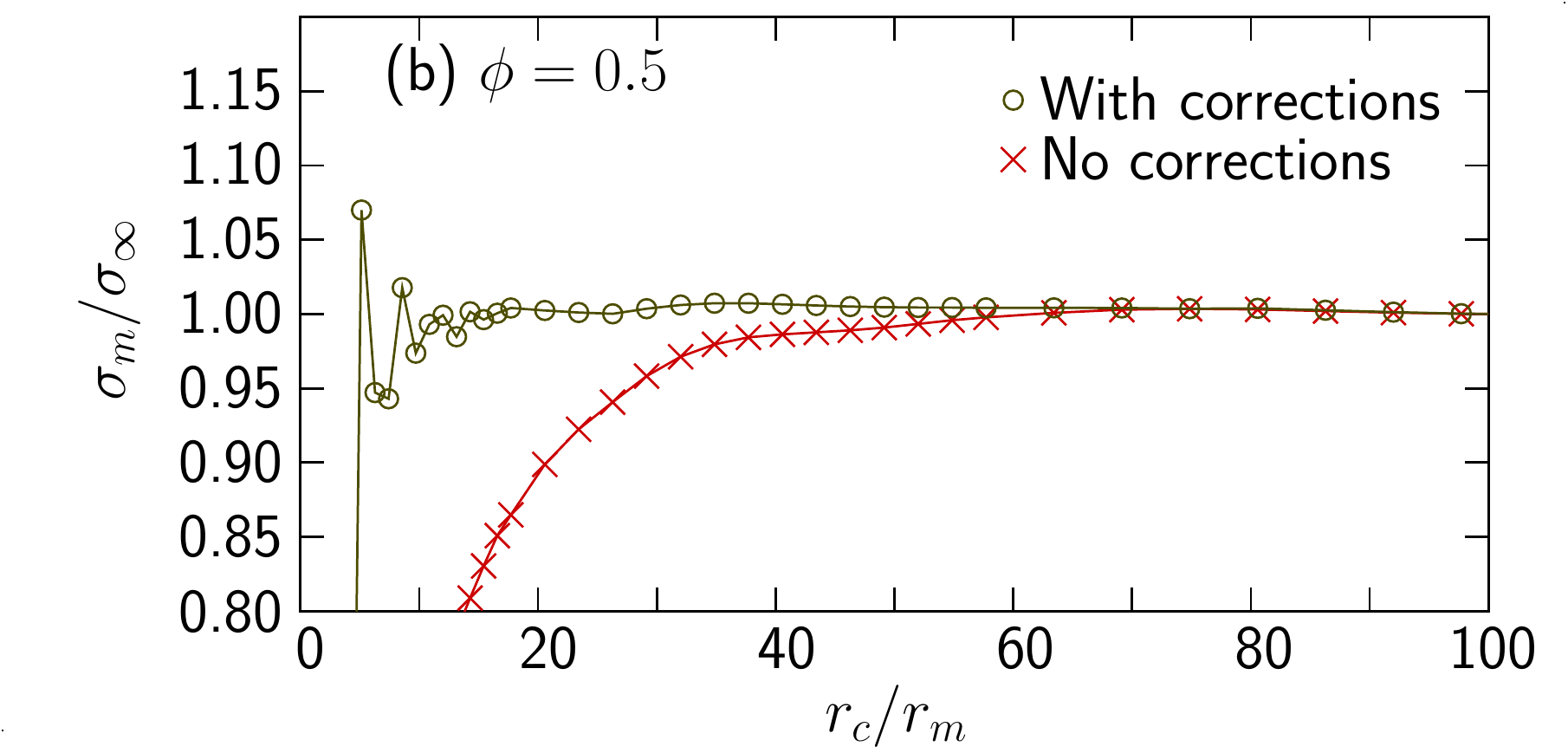}}
  \subfigure{\includegraphics[height=3.7cm] {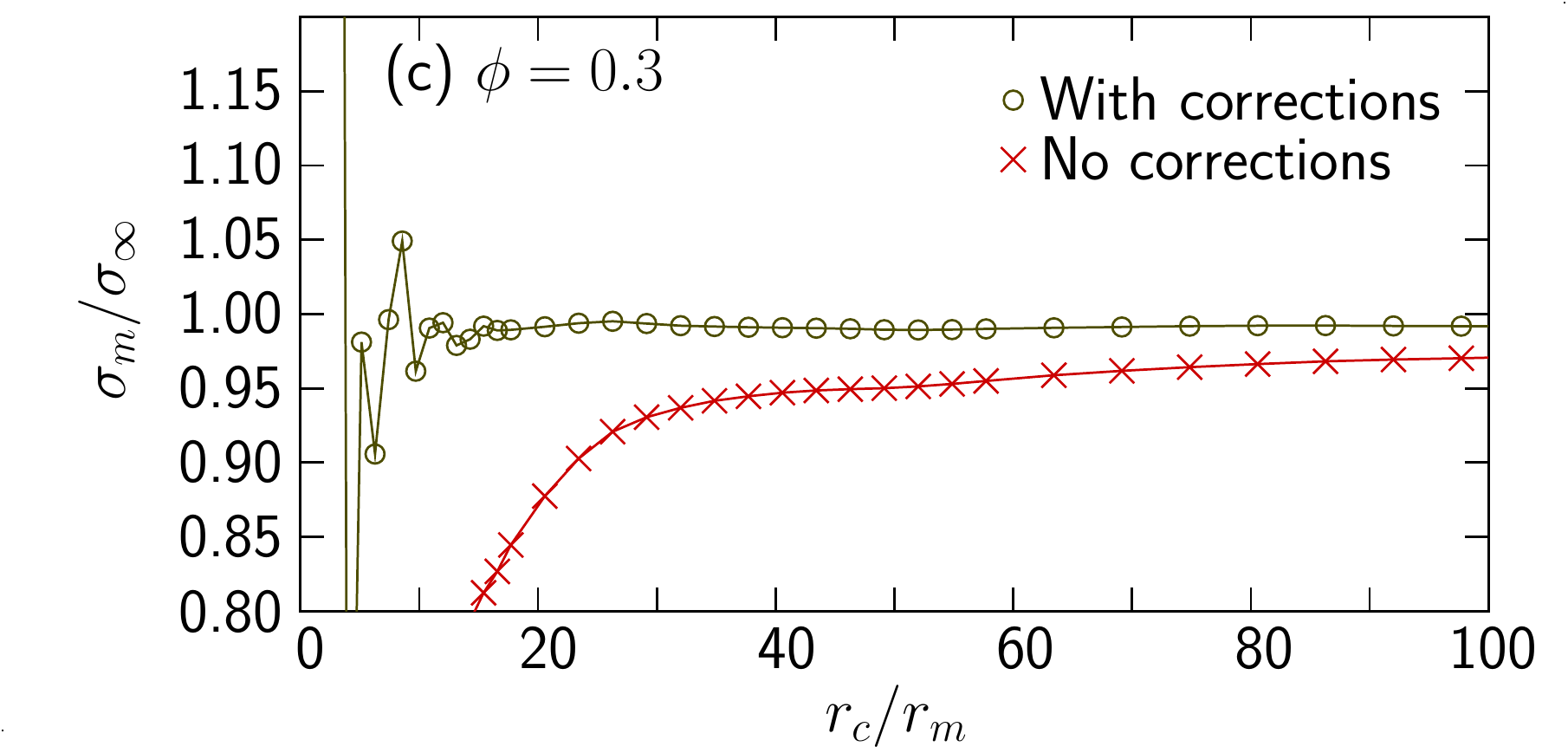}}
  \end{center}
   \caption{ The effect of different cut off distances on the magnetic pressure $p_m$, and stress $\sigma_m$ with and without the long range correction terms. The y-axis shows the relative change in the measured quantities relative to the most accurate value obtained using the highest possible $r_c$. The x-axis indicates the cut off distance in units of magnetic core radii $r_m$. The curves are obtained by analyzing a single RD configuration generated by simulating using a fixed value $r_c=15r_m$ at $\dot\gamma/H^2=10^{-5}$ but measured using different $r_c$. }
  \label{fig:cut}
\end{figure}

\begin{figure}[tb]
 \begin{center}
  \subfigure{\includegraphics[height=3.7cm] {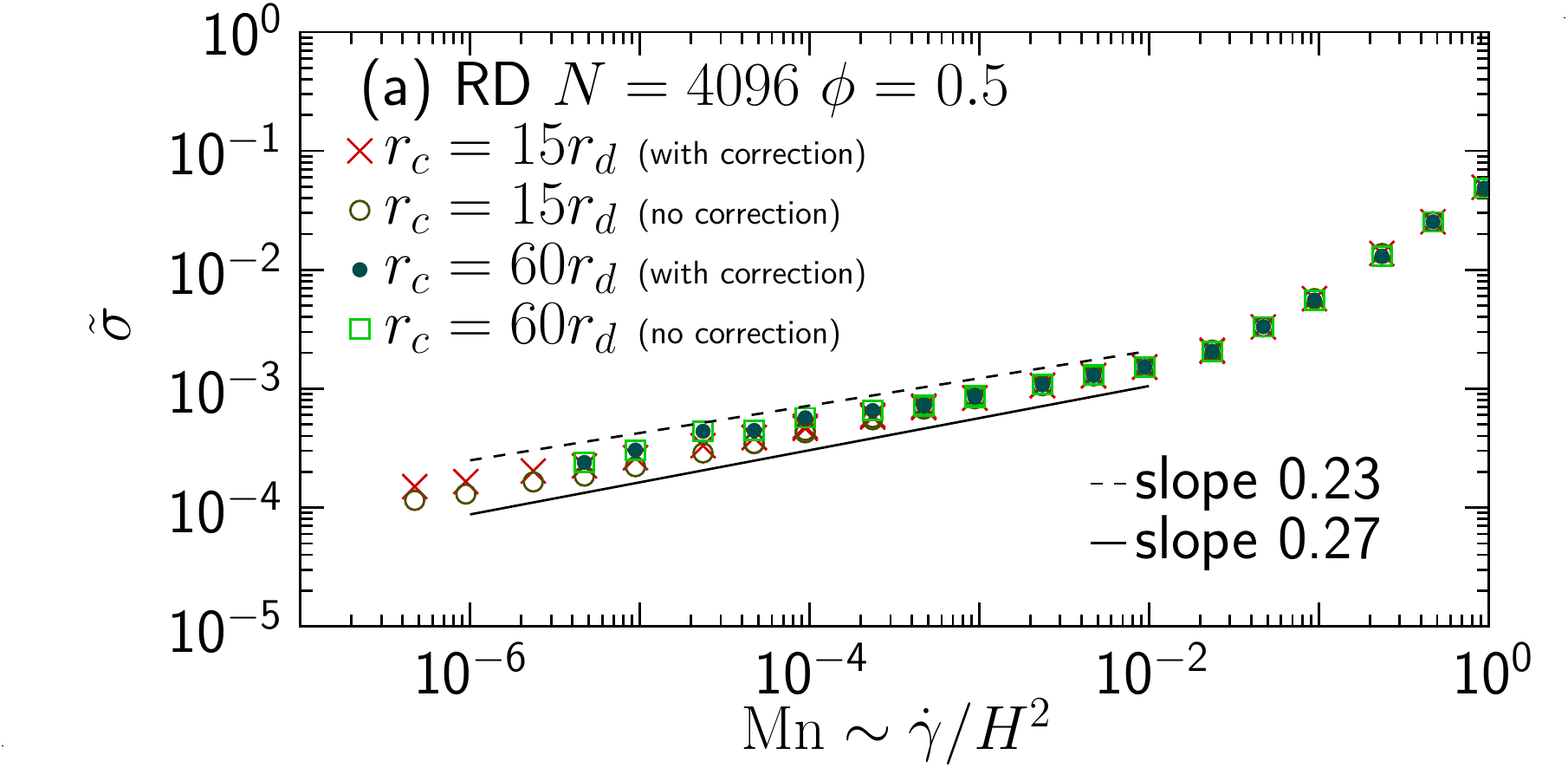}}
  \subfigure{\includegraphics[height=3.7cm] {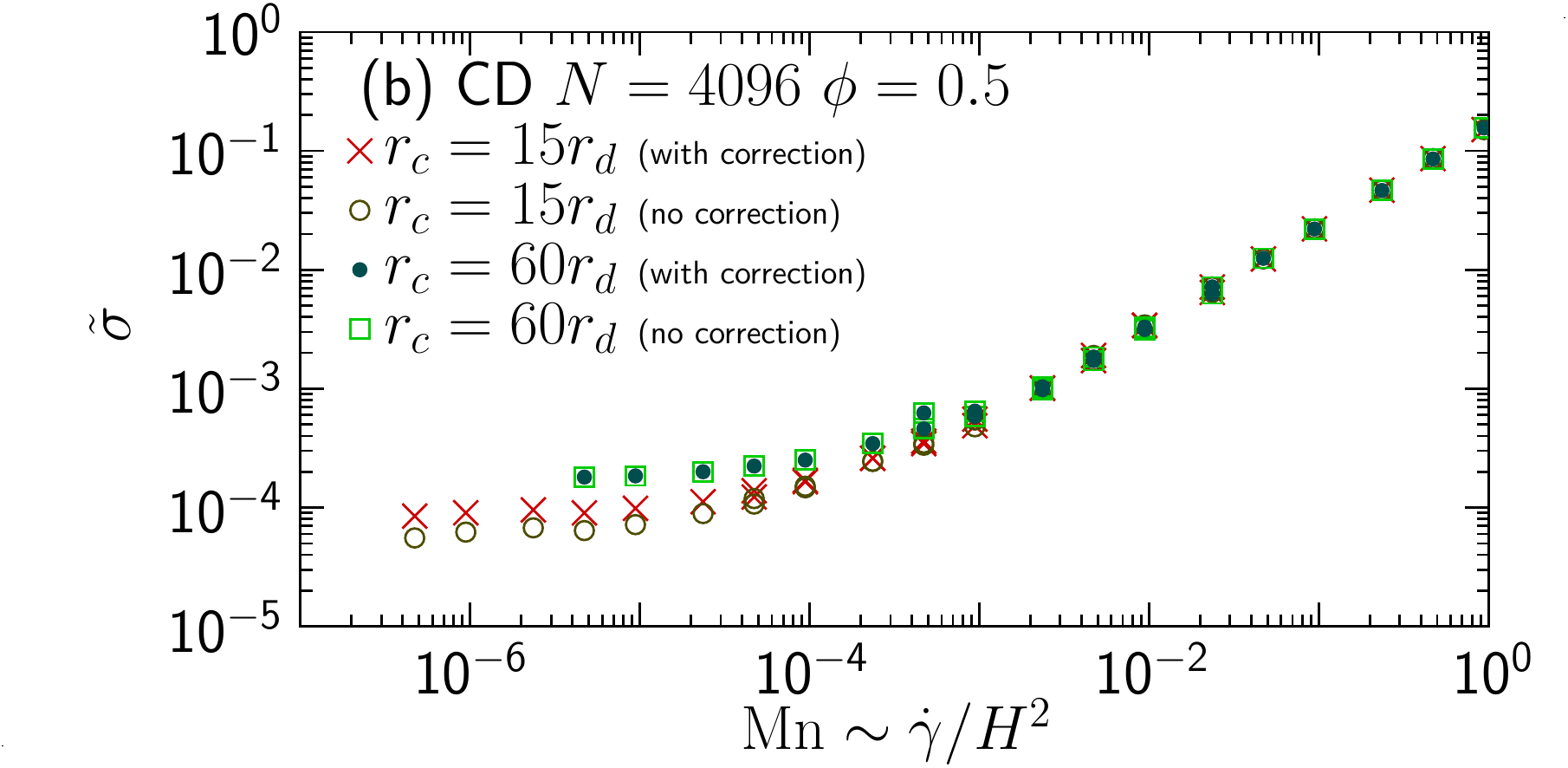}}
  \end{center}
   \caption{ The effect of different cut-off distances and with and without the correction term on the dimensionless flow curves. The data is generated by both simulating and analysing the configurations using the $r_c$ stated in the legend. }
  \label{fig:cut_flow}
\end{figure}

Figure \ref{fig:cut} shows the effect of the above mentioned correction terms. In our simulations we use $r_c=15r_m$ for $\phi>0.3$ and $r_c=60r_m$ for dilute systems with $0.1<\phi\le0.3$. Here $r_m$ is the radius of the magnetic core of the larger particles.

In general the the need for corrections is lower for isotropic packings, i.e packings with high $\Mn$ or high $\phi$, and their contribution is often insignificant at the  $r_c$ we use. At the other end in dilute low $\Mn$ packings the corrections play an important role as they can reduce the $r_c$ needed during simulation. 
In figure \ref{fig:cut_flow} we see the flow curve $\tilde\sigma$ vs $\Mn$ for $\phi=0.5$ with and without corrections. It is clear from the figure that the corrections are only important at the lowest $\Mn$. The use of the stress correction term shifts the onset of the yield stress plateau to higher $\Mn$, making the plateau easier to observe. However our main conclusions are not sensitive to the use of the correction term; most significantly, our observations regarding the presence or absence of a yield stress plateau at low $\Mn$ are also supported by looking at the raw stress without the correction term.






\begin{mcitethebibliography}{61}
\providecommand*{\natexlab}[1]{#1}
\providecommand*{\mciteSetBstSublistMode}[1]{}
\providecommand*{\mciteSetBstMaxWidthForm}[2]{}
\providecommand*{\mciteBstWouldAddEndPuncttrue}
  {\def\EndOfBibitem{\unskip.}}
\providecommand*{\mciteBstWouldAddEndPunctfalse}
  {\let\EndOfBibitem\relax}
\providecommand*{\mciteSetBstMidEndSepPunct}[3]{}
\providecommand*{\mciteSetBstSublistLabelBeginEnd}[3]{}
\providecommand*{\EndOfBibitem}{}
\mciteSetBstSublistMode{f}
\mciteSetBstMaxWidthForm{subitem}
{(\emph{\alph{mcitesubitemcount}})}
\mciteSetBstSublistLabelBeginEnd{\mcitemaxwidthsubitemform\space}
{\relax}{\relax}

\bibitem[de~Vicente \emph{et~al.}(2011)de~Vicente, Klingenberg, and
  Hidalgo-Alvarez]{vicente2011review}
J.~de~Vicente, D.~J. Klingenberg and R.~Hidalgo-Alvarez, \emph{Soft Matter},
  2011, \textbf{7}, 3701--3710\relax
\mciteBstWouldAddEndPuncttrue
\mciteSetBstMidEndSepPunct{\mcitedefaultmidpunct}
{\mcitedefaultendpunct}{\mcitedefaultseppunct}\relax
\EndOfBibitem
\bibitem[Ghaffari \emph{et~al.}(2015)Ghaffari, Hashemabadi, and
  Ashtiani]{ghaffari2015review}
A.~Ghaffari, S.~H. Hashemabadi and M.~Ashtiani, \emph{Journal of Intelligent
  Material Systems and Structures}, 2015, \textbf{26}, 881--904\relax
\mciteBstWouldAddEndPuncttrue
\mciteSetBstMidEndSepPunct{\mcitedefaultmidpunct}
{\mcitedefaultendpunct}{\mcitedefaultseppunct}\relax
\EndOfBibitem
\bibitem[Marshall \emph{et~al.}(1989)Marshall, Zukoski, and
  Goodwin]{marshall1989}
L.~Marshall, C.~F. Zukoski and J.~W. Goodwin, \emph{Journal of the Chemical
  Society, Faraday Transactions 1: Physical Chemistry in Condensed Phases},
  1989, \textbf{85}, 2785--2795\relax
\mciteBstWouldAddEndPuncttrue
\mciteSetBstMidEndSepPunct{\mcitedefaultmidpunct}
{\mcitedefaultendpunct}{\mcitedefaultseppunct}\relax
\EndOfBibitem
\bibitem[Klingenberg and Zukoski~IV(1990)]{klingenberg1990model}
D.~J. Klingenberg and C.~F. Zukoski~IV, \emph{Langmuir}, 1990, \textbf{6},
  15--24\relax
\mciteBstWouldAddEndPuncttrue
\mciteSetBstMidEndSepPunct{\mcitedefaultmidpunct}
{\mcitedefaultendpunct}{\mcitedefaultseppunct}\relax
\EndOfBibitem
\bibitem[Martin and Anderson(1996)]{martin1996}
J.~E. Martin and R.~A. Anderson, \emph{J. Chem. Phys.}, 1996, \textbf{104},
  4814--4827\relax
\mciteBstWouldAddEndPuncttrue
\mciteSetBstMidEndSepPunct{\mcitedefaultmidpunct}
{\mcitedefaultendpunct}{\mcitedefaultseppunct}\relax
\EndOfBibitem
\bibitem[de~Vicente \emph{et~al.}(2004)de~Vicente, L{\'o}pez-L{\'o}pez,
  Dur{\'a}n, and Gonz{\'a}lez-Caballero]{vicente2004}
J.~de~Vicente, M.~T. L{\'o}pez-L{\'o}pez, J.~D. Dur{\'a}n and
  F.~Gonz{\'a}lez-Caballero, \emph{Rheologica acta}, 2004, \textbf{44},
  94--103\relax
\mciteBstWouldAddEndPuncttrue
\mciteSetBstMidEndSepPunct{\mcitedefaultmidpunct}
{\mcitedefaultendpunct}{\mcitedefaultseppunct}\relax
\EndOfBibitem
\bibitem[de~Gans \emph{et~al.}(1999)de~Gans, Hoekstra, and
  Mellema]{gans1999nonlinear}
B.-J. de~Gans, H.~Hoekstra and J.~Mellema, \emph{Faraday Discuss.}, 1999,
  \textbf{112}, 209--224\relax
\mciteBstWouldAddEndPuncttrue
\mciteSetBstMidEndSepPunct{\mcitedefaultmidpunct}
{\mcitedefaultendpunct}{\mcitedefaultseppunct}\relax
\EndOfBibitem
\bibitem[Halsey \emph{et~al.}(1992)Halsey, Martin, and Adolf]{halsey1992}
T.~C. Halsey, J.~E. Martin and D.~Adolf, \emph{Phys. Rev. Lett.}, 1992,
  \textbf{68}, 1519--1522\relax
\mciteBstWouldAddEndPuncttrue
\mciteSetBstMidEndSepPunct{\mcitedefaultmidpunct}
{\mcitedefaultendpunct}{\mcitedefaultseppunct}\relax
\EndOfBibitem
\bibitem[Bossis \emph{et~al.}(2002)Bossis, Volkova, Lacis, and
  Meunier]{bossis2002}
G.~Bossis, O.~Volkova, S.~Lacis and A.~Meunier, \emph{Ferrofluids}, Springer,
  2002, pp. 202--230\relax
\mciteBstWouldAddEndPuncttrue
\mciteSetBstMidEndSepPunct{\mcitedefaultmidpunct}
{\mcitedefaultendpunct}{\mcitedefaultseppunct}\relax
\EndOfBibitem
\bibitem[Ruiz-L{\'o}pez \emph{et~al.}(2016)Ruiz-L{\'o}pez,
  Fern{\'a}ndez-Toledano, Klingenberg, Hidalgo-Alvarez, and
  de~Vicente]{ruiz2016}
J.~A. Ruiz-L{\'o}pez, J.~C. Fern{\'a}ndez-Toledano, D.~J. Klingenberg,
  R.~Hidalgo-Alvarez and J.~de~Vicente, \emph{Journal of Rheology}, 2016,
  \textbf{60}, 61--74\relax
\mciteBstWouldAddEndPuncttrue
\mciteSetBstMidEndSepPunct{\mcitedefaultmidpunct}
{\mcitedefaultendpunct}{\mcitedefaultseppunct}\relax
\EndOfBibitem
\bibitem[Baxter-Drayton and Brady(1996)]{baxter1996brownian}
Y.~Baxter-Drayton and J.~F. Brady, \emph{Journal of Rheology}, 1996,
  \textbf{40}, 1027--1056\relax
\mciteBstWouldAddEndPuncttrue
\mciteSetBstMidEndSepPunct{\mcitedefaultmidpunct}
{\mcitedefaultendpunct}{\mcitedefaultseppunct}\relax
\EndOfBibitem
\bibitem[You \emph{et~al.}(2007)You, Park, Choi, Choi, and Jhon]{you2007}
J.~You, B.~Park, H.~Choi, S.~Choi and M.~Jhon, \emph{Int. J. Mod. Phys. B},
  2007, \textbf{21}, 4996--5002\relax
\mciteBstWouldAddEndPuncttrue
\mciteSetBstMidEndSepPunct{\mcitedefaultmidpunct}
{\mcitedefaultendpunct}{\mcitedefaultseppunct}\relax
\EndOfBibitem
\bibitem[Choi \emph{et~al.}(2007)Choi, Park, Cho, and You]{choi2007}
H.~Choi, B.~Park, M.~Cho and J.~You, \emph{Journal of Magnetism and Magnetic
  Materials}, 2007, \textbf{310}, 2835--2837\relax
\mciteBstWouldAddEndPuncttrue
\mciteSetBstMidEndSepPunct{\mcitedefaultmidpunct}
{\mcitedefaultendpunct}{\mcitedefaultseppunct}\relax
\EndOfBibitem
\bibitem[Ko \emph{et~al.}(2009)Ko, Lim, Park, Yang, and Choi]{ko2009}
S.~Ko, J.~Lim, B.~Park, M.~Yang and H.~Choi, \emph{J. Appl. Phys.}, 2009,
  \textbf{105}, 07E703\relax
\mciteBstWouldAddEndPuncttrue
\mciteSetBstMidEndSepPunct{\mcitedefaultmidpunct}
{\mcitedefaultendpunct}{\mcitedefaultseppunct}\relax
\EndOfBibitem
\bibitem[Fang \emph{et~al.}(2010)Fang, Choi, and Choi]{fang2010}
F.~F. Fang, H.~J. Choi and W.~Choi, \emph{Colloid and Polymer Science}, 2010,
  \textbf{288}, 359--363\relax
\mciteBstWouldAddEndPuncttrue
\mciteSetBstMidEndSepPunct{\mcitedefaultmidpunct}
{\mcitedefaultendpunct}{\mcitedefaultseppunct}\relax
\EndOfBibitem
\bibitem[Fang \emph{et~al.}(2009)Fang, Choi, and Seo]{fang2009}
F.~F. Fang, H.~J. Choi and Y.~Seo, \emph{ACS Applied Materials \& Interfaces},
  2009, \textbf{2}, 54--60\relax
\mciteBstWouldAddEndPuncttrue
\mciteSetBstMidEndSepPunct{\mcitedefaultmidpunct}
{\mcitedefaultendpunct}{\mcitedefaultseppunct}\relax
\EndOfBibitem
\bibitem[Cox \emph{et~al.}(2016)Cox, Wang, Bar{\'e}s, and Behringer]{cox2016}
M.~Cox, D.~Wang, J.~Bar{\'e}s and R.~P. Behringer, \emph{EPL}, 2016,
  \textbf{115}, 64003\relax
\mciteBstWouldAddEndPuncttrue
\mciteSetBstMidEndSepPunct{\mcitedefaultmidpunct}
{\mcitedefaultendpunct}{\mcitedefaultseppunct}\relax
\EndOfBibitem
\bibitem[V{\aa}gberg \emph{et~al.}(2014)V{\aa}gberg, Olsson, and
  Teitel]{vagberg2014dissipation}
D.~V{\aa}gberg, P.~Olsson and S.~Teitel, \emph{Phys. Rev. Lett.}, 2014,
  \textbf{112}, 208303\relax
\mciteBstWouldAddEndPuncttrue
\mciteSetBstMidEndSepPunct{\mcitedefaultmidpunct}
{\mcitedefaultendpunct}{\mcitedefaultseppunct}\relax
\EndOfBibitem
\bibitem[Durian(1995)]{durian1995}
D.~Durian, \emph{Phys. Rev. Lett.}, 1995, \textbf{75}, 4780\relax
\mciteBstWouldAddEndPuncttrue
\mciteSetBstMidEndSepPunct{\mcitedefaultmidpunct}
{\mcitedefaultendpunct}{\mcitedefaultseppunct}\relax
\EndOfBibitem
\bibitem[Liu and Nagel(1998)]{liu1998}
A.~J. Liu and S.~R. Nagel, \emph{Nature}, 1998, \textbf{396}, 21--22\relax
\mciteBstWouldAddEndPuncttrue
\mciteSetBstMidEndSepPunct{\mcitedefaultmidpunct}
{\mcitedefaultendpunct}{\mcitedefaultseppunct}\relax
\EndOfBibitem
\bibitem[O'Hern \emph{et~al.}(2003)O'Hern, Silbert, Liu, and Nagel]{ohern2003}
C.~S. O'Hern, L.~E. Silbert, A.~J. Liu and S.~R. Nagel, \emph{Phys. Rev. E},
  2003, \textbf{68}, 011306\relax
\mciteBstWouldAddEndPuncttrue
\mciteSetBstMidEndSepPunct{\mcitedefaultmidpunct}
{\mcitedefaultendpunct}{\mcitedefaultseppunct}\relax
\EndOfBibitem
\bibitem[Olsson and Teitel(2007)]{olsson2007}
P.~Olsson and S.~Teitel, \emph{Phys. Rev. Lett.}, 2007, \textbf{99},
  178001\relax
\mciteBstWouldAddEndPuncttrue
\mciteSetBstMidEndSepPunct{\mcitedefaultmidpunct}
{\mcitedefaultendpunct}{\mcitedefaultseppunct}\relax
\EndOfBibitem
\bibitem[Heussinger and Barrat(2009)]{heussinger2009}
C.~Heussinger and J.-L. Barrat, \emph{Phys. Rev. Lett.}, 2009, \textbf{102},
  218303\relax
\mciteBstWouldAddEndPuncttrue
\mciteSetBstMidEndSepPunct{\mcitedefaultmidpunct}
{\mcitedefaultendpunct}{\mcitedefaultseppunct}\relax
\EndOfBibitem
\bibitem[Hatano(2009)]{hatano09}
T.~Hatano, \emph{Phys. Rev. E}, 2009, \textbf{79}, 050301\relax
\mciteBstWouldAddEndPuncttrue
\mciteSetBstMidEndSepPunct{\mcitedefaultmidpunct}
{\mcitedefaultendpunct}{\mcitedefaultseppunct}\relax
\EndOfBibitem
\bibitem[Tighe \emph{et~al.}(2010)Tighe, Woldhuis, Remmers, van Saarloos, and
  van Hecke]{tighe2010}
B.~P. Tighe, E.~Woldhuis, J.~J.~C. Remmers, W.~van Saarloos and M.~van Hecke,
  \emph{Phys. Rev. Lett.}, 2010, \textbf{105}, 088303\relax
\mciteBstWouldAddEndPuncttrue
\mciteSetBstMidEndSepPunct{\mcitedefaultmidpunct}
{\mcitedefaultendpunct}{\mcitedefaultseppunct}\relax
\EndOfBibitem
\bibitem[Tighe(2011)]{tighe2011}
B.~P. Tighe, \emph{Phys. Rev. Lett.}, 2011, \textbf{107}, 158303\relax
\mciteBstWouldAddEndPuncttrue
\mciteSetBstMidEndSepPunct{\mcitedefaultmidpunct}
{\mcitedefaultendpunct}{\mcitedefaultseppunct}\relax
\EndOfBibitem
\bibitem[Ikeda \emph{et~al.}(2012)Ikeda, Berthier, and Sollich]{ikeda2012}
A.~Ikeda, L.~Berthier and P.~Sollich, \emph{Phys. Rev. Lett.}, 2012,
  \textbf{109}, 018301\relax
\mciteBstWouldAddEndPuncttrue
\mciteSetBstMidEndSepPunct{\mcitedefaultmidpunct}
{\mcitedefaultendpunct}{\mcitedefaultseppunct}\relax
\EndOfBibitem
\bibitem[Boschan \emph{et~al.}(2016)Boschan, V{\aa}gberg, Somfai, and
  Tighe]{boschan2016}
J.~Boschan, D.~V{\aa}gberg, E.~Somfai and B.~P. Tighe, \emph{Soft Matter},
  2016, \textbf{12}, 5450--5460\relax
\mciteBstWouldAddEndPuncttrue
\mciteSetBstMidEndSepPunct{\mcitedefaultmidpunct}
{\mcitedefaultendpunct}{\mcitedefaultseppunct}\relax
\EndOfBibitem
\bibitem[Baumgarten \emph{et~al.}(2017)Baumgarten, V{\aa}gberg, and
  Tighe]{baumgarten2017}
K.~Baumgarten, D.~V{\aa}gberg and B.~P. Tighe, \emph{Phys. Rev. Lett.}, 2017,
  \textbf{118}, 098001\relax
\mciteBstWouldAddEndPuncttrue
\mciteSetBstMidEndSepPunct{\mcitedefaultmidpunct}
{\mcitedefaultendpunct}{\mcitedefaultseppunct}\relax
\EndOfBibitem
\bibitem[Felt \emph{et~al.}(1996)Felt, Hagenbuchle, Liu, and Richard]{felt1996}
D.~W. Felt, M.~Hagenbuchle, J.~Liu and J.~Richard, \emph{Journal of Intelligent
  Material Systems and Structures}, 1996, \textbf{7}, 589--593\relax
\mciteBstWouldAddEndPuncttrue
\mciteSetBstMidEndSepPunct{\mcitedefaultmidpunct}
{\mcitedefaultendpunct}{\mcitedefaultseppunct}\relax
\EndOfBibitem
\bibitem[Martin \emph{et~al.}(1994)Martin, Odinek, and Halsey]{martin1994}
J.~E. Martin, J.~Odinek and T.~C. Halsey, \emph{Phys. Rev. E}, 1994,
  \textbf{50}, 3263\relax
\mciteBstWouldAddEndPuncttrue
\mciteSetBstMidEndSepPunct{\mcitedefaultmidpunct}
{\mcitedefaultendpunct}{\mcitedefaultseppunct}\relax
\EndOfBibitem
\bibitem[De~Gans \emph{et~al.}(2000)De~Gans, Duin, Van~den Ende, and
  Mellema]{gans2000size}
B.~De~Gans, N.~Duin, D.~Van~den Ende and J.~Mellema, \emph{J. Chem. Phys.},
  2000, \textbf{113}, 2032--2042\relax
\mciteBstWouldAddEndPuncttrue
\mciteSetBstMidEndSepPunct{\mcitedefaultmidpunct}
{\mcitedefaultendpunct}{\mcitedefaultseppunct}\relax
\EndOfBibitem
\bibitem[Volkova \emph{et~al.}(2000)Volkova, Bossis, Guyot, Bashtovoi, and
  Reks]{volkova2000}
O.~Volkova, G.~Bossis, M.~Guyot, V.~Bashtovoi and A.~Reks, \emph{Journal of
  Rheology (1978-present)}, 2000, \textbf{44}, 91--104\relax
\mciteBstWouldAddEndPuncttrue
\mciteSetBstMidEndSepPunct{\mcitedefaultmidpunct}
{\mcitedefaultendpunct}{\mcitedefaultseppunct}\relax
\EndOfBibitem
\bibitem[Sherman \emph{et~al.}(2015)Sherman, Becnel, and Wereley]{sherman2015}
S.~G. Sherman, A.~C. Becnel and N.~M. Wereley, \emph{Journal of Magnetism and
  Magnetic Materials}, 2015, \textbf{380}, 98--104\relax
\mciteBstWouldAddEndPuncttrue
\mciteSetBstMidEndSepPunct{\mcitedefaultmidpunct}
{\mcitedefaultendpunct}{\mcitedefaultseppunct}\relax
\EndOfBibitem
\bibitem[Bonnecaze and Brady(1992)]{bonnecaze1992dynamic}
R.~Bonnecaze and J.~Brady, \emph{J. Chem. Phys.}, 1992, \textbf{96},
  2183--2202\relax
\mciteBstWouldAddEndPuncttrue
\mciteSetBstMidEndSepPunct{\mcitedefaultmidpunct}
{\mcitedefaultendpunct}{\mcitedefaultseppunct}\relax
\EndOfBibitem
\bibitem[Melrose(1992)]{melrose1992brownian}
J.~Melrose, \emph{Mol. Phys.}, 1992, \textbf{76}, 635--660\relax
\mciteBstWouldAddEndPuncttrue
\mciteSetBstMidEndSepPunct{\mcitedefaultmidpunct}
{\mcitedefaultendpunct}{\mcitedefaultseppunct}\relax
\EndOfBibitem
\bibitem[Evans and Morris(1990)]{evans1990}
D.~Evans and G.~P. Morris, \emph{Academic, London}, 1990\relax
\mciteBstWouldAddEndPuncttrue
\mciteSetBstMidEndSepPunct{\mcitedefaultmidpunct}
{\mcitedefaultendpunct}{\mcitedefaultseppunct}\relax
\EndOfBibitem
\bibitem[V{\aa}gberg \emph{et~al.}(2017)V{\aa}gberg, Olsson, and
  Teitel]{vaagberg2017shear}
D.~V{\aa}gberg, P.~Olsson and S.~Teitel, \emph{arXiv preprint
  arXiv:1703.01652}, 2017\relax
\mciteBstWouldAddEndPuncttrue
\mciteSetBstMidEndSepPunct{\mcitedefaultmidpunct}
{\mcitedefaultendpunct}{\mcitedefaultseppunct}\relax
\EndOfBibitem
\bibitem[V{\aa}gberg \emph{et~al.}(2014)V{\aa}gberg, Olsson, and
  Teitel]{vagberg2014universality}
D.~V{\aa}gberg, P.~Olsson and S.~Teitel, \emph{Phys. Rev. Lett.}, 2014,
  \textbf{113}, 148002\relax
\mciteBstWouldAddEndPuncttrue
\mciteSetBstMidEndSepPunct{\mcitedefaultmidpunct}
{\mcitedefaultendpunct}{\mcitedefaultseppunct}\relax
\EndOfBibitem
\bibitem[Klingenberg \emph{et~al.}(2007)Klingenberg, Ulicny, and
  Golden]{klingenberg2007}
D.~J. Klingenberg, J.~C. Ulicny and M.~A. Golden, \emph{Journal of Rheology
  (1978-present)}, 2007, \textbf{51}, 883--893\relax
\mciteBstWouldAddEndPuncttrue
\mciteSetBstMidEndSepPunct{\mcitedefaultmidpunct}
{\mcitedefaultendpunct}{\mcitedefaultseppunct}\relax
\EndOfBibitem
\bibitem[Koeze \emph{et~al.}(2016)Koeze, V{\aa}gberg, Tjoa, and
  Tighe]{koeze2016}
D.~Koeze, D.~V{\aa}gberg, B.~Tjoa and B.~Tighe, \emph{EPL}, 2016, \textbf{113},
  54001\relax
\mciteBstWouldAddEndPuncttrue
\mciteSetBstMidEndSepPunct{\mcitedefaultmidpunct}
{\mcitedefaultendpunct}{\mcitedefaultseppunct}\relax
\EndOfBibitem
\bibitem[V{\aa}gberg \emph{et~al.}(2011)V{\aa}gberg, Olsson, and
  Teitel]{vaagberg2011glassiness}
D.~V{\aa}gberg, P.~Olsson and S.~Teitel, \emph{Phys. Rev. E}, 2011,
  \textbf{83}, 031307\relax
\mciteBstWouldAddEndPuncttrue
\mciteSetBstMidEndSepPunct{\mcitedefaultmidpunct}
{\mcitedefaultendpunct}{\mcitedefaultseppunct}\relax
\EndOfBibitem
\bibitem[Olsson and Teitel(2012)]{olsson2012hb}
P.~Olsson and S.~Teitel, \emph{Phys. Rev. Lett.}, 2012, \textbf{109},
  108001\relax
\mciteBstWouldAddEndPuncttrue
\mciteSetBstMidEndSepPunct{\mcitedefaultmidpunct}
{\mcitedefaultendpunct}{\mcitedefaultseppunct}\relax
\EndOfBibitem
\bibitem[Maxwell(1864)]{maxwell1864}
J.~C. Maxwell, \emph{The London, Edinburgh, and Dublin Philosophical Magazine
  and Journal of Science}, 1864, \textbf{27}, 294--299\relax
\mciteBstWouldAddEndPuncttrue
\mciteSetBstMidEndSepPunct{\mcitedefaultmidpunct}
{\mcitedefaultendpunct}{\mcitedefaultseppunct}\relax
\EndOfBibitem
\bibitem[Tighe and Vlugt(2011)]{tighe2011b}
B.~P. Tighe and T.~J.~H. Vlugt, \emph{J. Stat. Mech.}, 2011,  P04002\relax
\mciteBstWouldAddEndPuncttrue
\mciteSetBstMidEndSepPunct{\mcitedefaultmidpunct}
{\mcitedefaultendpunct}{\mcitedefaultseppunct}\relax
\EndOfBibitem
\bibitem[Goodrich \emph{et~al.}(2012)Goodrich, Liu, and Nagel]{goodrich2012}
C.~P. Goodrich, A.~J. Liu and S.~R. Nagel, \emph{Phys. Rev. Lett.}, 2012,
  \textbf{109}, 095704\relax
\mciteBstWouldAddEndPuncttrue
\mciteSetBstMidEndSepPunct{\mcitedefaultmidpunct}
{\mcitedefaultendpunct}{\mcitedefaultseppunct}\relax
\EndOfBibitem
\bibitem[Dagois-Bohy \emph{et~al.}(2012)Dagois-Bohy, Tighe, Simon, Henkes, and
  van Hecke]{dagois-bohy2012}
S.~Dagois-Bohy, B.~P. Tighe, J.~Simon, S.~Henkes and M.~van Hecke, \emph{Phys.
  Rev. Lett.}, 2012, \textbf{109}, 095703\relax
\mciteBstWouldAddEndPuncttrue
\mciteSetBstMidEndSepPunct{\mcitedefaultmidpunct}
{\mcitedefaultendpunct}{\mcitedefaultseppunct}\relax
\EndOfBibitem
\bibitem[Goodrich \emph{et~al.}(2014)Goodrich, Dagois-Bohy, Tighe, van Hecke,
  Liu, and Nagel]{goodrich2014}
C.~P. Goodrich, S.~Dagois-Bohy, B.~P. Tighe, M.~van Hecke, A.~J. Liu and S.~R.
  Nagel, \emph{Phys. Rev. E}, 2014, \textbf{90}, 022138\relax
\mciteBstWouldAddEndPuncttrue
\mciteSetBstMidEndSepPunct{\mcitedefaultmidpunct}
{\mcitedefaultendpunct}{\mcitedefaultseppunct}\relax
\EndOfBibitem
\bibitem[Calladine(1978)]{calladine1978}
C.~Calladine, \emph{International Journal of Solids and Structures}, 1978,
  \textbf{14}, 161--172\relax
\mciteBstWouldAddEndPuncttrue
\mciteSetBstMidEndSepPunct{\mcitedefaultmidpunct}
{\mcitedefaultendpunct}{\mcitedefaultseppunct}\relax
\EndOfBibitem
\bibitem[Lois \emph{et~al.}(2008)Lois, Blawzdziewicz, and O'Hern]{lois2008}
G.~Lois, J.~Blawzdziewicz and C.~S. O'Hern, \emph{Physical review letters},
  2008, \textbf{100}, 028001\relax
\mciteBstWouldAddEndPuncttrue
\mciteSetBstMidEndSepPunct{\mcitedefaultmidpunct}
{\mcitedefaultendpunct}{\mcitedefaultseppunct}\relax
\EndOfBibitem
\bibitem[Wyart \emph{et~al.}(2005)Wyart, Silbert, Nagel, and Witten]{wyart2005}
M.~Wyart, L.~E. Silbert, S.~R. Nagel and T.~A. Witten, \emph{Phys. Rev. E},
  2005, \textbf{72}, 051306\relax
\mciteBstWouldAddEndPuncttrue
\mciteSetBstMidEndSepPunct{\mcitedefaultmidpunct}
{\mcitedefaultendpunct}{\mcitedefaultseppunct}\relax
\EndOfBibitem
\bibitem[Head(2007)]{head2007}
D.~Head, \emph{Eur. Phys. J. E}, 2007, \textbf{22}, 151--155\relax
\mciteBstWouldAddEndPuncttrue
\mciteSetBstMidEndSepPunct{\mcitedefaultmidpunct}
{\mcitedefaultendpunct}{\mcitedefaultseppunct}\relax
\EndOfBibitem
\bibitem[Lerner \emph{et~al.}(2012)Lerner, D{\"u}ring, and Wyart]{lerner2012}
E.~Lerner, G.~D{\"u}ring and M.~Wyart, \emph{Proc. Nat. Acad. Sci.}, 2012,
  \textbf{109}, 4798--4803\relax
\mciteBstWouldAddEndPuncttrue
\mciteSetBstMidEndSepPunct{\mcitedefaultmidpunct}
{\mcitedefaultendpunct}{\mcitedefaultseppunct}\relax
\EndOfBibitem
\bibitem[Goldhirsch and Zanetti(1993)]{goldhirsch1993}
I.~Goldhirsch and G.~Zanetti, \emph{Phys. Rev. Lett.}, 1993, \textbf{70},
  1619\relax
\mciteBstWouldAddEndPuncttrue
\mciteSetBstMidEndSepPunct{\mcitedefaultmidpunct}
{\mcitedefaultendpunct}{\mcitedefaultseppunct}\relax
\EndOfBibitem
\bibitem[McNamara and Young(1996)]{mcnamara1996}
S.~McNamara and W.~Young, \emph{Phys. Rev. E}, 1996, \textbf{53}, 5089\relax
\mciteBstWouldAddEndPuncttrue
\mciteSetBstMidEndSepPunct{\mcitedefaultmidpunct}
{\mcitedefaultendpunct}{\mcitedefaultseppunct}\relax
\EndOfBibitem
\bibitem[Essam \emph{et~al.}(1988)Essam, Guttmann, and De'Bell]{essam1988}
J.~W. Essam, A.~J. Guttmann and K.~De'Bell, \emph{J. Phys. A}, 1988,
  \textbf{21}, 3815\relax
\mciteBstWouldAddEndPuncttrue
\mciteSetBstMidEndSepPunct{\mcitedefaultmidpunct}
{\mcitedefaultendpunct}{\mcitedefaultseppunct}\relax
\EndOfBibitem
\bibitem[Vesely(1977)]{vesely1977}
F.~J. Vesely, \emph{J. Comp. Phys.}, 1977, \textbf{24}, 361--371\relax
\mciteBstWouldAddEndPuncttrue
\mciteSetBstMidEndSepPunct{\mcitedefaultmidpunct}
{\mcitedefaultendpunct}{\mcitedefaultseppunct}\relax
\EndOfBibitem
\bibitem[Ewald(1921)]{ewald1921berechnung}
P.~P. Ewald, \emph{Annalen der Physik}, 1921, \textbf{369}, 253--287\relax
\mciteBstWouldAddEndPuncttrue
\mciteSetBstMidEndSepPunct{\mcitedefaultmidpunct}
{\mcitedefaultendpunct}{\mcitedefaultseppunct}\relax
\EndOfBibitem
\bibitem[Barker and Watts(1973)]{barker1973monte}
J.~Barker and R.~Watts, \emph{Mol. Phys.}, 1973, \textbf{26}, 789--792\relax
\mciteBstWouldAddEndPuncttrue
\mciteSetBstMidEndSepPunct{\mcitedefaultmidpunct}
{\mcitedefaultendpunct}{\mcitedefaultseppunct}\relax
\EndOfBibitem
\bibitem[Allen and Tildesley(1989)]{allen1989computer}
M.~P. Allen and D.~J. Tildesley, \emph{Computer Simulation of Liquids}, Oxford
  University Press, 1989\relax
\mciteBstWouldAddEndPuncttrue
\mciteSetBstMidEndSepPunct{\mcitedefaultmidpunct}
{\mcitedefaultendpunct}{\mcitedefaultseppunct}\relax
\EndOfBibitem
\bibitem[Fennell and Gezelter(2006)]{fennell2006ewald}
C.~J. Fennell and J.~D. Gezelter, \emph{J. Chem. Phys.}, 2006, \textbf{124},
  234104\relax
\mciteBstWouldAddEndPuncttrue
\mciteSetBstMidEndSepPunct{\mcitedefaultmidpunct}
{\mcitedefaultendpunct}{\mcitedefaultseppunct}\relax
\EndOfBibitem
\end{mcitethebibliography}

\providecommand*{\mcitethebibliography}{\thebibliography}
\csname @ifundefined\endcsname{endmcitethebibliography}
{\let\endmcitethebibliography\endthebibliography}{}

\end{document}